\documentclass[useAMS,usenatbib,usegraphicx]{mn2e}
\usepackage{bm}
\usepackage{psfig}
\usepackage{amssymb}
\usepackage{amsmath}
\usepackage{multirow}
\usepackage{twoopt}
\usepackage{times}
\usepackage{color}
\usepackage{units}
\usepackage{subfigure}
\usepackage{version}

\usepackage[pdftex,             % ATTENTION : dans ce cas lance pdftex
            breaklinks=true,%
            colorlinks=true,%
            pdfauthor={Bonnivard et al.},%
            pdftitle={Template for manuscripts MNRAS}%
           ]{hyperref}

\makeatletter
\newcommand{\eq}[1]{Eq.~(\ref{#1})}
\newcommand{\beq}{\begin{equation}}
\newcommand{\eeq}{\end{equation}}
\newcommandtwoopt{\citeads}[3][][]{\href{http://adsabs.harvard.edu/abs/#3}{\def\hyper@linkstart##1##2{}\let\hyper@linkend\@empty\citealp[#1][#2]{#3}}}
\newcommandtwoopt{\citepads}[3][][]{\href{http://adsabs.harvard.edu/abs/#3}{\def\hyper@linkstart##1##2{}\let\hyper@linkend\@empty\citep[#1][#2]{#3}}}
\newcommandtwoopt{\citetads}[3][][]{\href{http://adsabs.harvard.edu/abs/#3}{\def\hyper@linkstart##1##2{}\let\hyper@linkend\@empty\citet[#1][#2]{#3}}}
\newcommandtwoopt{\citealpads}[3][][]{\href{http://adsabs.harvard.edu/abs/#3}{\def\hyper@linkstart##1##2{}\let\hyper@linkend\@empty\citealp[#1][#2]{#3}}}
\newcommandtwoopt{\citealtads}[3][][]{\href{http://adsabs.harvard.edu/abs/#3}{\def\hyper@linkstart##1##2{}\let\hyper@linkend\@empty\citealt[#1][#2]{#3}}}
\newcommandtwoopt{\citeyearads}[3][][]{\href{http://adsabs.harvard.edu/abs/#3}{\def\hyper@linkstart##1##2{}\let\hyper@linkend\@empty\citeyear[#1][#2]{#3}}}
\newcommandtwoopt{\citeadsstar}[3][][]{\href{http://adsabs.harvard.edu/abs/#3}{\def\hyper@linkstart##1##2{}\let\hyper@linkend\@empty\citealp*[#1][#2]{#3}}}
\newcommandtwoopt{\citepadsstar}[3][][]{\href{http://adsabs.harvard.edu/abs/#3}{\def\hyper@linkstart##1##2{}\let\hyper@linkend\@empty\citep*[#1][#2]{#3}}}
\newcommandtwoopt{\citetadsstar}[3][][]{\href{http://adsabs.harvard.edu/abs/#3}{\def\hyper@linkstart##1##2{}\let\hyper@linkend\@empty\citet*[#1][#2]{#3}}}
\newcommandtwoopt{\citeyearadsstar}[3][][]{\href{http://adsabs.harvard.edu/abs/#3}{\def\hyper@linkstart##1##2{}\let\hyper@linkend\@empty\citeyear*[#1][#2]{#3}}}
\newcommandtwoopt{\citeauthoradsstar}[3][][]{\href{http://adsabs.harvard.edu/abs/#3}{\def\hyper@linkstart##1##2{}\let\hyper@linkend\@empty\citeauthor*[#1][#2]{#3}}}
\newcommandtwoopt{\citepthesis}[3][][]{\href{http://tel.archives-ouvertes.fr/docs/#3}{\def\hyper@linkstart##1##2{}\let\hyper@linkend\@empty\citep[#1][#2]{#3}}}
\newcommandtwoopt{\citetthesis}[3][][]{\href{http://tel.archives-ouvertes.fr/docs/#3}{\def\hyper@linkstart##1##2{}\let\hyper@linkend\@empty\citet[#1][#2]{#3}}}
\makeatother
\def \new#1{{\textcolor{blue}{#1}}}

\begin{document}
%\title[Short title]{Title}
\title[Jeans analysis: biases and uncertainties]{Spherical Jeans analysis for dark matter indirect detection in dwarf spheroidal galaxies - Impact of physical parameters and triaxiality} 
\author[Bonnivard, Combet, Maurin, Walker]{V. Bonnivard$^{1}$\thanks{E-mails:bonnivard@lpsc.in2p3.fr (VB), combet@lpsc.in2p3.fr (CC), dmaurin@lpsc.in2p3.fr (DM), mgwalker@andrew.cmu.edu (MGW)}, C. Combet$^{1}$, D. Maurin$^{1}$, M. G. Walker$^{2,3}$ \\
  $^1$LPSC, Universit\'e Grenoble-Alpes, CNRS/IN2P3,
     53 avenue des Martyrs, 38026 Grenoble, France\\
  $^2$Department of Physics, Carnegie Mellon University, Pittsburgh, PA 15213, USA\\
  $^3$McWilliams Center for Cosmology, 5000 Forbes Avenue Pittsburgh, PA 15213, USA\\
}
%\vspace{-0.7cm}

\pagerange{\pageref{firstpage}--\pageref{lastpage}} \pubyear{Xxxx}
\date{Accepted Xxxx. Received Xxxx; in original form Xxxx}
\label{firstpage}

\setcounter{tocdepth}{4}
\maketitle
%\tableofcontents

\begin{abstract}
Dwarf spheroidal (dSph) galaxies are among the most promising targets
for the indirect detection of dark matter (DM) from annihilation and/or
decay products.  Empirical estimates of their DM content---and
hence the magnitudes of expected signals---rely on
inferences from stellar-kinematic data.  However, various kinematic analyses can
give different results and it is not obvious which are most reliable.  Using extensive sets of mock data of various
sizes (mimicking `ultra-faint' and `classical' dSphs) and
an MCMC engine, here we investigate biases, uncertainties,
and limitations of analyses based on parametric solutions to the
spherical Jeans equation.  For a variety of functional forms
for the tracer and DM density profiles, as well as the orbital
anisotropy profile, we examine reliability of
estimates for the astrophysical $J$- and $D$-factors for annihilation
and decay, respectively.  For large ($N \ga 1000$) stellar-kinematic samples typical
of `classical' dSphs, errors tend to be dominated by systematics, which
can be reduced through the use of sufficiently general and flexible
functional forms.  For small ($N\la 100$) samples typical of `ultrafaints', statistical uncertainties
tend to dominate systematic errors and flexible models are less
necessary. We define an optimal
strategy that would mitigate sensitivity to priors and other aspects
of analyses based on the spherical Jeans equation. We also find that the assumption of spherical symmetry
can bias estimates of $J$ (with the 95\% credibility intervals not encompassing the true $J$-factor) when the object
is mildly triaxial (axis ratios $b/a=0.8$, $c/a=0.6$). A concluding table summarises the typical error budget and biases for
the different sample sizes considered. 
\end{abstract}


%________________________________________________________________________
\section{Introduction\label{sec:intro}}

Gamma-ray observations of dark matter (DM)-rich systems have proven
a competitive and complementary path to other DM indirect detection
approaches. Exotic $\gamma$-ray signals have been looked for at the Galactic
centre \citepads{2011PhLB..697..412H,2011PhRvL.106p1301A,2012JCAP...08..007W,2014arXiv1402.6703D}, in clusters of galaxies \citepads{2010JCAP...05..025A,2012ApJ...757..123A}, in Galactic dark
clumps \citepads{2011arXiv1109.5935N,2012ApJ...747..121A} and in dwarf
spheroidal (dSph) galaxies orbiting the Milky Way \citepads{2011PhRvL.107x1302A,2011PhRvL.107x1303G,2014PhRvD..89d2001A}.
The latter are interesting targets because of their proximity, potentially high DM densities and small astrophysical $\gamma$-ray
backgrounds \citepads{1990Natur.346...39L,2004PhRvD..69l3501E}.  As a
result, dSphs can provide a crucial check on dark-matter interpretations of gamma-ray signals from high-background environments such as the Galactic
centre.

Recent Fermi-LAT results \citepads{2011PhRvL.107x1303G,2011PhRvL.107x1302A,2014PhRvD..89d2001A} show that indirect detection
in dSph galaxies can play a significant role in constraining the
nature of DM.  Given the instrument sensitivity and lack of an obvious
signal, limits on $\langle\sigma v\rangle$ are now reaching the canonical $3\times
10^{-26}$\;cm$^3$\;s$^{-1}$ below which most supersymmetric DM models
lie \citepads{2010ARA&A..48..495F}. It therefore becomes critical to review how the most
common underlying assumptions made in deriving these limits may
affect or bias the results. Rather than on particle physics aspects,
this paper will focus solely on the
astrophysical assumptions (e.g., DM profile parametrisation, velocity anisotropy and light profile modelling) needed to compute
astrophysical $J$ and $D$-factors (and their uncertainties). The $D$-factor allows
constraints on the lifetime $\tau$ of decaying DM \citepads{2009PhRvD..80b3506E,2010JCAP...07..023P}, while the $J$-factor is required in computing limits on $\langle\sigma v\rangle$ for annihilating DM. The body of this paper
focuses mainly on the latter, while the corresponding results for the
$D$-factor are similar and discussed in appendices.  

To estimate the $J$ and $D$-factors, different authors rely on different priors in order to recover the mass and density
profiles of the dSph galaxies. In many studies, dark matter density
profiles are fitted directly to the kinematic data of the dSph under
scrutiny \citepads{2006PhRvD..73f3510B,2007PhRvD..76l3509S,2009JCAP...01..016B,2009MNRAS.399.2033P,2009A&A...496..351P}
while others use `cosmological priors' from structure formation
simulations \citepads{2007PhRvD..75h3526S,2009JCAP...06..014M}. These priors are uncertain in the absence of a complete understanding of the role of baryons, and can bias the results
for systems in which
little kinematic information is
available, such as for ultra-faint dSph galaxies. The latter, such as
Segue~I, Willman~I or Coma Berenices, are playing an increasing role in setting limits from $\gamma-$ray
indirect searches, their short distances ($\sim 20-40$\;kpc) allowing for higher values
of $J$ and therefore more stringent constraints on $\langle \sigma v
\rangle$. Only a few studies do \emph{not} use strong priors for the DM profiles
\citepads{2009PhRvD..80b3506E,2011ApJ...733L..46W,2011MNRAS.418.1526C}, allowing for a
data-driven analysis which provides larger, hence more conservative,
error bars.

In this paper, we focus on
data-driven analyses that rely on parametric solutions to the
spherical Jeans equation (e.g., \citealtads{2006ApJ...652..306S,2010ApJ...710..886W,2011MNRAS.418.1526C}). While
 \citetads{2011MNRAS.418.1526C} have previously examined some limitations
 inherent to this approach regarding the analysis of relatively
 luminous, `classical' dSphs, here we use simulated data sets of
 various sizes in order to compare the relative importance of
 systematic errors for analyses of classical and `ultra-faint' dSphs.  
In each case, we identify which modelling assumptions are most
critical in terms of $J$ and $D$-factor determination, and suggest {\it safer}
options whenever possible. In a forthcoming article, we will apply the
findings of this analysis to real data in order to provide 
robust $J$ and $D$-factor values for classical and ultra-faint dSph galaxies.

This paper is organised as follows. In Section~\ref{sec:method}, we cover all the ingredients needed
in this study, namely the spherical Jeans equation, the
computation of astrophysical $J$ and $D$-factors,
the Markov Chain Monte Carlo (MCMC) algorithm and the description of the simulated
data used. In Section~\ref{sec:starter}, we highlight differences of our analysis
w.r.t. that of \citetads{2011MNRAS.418.1526C}. 
In Section~\ref{sec:jeans_ref}, we run the Jeans analysis
in the ideal case where the light and velocity anisotropy profiles are
known, allowing us to study the impact of the DM density profile
parametrisation and evaluate the minimal uncertainties related to the
sample size. The impact of the modelling of the velocity anisotropy and the light profile are then 
discussed in Sections~\ref{sec:anisotropy} and \ref{sec:light}, as well as the biases
introduced when assuming spherical symmetry for triaxial DM halos
(Section~\ref{sec:triaxiality}). Finally, we discuss our results and
conclude in Section~\ref{sec:conclusions}.

%________________________________________________________________________
\section{Jeans analysis, $J$ factors, MCMC, and mock data}
\label{sec:method}

%________________________________________________________________________
\subsection{Jeans analysis of dSph kinematics}
\label{sec:jeans}

Estimation of DM indirect detection signals from dSph galaxies requires
knowledge of their mass density profiles, which have been
particularly investigated in the last decade thanks to the increase of
kinematic measurements. Different techniques
(Jeans models, Schwarzschild models, distribution function
models, Made-to-Measure models, etc.)
have been developed to infer the mass profile from stellar kinematic
data and we refer the reader to recent reviews (and references
therein) by \citetads{2013pss5.book.1039W,2013NewAR..57...52B,2013PhR...531....1S}
for comprehensive descriptions of these methods. Here, we focus
solely on analyses that use parametric functions for velocity
anisotropy, tracer and DM
density profiles in order to solve the spherically-symmetric Jeans
equation, which has been widely applied to dSph stellar 
kinematics \citepads{2007PhRvD..75h3526S,2010PhRvD..82l3503E,2011ApJ...733L..46W,2011MNRAS.418.1526C}.

\paragraph*{Spherical Jeans equation and solution}
Assuming that dSphs are collisionless systems in steady-state, we can apply the collisionless Boltzmann equation to
their stellar phase space distribution function.  The Jeans equation
is obtained by integrating moments of this distribution function.  Further assuming spherical
symmetry and negligible rotational support, the second-order Jeans
equation reads \citepads{2008gady.book.....B}:
\begin{equation}
\frac{1}{\nu}\frac{d}{dr}(\nu \bar{v_r^2})+2\frac{\beta_{\rm ani}(r)\bar{v_r^2}}{r}=-\frac{GM(r)}{r^2},
\label{eq:jeans}
\end{equation}
where $\nu(r)$, $\bar{v_r^2}(r)$, and $\beta_{\rm ani}(r)\equiv
1-\bar{v_{\theta}^2}/\bar{v_r^2}$ describe the 3-dimensional density,
the radial velocity dispersion, and the velocity anisotropy of
the stellar component, respectively. dSphs tend to be strongly DM-dominated, such that the contribution
of stars to the enclosed-mass, $M(r)$, can be neglected. It follows that
\begin{equation}
 \label{eq:mass}
 M(r) = 4\pi \int_{0}^{r}\rho_{\rm DM}(s)\,s^{2}ds,
\end{equation}
with $\rho_{\rm DM}(r)$ the DM density profile. The generic solution to the Jeans equation is
\begin{equation}
 \label{eq:jeans_sol}
 \nu(r)\bar{v_r^2}(r) = \frac{1}{f(r)}\times \int_r^{+\infty} f(s) \nu(s) \frac{GM(s)}{s^2}\, ds            
\end{equation}
with\footnote{Note that the variable $r_1$ in Eq.~(\ref{eq:fr}) is mute and leads
to a normalisation factor (after integration) that cancels out in the Jeans solution~(\ref{eq:jeans_sol}).}
\begin{equation}
\label{eq:fr}
     f(r) = f_{r_1} \exp\left[\int_{r_1}^r \frac{2}{t}\beta_{\rm ani}(t)\, dt \right]. 
\end{equation}

\paragraph*{Solution for projected quantities}
In practice, the observables are projected quantities on the sky. For
spherically symmetric systems, projection (resp. de-projection) of a quantity
$f(r)$ into $F(R)$ is given by the Abel (resp. inverse Abel) transform 
\begin{equation}
 \label{eq:abel}
   \!\!F(R)\!=\!2\! \int_R^{+\infty} \!\!\frac{f(r)\,rdr}{\!\sqrt{r^2\!-\!R^2}} 
   \quad\biggl(\!{\rm resp.~}
   f(r)\!=\!\!\int_r^{+\infty} \!\!\frac{dF}{dR} \frac{-dR}{\pi\sqrt{\!R^2\!-\!r^2\!}}\!\biggr ),\!\!
\end{equation}
where $R$ is the projected radius. Projecting Eq. (\ref{eq:jeans_sol}) along the line of sight (l.o.s.), the DM mass profile $M(r)$ relates to the projected velocity dispersion, $\sigma_p(R)$,
\begin{equation}
  \sigma_p^2(R)=\frac{2}{I(R)}\displaystyle \int_{R}^{\infty}\biggl (1-\beta_{\rm ani}(r)\frac{R^2}{r^2}\biggr ) 
  \frac{\nu(r)\, \bar{v_r^2}(r)\,r}{\sqrt{r^2-R^2}}\mathrm{d}r,
  \label{eq:jeansproject}
\end{equation}
with $I(R)$ the projected light profile (or surface brightness),
\begin{equation}
   I(R)\!=\!2\! \int_R^{+\infty} \!\!\!\!\frac{\nu(r)\,r\,dr}{\sqrt{r^2-R^2}} .
  \label{eq:lightproject}
\end{equation}
While $I(R)$ and $\sigma_p(R)$ can be directly measured, the l.o.s. velocity dispersion profiles
provide little information about the anisotropy $\beta_{\rm ani}(r)$. 
A common approach, which we examine critically here, is to compute $\sigma_p(R)$ via Eq.~(\ref{eq:jeansproject}), adopting parametric models for the projected
stellar density $I(R)$, the DM profile $\rho_{\rm DM}(r)$, and the
anisotropy profile $\beta(r)$, and to find the best-fit parameters that
reproduce the measured velocity dispersion
profile $\sigma_{obs}(R)$.

\subsubsection{Dark matter profiles}
\label{subsubsec:dm_prof}
We use two families of DM profiles in this study:
  \begin{itemize}
    \item {\em Zhao} profiles \citepads{1990ApJ...356..359H,1996MNRAS.278..488Z}.
     This family of profiles requires three slope parameters
     $(\alpha,\beta,\gamma)$, the values of which allow the
     recovery of several DM profiles used in the literature (e.g., core, NFW, Moore).
     It is parameterized as
        \begin{equation}
           \rho_{\rm DM}^{\rm Zhao}(r)=\frac{\rho_s}{(r/r_s)^\gamma \cdot [1+(r/r_s)^\alpha]^{(\beta-\gamma)/\alpha}}\;,
           \label{eq:rho_dm_zhao}
        \end{equation}
     with $\rho_s$ the normalisation, $r_s$ the scale radius,
     $\gamma$ the inner slope, $\beta$ the outer slope, and $\alpha$ the transition slope.
     Note that with this definition, $\rho_s = \rho(r_s) \cdot 2^{(\beta-\gamma)/\alpha}$.

    \item {\em Einasto} profiles (e.g., \citealtads{2006AJ....132.2685M}). This second family
    of profiles, using a logarithmic inner slope, was found to better
    fit DM halos in numerical simulations \citepads{2004MNRAS.349.1039N,2008MNRAS.391.1685S}:
        \begin{equation}
           \rho_{\rm DM}^{\rm Einasto}(r)=\rho_{-2} \exp\left\{-\frac{2}{\alpha}\left[\left(\frac{r}{r_{-2}}\right)^\alpha -1\right]\right\}\;,
        \label{eq:rho_dm_einasto}
        \end{equation}
     where $r_{-2}$ is the radius for which the logarithmic slope equals $-2$, and $\alpha$ controls the logarithmic slope
     sharpness.
  \end{itemize}

\subsubsection{Light profiles}
\label{subsubsec:light_prof}
Stellar surface brightnesses of dSphs are generally fitted using
\citetads{1911MNRAS..71..460P}, \citetads{1962AJ.....67..471K}, or \citetads{1968adga.book.....S} profiles
(e.g., \citealtads{1995MNRAS.277.1354I}), but exponential and Zhao (for the 3D stellar density) profiles
have also been considered. De-projection (or projection in the Zhao case) of these profiles rely
on the Abel transform of \eq{eq:abel}, which may be analytically
computed in some cases.
We give below the adopted parametrisations and refer the reader to the associated references for
de-projected analytical formulae (for the Plummer, exponential and
King cases).
   \begin{itemize}
      \item The {\em Plummer} profile
        \citepads{1911MNRAS..71..460P} reads
         \begin{equation}
            I^{\rm Plummer}(R)=\frac{L}{\pi r_{\rm half}^2}\frac{1}{[1+R^2/r_{\rm half}^2]^2},
            \label{eq:light_plummer2D}
         \end{equation}
         with $L$ the total luminosity, and $r_{\rm half}$ the projected half-light radius.  
      
      \item The {\em exponential} profile \citepads{2009MNRAS.393L..50E}
         is parameterized as
         \begin{equation}
            I^{\rm exp}(R)=I_0 \cdot \exp\left(-\frac{R}{r_c}\right)\,,
            \label{eq:light_exp2D}
         \end{equation}
         with $I_0$ the normalisation, and $r_c$ the scale radius of exponential decrease.

      \item The {\em King} profile
        \citepads{1962AJ.....67..471K} is written as
         \begin{equation}
            I^{\rm King}(R)=I_0 \cdot\left[\left(1+\frac{R^2}{r_c^2}\right)^{-1/2}
                   -\left(1+\frac{r_{\rm lim}^2}{r_c^2}\right)^{-1/2}\right],
            \label{eq:light_king2D}
         \end{equation}
         with $I_0$ the normalisation, $r_c$ the `core' radius and $r_{\rm
           lim}$ the maximum radius beyond which the density goes
         to zero.
       
      \item The {\em S\'ersic} profile
        \citepads{1968adga.book.....S,1997A&A...321..111P} reads
         \begin{equation}
            I^{\rm S\acute{e}rsic}(R)=I_0 \cdot \exp\left\{-b_n \cdot \left[ \left(\frac{R}{r_c}\right)^{1/n} - 1 \right]\right\},
            \label{eq:light_sersic2D}
         \end{equation}
         with $b_n = 2n - 1/3 + 0.009876/n$, $I_0$ the normalisation, 
         $n\gtrsim 0.5$ an irrational number (controlling the sharpness of the
         logarithmic decrease), and $r_c$ a scale radius.

      \item Finally, the {\em Zhao} profile
        \citepads{1990ApJ...356..359H,1996MNRAS.278..488Z} given by
        Eq.~(\ref{eq:rho_dm_zhao}) is here applied to the 3D
        (i.e. unprojected) light profile,
         \begin{equation}      
           \nu^{\rm Zhao}(r)=\rho_{\rm DM}^{\rm Zhao}(r)\;.
         \end{equation}
        In this case, the light profile is analytical for the 3D
        density profile $\nu(r)$ but has to be numerically projected
        along the l.o.s. to provide the surface brightness $I(r)$.
   \end{itemize}
   
As already mentioned, we assume that DM dominates the gravitational potential at all radii
(all measured dSphs have central mass-to-light ratios $\ga 10$, e.g., \citealtads{1998ARA&A..36..435M}), so that the value of the normalisation factor ($L$ or $I_0$) has no bearing
on the analysis.

\subsubsection{Velocity anisotropy profiles}
\label{subsubsec:ani_prof}
We recall that the velocity anisotropy profile is given by a combination of the radial and tangential
velocity dispersion:
   \begin{equation}
      \beta_{\rm ani}(r)\equiv 1-\frac{\bar{v_{\theta}^2}(r)}{\bar{v_r^2}(r)}\,.
      \label{eq:beta_ani}
   \end{equation}
Due to the lack of observational constraints on this quantity, the first anisotropy profiles discussed in
the literature were based on analytical studies aiming at building dynamical models (in spherical
symmetry) with self-consistent stellar phase space distribution
functions.  Many such models have simple anisotropy profiles that are
either constant or change from isotropic near the centre to radial at
large radius (e.g., \citealtads{1979PAZh....5...77O,1985AJ.....90.1027M}, see below).
More recently, indications of radial anisotropy
in the outer regions of DM halos have been obtained from numerical simulations
(e.g., \citealtads{2004MNRAS.352..535D}). In the inner region, a strong anisotropy 
can be generated by dynamical formation and evolution processes. To better describe these
profiles, \citetads{2007A&A...471..419B} introduced a technique to construct dynamical
models with arbitrary mass density and anisotropy profiles. These
three different families of anisotropy profiles are described below and will be explored in
\S\ref{sec:anisotropy}.
   \begin{itemize}
      \item The constant anisotropy modelling (e.g., used by
        \citealtads{2011MNRAS.418.1526C}) simply reads
         \begin{equation}
            \beta_{\rm ani}^{\rm Cst}(r)=\beta_0.
            \label{eq:beta_constant}
         \end{equation}
      
       \item The {\em Osipkov - Merritt} profile
         \citepads{1979PAZh....5...77O,1985AJ.....90.1027M} is
         parameterized as
         \begin{equation}
          \beta_{\rm ani}^{\rm Osipkov}(r)=\frac{r^2}{r^2+r_a^2},
            \label{eq:beta_osipkov}
         \end{equation}
         with a single free scale parameter $r_a$ which locates the transition 
         from $\beta_{\rm ani}=0$ in the inner parts (isotropic) to 1 at large radii (full radial
         anisotropy).
    
       \item The {\em Baes \& van Hese} profile
         \citepads{2007A&A...471..419B} is more general and is written as
         \begin{equation}
             \beta_{\rm ani}^{\rm Baes}(r) =\frac{\beta_0 + \beta_\infty (r/r_a)^\eta}{1+(r/r_a)^\eta}\,,
            \label{eq:beta_baes}
         \end{equation}
         where the four parameters are the central anisotropy $\beta_0$, the
         anisotropy at large radii $\beta_\infty$, and the sharpness of the transition
         $\eta$ at the scale radius $r_a$. The Osipkov-Merritt
         profile is recovered when using $\beta_0=0$, $\beta_\infty=1$, and $\eta=2$.
      \end{itemize}
 
\subsubsection{Technicalities (for the Jeans solution)}
As seen from Eqs.~(\ref{eq:jeans_sol}), (\ref{eq:fr}), and (\ref{eq:jeansproject}), the
solution of the projected Jeans equation requires in principle three
successive integrations.
However, as shown by \citetads{2005MNRAS.363..705M,2006MNRAS.370.1582M}, the calculation of
Eq.~(\ref{eq:jeansproject}) for constant and Osipkov-Merritt anisotropy profiles (and many
others) can be cast as a single integration. This tremendously speeds
up the calculation with respect to the Baes \& van Hese profile case, for which no shortcut was found in the literature.

All the Jeans analyses presented here are performed with a new module of the 
{\tt CLUMPY}\footnote{\url{http://lpsc.in2p3.fr/clumpy}} code~\citepads{2012CoPhC.183..656C},
which was developed and used to calculate $J$ and $D$ factors  
\citepads{2011ApJ...733L..46W,2011MNRAS.418.1526C,2012MNRAS.425..477N,2012PhRvD..85f3517C,2012A&A...547A..16M}. 
The new Jeans analysis module was validated by systematic cross-checks with the results
(obtained with a different code and MCMC engine) of \citetads{2011MNRAS.418.1526C}; it
will be made publicly available in the second release of {\tt CLUMPY} (Bonnivard et al., in prep.).

%________________________________________________________________________
\subsection{Astrophysical factor for annihilation and decay}
\label{sec:mcmc}

For a given dSph galaxy, using the DM density profile obtained from the Jeans analysis, we
calculate the astrophysical $J$-factor
(resp. $D$-factor) needed in the computation of the exotic
$\gamma$-ray signal from DM annihilation (resp. decay). This
astrophysical factor corresponds to the integration along the
l.o.s. of the DM density squared (resp. DM density) over the solid
angle $\Delta\Omega$ and reads \citepads{1998APh.....9..137B}
 \begin{equation}
      \!\!J \!=\! \!\int\!\!\!\!\!\int\!\! \rho_{\rm DM}^2 (l,\Omega) \,dld\Omega
      \quad\left({\rm \!resp.~} D \!=\!\! \int\!\!\!\!\!\int \!\!\rho_{\rm DM}(l,\Omega) \,dld\Omega\right),
      \label{eq:J}
 \end{equation}
where the solid angle and integration angle $\alpha_{\rm int}$ are related by
\[
  \Delta\Omega = 2\pi\times[1-\cos(\alpha_{\rm int})] \,.
\]

The $J$-factor (resp. $D$-factor) is useful knowledge as it allows us
to rank the DM targets (in terms of their expected $\gamma$-ray flux) independently of the details
and couplings of the underlying particle physics model. 

For annihilating DM, the well-established presence of substructures in DM halos
\citepads[e.g.,][]{2008MNRAS.391.1685S} can boost the signal.
However, the smaller the host halo mass, the less boosted the signal is \citepads[e.g.,][]{2014MNRAS.442.2271S}.
Taking generic configurations of the substructure distribution and the host halo parameters,
\citetads{2011MNRAS.418.1526C} found that DM substructures in dSph galaxies do not significantly boost
the annihilation signal. In a completely different approach and in the context of the Milky Way DM halo,
the role of fine-grained phase-space structures (halos and streams) was studied in details in \citetads{2011MNRAS.413.1419V},
who found a very small impact on the annihilation signal and also on the variance of the mean density. These results
motivate the description of the total DM halo of dSph galaxies as a smooth density, whose reconstructed profile
(by means of the Jeans analysis) can be interpreted as the sum of the smooth halo plus the mean density of
substructures.

%__________
\subsection{Likelihood, MCMC, posteriors, and credibility intervals}
\label{subsec:mcmc}
For a given choice of DM density and velocity anisotropy parameters,
we compare the projected velocity dispersion profile
$\sigma_{p}(R)$ calculated from Eq. (\ref{eq:jeansproject}) to the
observed one $\sigma_{\rm obs}(R)$. The latter is reconstructed from
individual stellar velocities (see next section), while the projected light profile $I(R)$|used to compute the velocity dispersion|is fitted separately (see section \ref{sec:light}). 

\paragraph*{Binned analysis} We use a binned likelihood function
\begin{equation}
  \mathcal{L}\!=\! \prod_{i=1}^N \frac{1}{\sqrt{2\pi}\,\Delta_{\sigma_{\rm obs}}(R_i)}\exp\biggl [\!-\frac{1}{2}\biggl (\!\frac{\sigma_{\rm obs}(R_{i}) \!-\!\sigma_p(R_i)}{\Delta_{\sigma_{\rm obs}}(R_i)}\biggr )^{2}\biggr ],
  \label{eq:likelihood}
\end{equation}
where $\Delta_{\sigma_{\rm obs}}(R_i)$ is the error on the velocity dispersion at the projected radius $R_{i}$. Other likelihood functions for binned or unbinned analyses have also been considered in the literature (e.g., \citealtads{2007PhRvD..75h3526S}), and their impact will be discussed elsewhere.

\paragraph*{Markov chain Monte Carlo (MCMC)}
In order to efficiently explore the large parameter space (up to 9
free parameters for a Zhao DM profile and a Baes \& van Hese
anisotropy profile), we employ an MCMC technique \citep{Neal93}. Based
on Bayesian parameter inference, this method allows for an efficient
sampling of the
posterior probability density function (PDF) of a vector of parameters
$\boldsymbol{\theta}$ using Markov chains. From the PDFs, credibility
intervals for any quantity of interest are easily computed. In this
analysis, we use the Grenoble Analysis Toolkit ({\tt GreAT}): it is a
modular {\tt C++} framework dedicated to statistical data analysis
(\citealtads{2011ICRC....6..260P}, \citealt{Putze:2014aba}), originally developed for cosmic-ray physics
studies
\citepads{2009A&A...497..991P,2010A&A...516A..66P,2011A&A...526A.101P}.
It relies on the Metropolis-Hastings algorithm
to sample the
posterior distributions
(\citealtads{1953JChPh..21.1087M};~\citealt{hastings70}). The number of chains used (with typically
25000 points/chain) depends on the number of free parameters and on
the size of the mock sample (see next section); it typically varies
from a few to more than a hundred chains, in order to gather a
sufficient number of points to calculate credibility intervals. 
The proposal function used in this work is a multivariate Gaussian distribution.
We provide in Appendix~\ref{app:MCMC_opt} specific combinations of the DM parameters
to improve the efficiency of the MCMC in the context of the Jeans analysis.

\paragraph*{Posterior distributions}
The posterior distributions are obtained after several post-processing
steps (burn-in length removal, correlation length
estimation and thinning of the chains) required to ensure the
insensitivity of the result to the initial conditions and independent sample selection.
Note that in a Bayesian analysis, the priors used for each
parameter can strongly impact the results, especially if the
parameters are loosely  constrained. We restrict this study to
uniform priors, and the extensive use of mock data allows us to
define `optimal' ranges, for instance for DM profile parameters (see Table~\ref{table:priors_DM}), as further discussed in
Sections \ref{sec:jeans_ref} and \ref{sec:anisotropy}.
\begin{table}
\begin{center}
\caption{Range of uniform priors used for the DM profile parameters. Adding the conditions reported in the last column leads to better (more constrained) credibility intervals.}
\label{table:priors_DM}
\begin{tabular}{ccccc}\hline\hline
 DM profile & Parameter & \!\!Prior\!\! & \multicolumn{2}{c}{Added condition} \\
 \hline
\multirow{4}{*}{``Zhao"} & $\log_{10}(\rho_{s}/\text{M}_{\odot}~\text{kpc}^{-3})$ & $[5,13]$ & - & \\
\multirow{4}{*}{Eq.~(\ref{eq:rho_dm_zhao})} & \!\!\!\!\!\!$\log_{10}(r_{s}/\text{kpc})$\!\!\!\!\!\! & $[-3,1]$ & $r_{s} \geq r_{s}^{*}$ & \!\!\!\!\!\!(\S \ref{sec:rs_cut}) \\
                   & $\alpha$               & $[0.5,3]$ & - &  \\
                   & $\beta$                & $[3,7]$   & - & \\\vspace{5mm}
                   & $\gamma$               & $[0,1.5]$ & $\gamma \leq 1$ & \!\!\!\!\!\!(\S \ref{sec:anisotropy}) \\%\cline{2-5}
\multirow{2}{*}{``Einasto"} & \!\!\!\!\!\!$\log_{10}(\rho_{-2}/\text{M}_{\odot}~\text{kpc}^{-3})$\!\!\!\!\!\! & $[5,13]$ & - & \\
\multirow{2}{*}{Eq.~(\ref{eq:rho_dm_einasto})}  & $\log_{10}(\text{r}_{-2}/\text{kpc})$ & $[$-$3,1]$ & $r_{-2} \geq r_{s}^{*}$ & \!\!\!\!\!\!(\S \ref{sec:rs_cut}) \\
                   & $\alpha$               & \!\!\!$[0.05,1]$\!\!\! & $\alpha \geq 0.12$ & \!\!\!\!\!\!(\S \ref{sec:anisotropy}) \\\hline
\end{tabular}
\end{center}
%\vspace{-3mm} 
\end{table}

\paragraph*{PDF and credibility intervals (CIs)}
The outputs of the MCMC analysis described above are PDFs and
correlations of the free parameters of the study. Credibility
intervals (CIs) for any quantity deriving from these parameters are obtained by filling an histogram of this quantity for each independent sample of the Markov chains. The credibility limit $X^{1-\theta}$ of the quantity $X$ with a probability $1-\theta$ is defined to be 
   \begin{equation}
      \int_{-\infty}^{X^{1-\theta}}\mathcal{P}(X)dX = 1 - \theta.
      \label{eq:cl}
   \end{equation}
The value $(\theta = 0.5)$ gives the median. We also use the $95\%$ quantile of the PDFs to display CIs on $\sigma_p(R)$, $\rho_{\rm DM} (r)$, $J(\alpha_{\rm int})$, and  $D(\alpha_{\rm int})$, for any radius/integration angle.
In the absence of a DM signal from a dSph galaxy, the most conservative choice when putting limits on an annihilation cross-section is to use the lower CIs on the $J$-factor; however the entire credibility interval is often used to propagate the $J$-factor uncertainties to the limits (see e.g. \citealtads{2014PhRvD..89d2001A}). After a possible discovery, the entire credibility interval is also needed to recover the uncertainties on the reconstructed cross-section. We will mainly focus here on the upper CIs, which are more sensitive to the different assumptions of the Jeans analysis, but will also comment on the behaviour of the lower CIs.

\paragraph*{Criterion to select the best analysis setup}
We remind that the ultimate goal of the analysis is to determine the
best setup (or best estimator) to calculate the $J$-factors of Milky Way's
dSph galaxies. This selection is made on mock data (see next section). The different estimators used are associated to different assumptions made for the input parameters (dark matter profile, light and anisotropy profile, see previous section). A statistically sound approach would be to characterise each estimator (bias, mean square error, consistency, etc.), by running many realisations of the data for each setup. This is however too computationally demanding. The approach followed here is to use a restricted (but still sizeable) set of samples to study several setups and the performances of the associated estimators (namely the binned likelihood from Eq.~(\ref{eq:likelihood}) for different models/configurations). Following \citetads{2011MNRAS.418.1526C}, we use the median values as the estimator of the true value (see their Appendix~F). The relative merit of the different setups tested is assessed by comparing the distribution of distances of the median and CIs to the true value of the mock data at hand. In particular, any trend for a systematic offset of the median distribution to the true values will be denoted `bias' (which is a small misuse of the word in the statistical context). A particular setup for an analysis will be said to be strongly biased if the reconstructed quantities do not encompass the true value at the 95\% level. In the following, we choose the best estimator to be the least biased (if possible) and the one leading to the smallest CIs.

%__________
\subsection{Mock data}
\label{subsec:mock}

\begin{figure}
\begin{center}
\includegraphics[width=\linewidth]{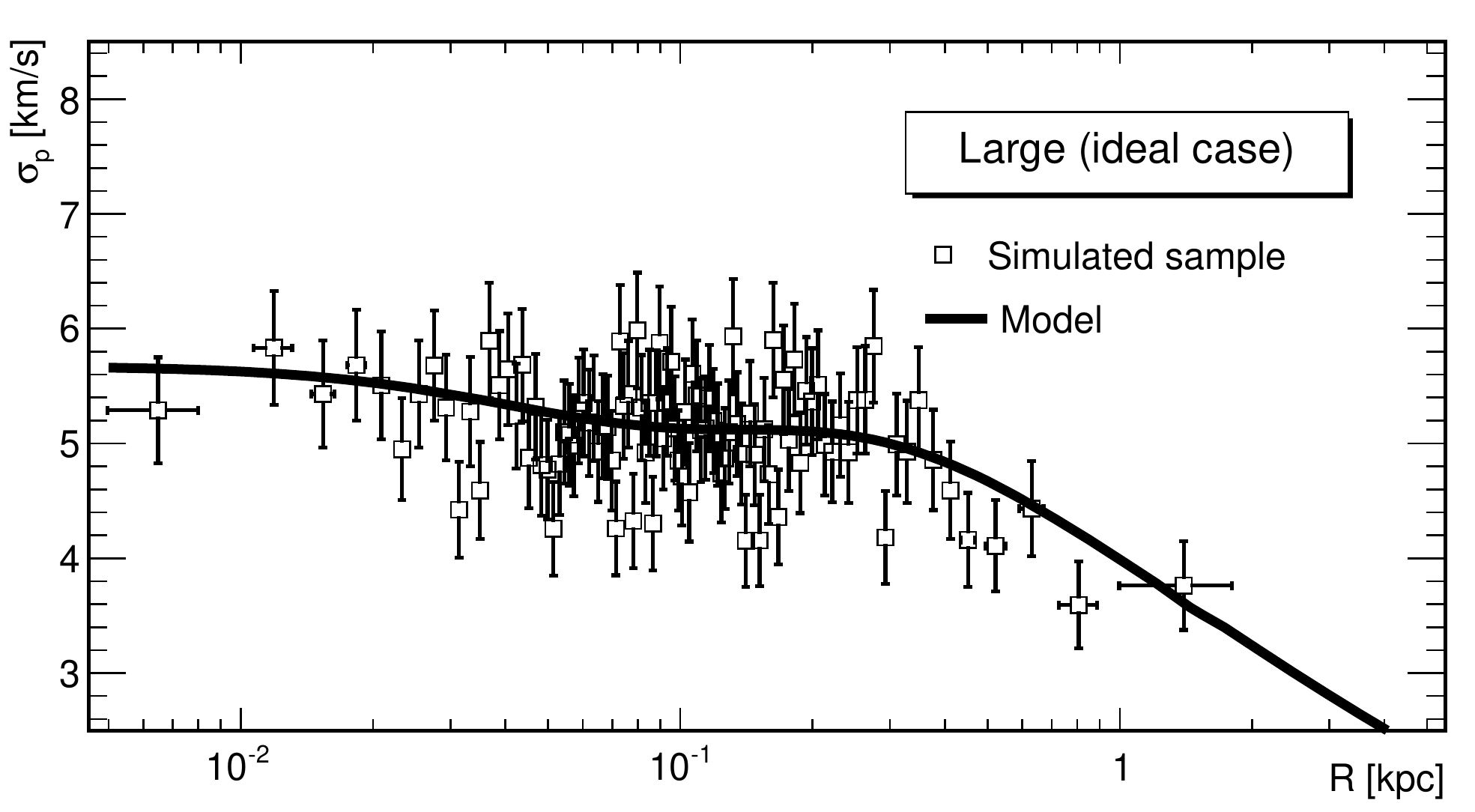}
\includegraphics[width=\linewidth]{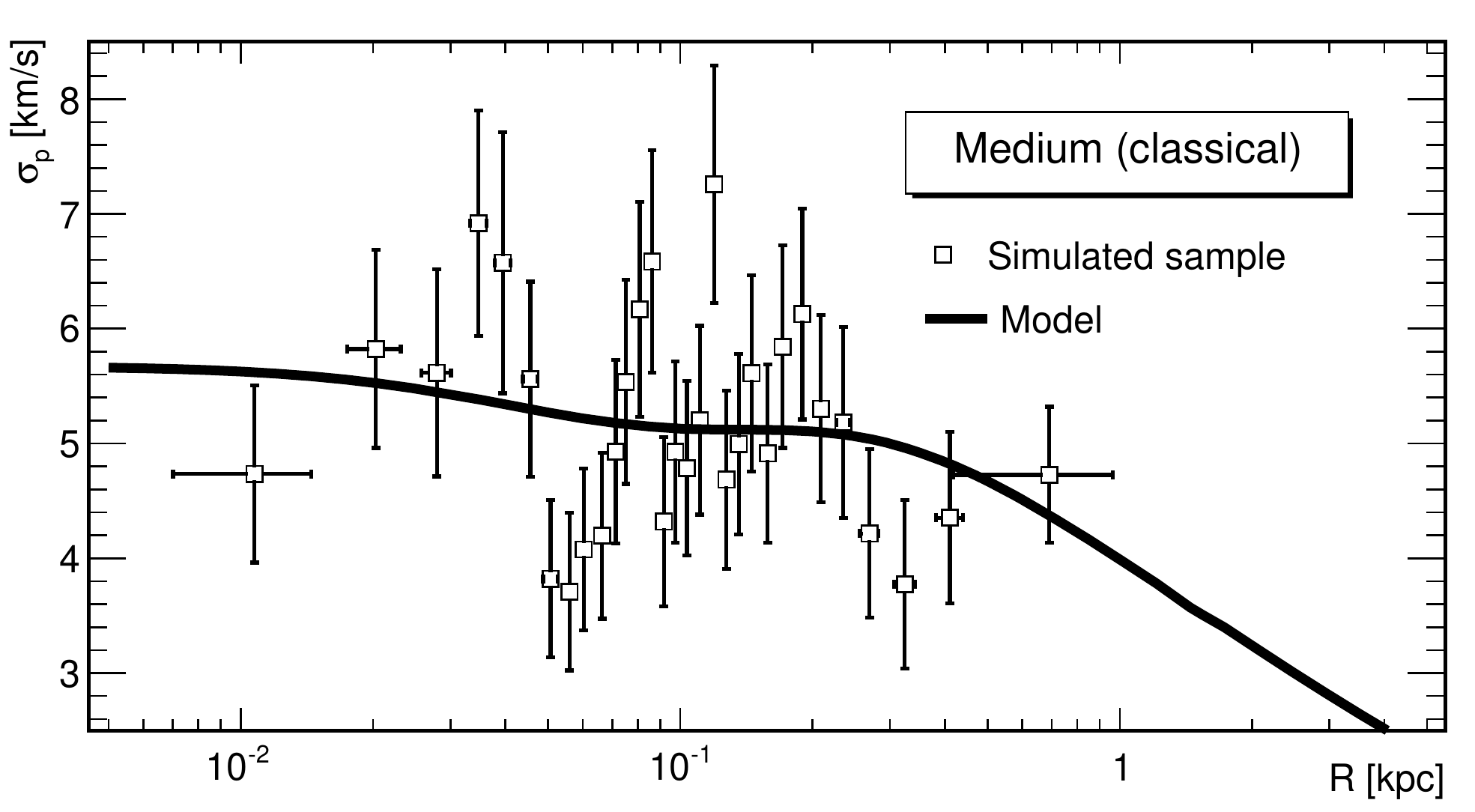}
\includegraphics[width=\linewidth]{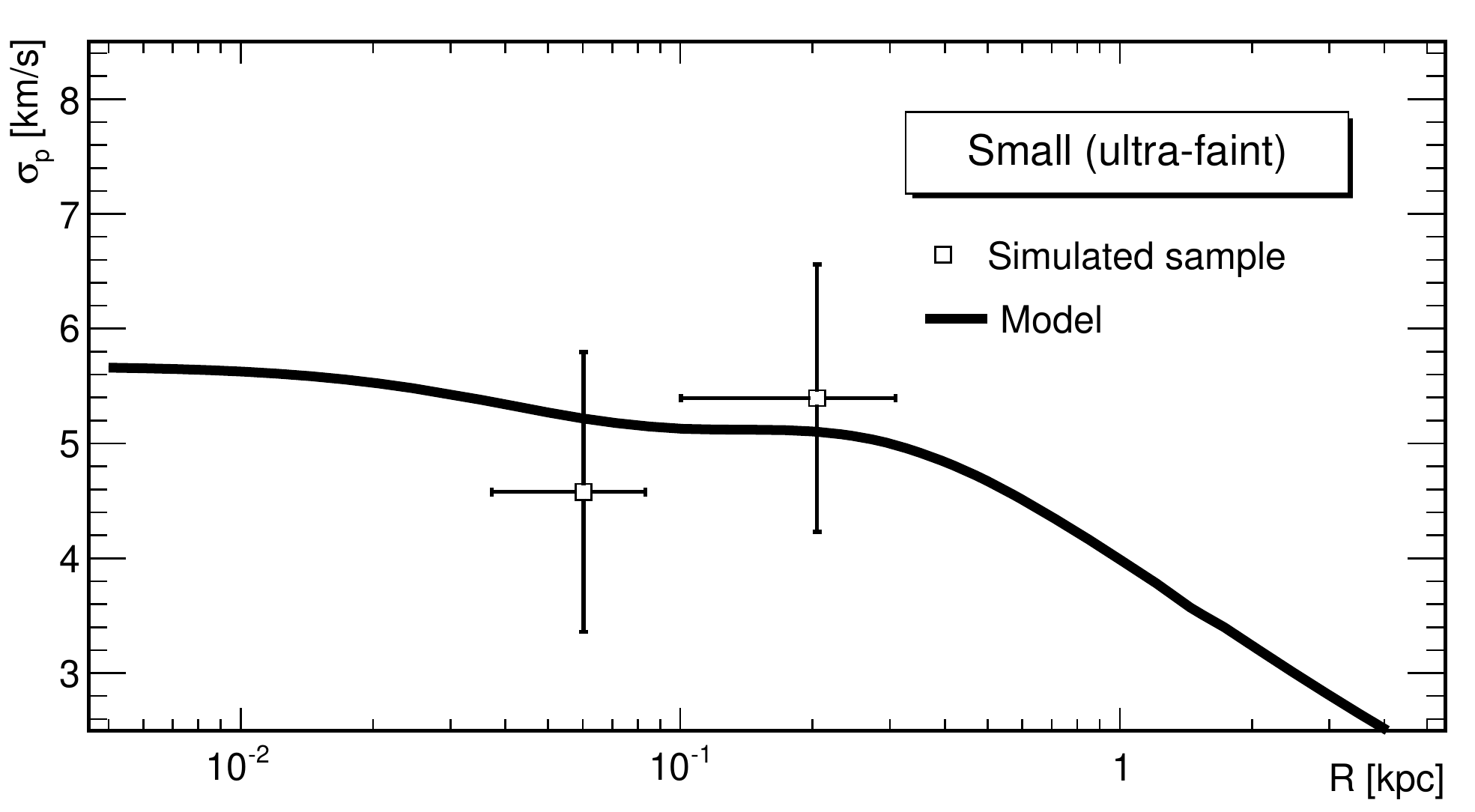}
\caption{Velocity dispersion profile obtained for 3 sample sizes of a same model. {\em Top}: large sample, with 10000 stars (ideal case). {\em Middle}: medium sample, with 1000 stars, corresponding to a classical dSph. {\em Bottom}: small sample, with 30 stars, mimicking an ultra-faint dSph.}
\label{fig:dispprofiles}
\end{center}
%\vspace{-3mm}
\end{figure}
In order to examine the performance of analyses that employ various
assumptions, we analyse three suites of mock data sets that consist of stellar positions
and velocities drawn from static distribution functions that satisfy the
collisionless Boltzmann equation.  The first suite is the same one
used previously by \citetads{2011ApJ...733L..46W} and
\citetads{2011MNRAS.418.1526C}.  Briefly, it randomly samples 
distribution functions of the form $L^{-2\beta_{\rm ani}}f(\epsilon)$, where
$\epsilon$ is energy and $L$ is angular momentum.  For given choices
of $\rho_{\rm DM}(r)$ and $\nu(r)$, the distribution functions are
computed numerically using the method of \citetads{1991MNRAS.253..414C}. All models
in this suite have stellar components described by\footnote{Parameters
with subscript `$_*$' refer to 3D stellar density profiles that
take the same functional form as Eq.~(\ref{eq:rho_dm_zhao}). For each suite, the DM profile is Zhao.} $\alpha_*=2$, $\beta_*=5$, and DM
halos described by $\alpha_{\rm DM}=2$, $\beta_{\rm DM}=3.1$, with normalisation
$\rho_s$ chosen such that the mass enclosed within the central 300 pc is $M_{300}\sim
10^7M_{\odot}$.  Other than this normalisation, the parameters that vary from model to model are then
$\gamma_{\rm DM}$, $r_*$, $r_s$, $\beta_{\rm ani}$ and
$\gamma_*$. We
consider cases with $\gamma_{\rm DM}$ between $0-1$, $\gamma_*$ between $0-0.7$, $r_s$
between $0.2-1$ kpc, and
$r_*/r_s$ between 0.1 and 1, allowing for cored profiles,
NFW-like cusps and a large range of stellar concentrations.
For each potential, the anisotropy is constant, with values between $\beta_{\rm ani}=-0.45$
(tangential anisotropy) and $\beta_{\rm ani}=+0.3$ (radial). 
These combinations of parameters
yield a suite of 64 unique models.  

The second suite of mock data sets is similar to the one used by
\citetads{2011ApJ...742...20W} and is available and described in detail as part of
\textit{The Gaia Challenge}, a community-wide effort to examine the
performance of various methods on common test problems\footnote{\url{http://astrowiki.ph.surrey.ac.uk/dokuwiki}}. Briefly, these samples
are generated from the family of spherical, anisotropic distribution
functions originally proposed by \citetads{1979PAZh....5...77O} and
\citetads{1985AJ.....90.1027M}. Thus they have anisotropy profiles of the form
given by Eq.~(\ref{eq:beta_osipkov}).  For given $\rho_{\rm DM}(r)$ and
$\nu(r)$, the distribution functions are calculated using Eq.~(11)
of \citetads{1985AJ.....90.1027M} and then sampled using an accept-reject algorithm.  
All models in this suite have stellar components with $\alpha_*=2$,
$\beta_*=5$, and  DM halos with $\alpha_{\rm DM}=1$, $\beta_{\rm DM}=3$,
and $r_s=1$ kpc, again with normalisation $\rho_s$
chosen such that $M_{300}\sim 10^7M_{\odot}$.  Other parameters that vary from model to model are
$r_*$, $\gamma_*$, $\gamma_{\rm DM}$, and the anisotropy radius $r_a$. We
consider cases with $\gamma_{\rm DM}=0, 1$, $\gamma_*=0.1, 1$, $r_*/r_s=0.1, 0.25, 0.5, 1$
and $r_a=r_*,\infty$, allowing 32 unique models. Note that for $r_a=\infty$, the anisotropy profile is equivalent to a constant profile with $\beta_{\rm ani} = 0$.

\begin{table}
\begin{center}
\caption{Properties of the 3 sets of simulated data used in this study. Two of them come from {\em The Gaia Challenge} (\url{astrowiki.ph.surrey.ac.uk/dokuwiki}). DM and light profiles are Zhao - Eq. (\ref{eq:rho_dm_zhao}). $\gamma$ refers to the logarithmical inner slope of the DM and light profiles of the models, $r_{s}$ to their scale radii, and $\beta_{\text{ani}}$ to their velocity anisotropy.}
\label{table:mock_data}
\begin{tabular}{lccc}\hline\hline
  Mock data                    &  Spherical$^\star$ & Spherical$^\circ$  & Triaxial$^\dagger$\\
  \hline
  \# of models                 &  64             & 32                   & 2                \vspace{2.mm}\\
  $\gamma$         &  $[0,1]$      & $0 - 1$               & $0.23 - 1$       \vspace{0mm}\\
  $r_{s}$ [kpc]    &  $[0.2,1]$      & 1                       & 1.5              \vspace{2.mm}\\
  $\gamma^{*}$                 &  $[0,0.7]$      & $0.1 - 1$               & 0.23             \vspace{0mm}\\
  $r_{s}^{*}$ [kpc]            &  $[0.1,1]$      & $[0.1,1]$               & 0.81             \vspace{2.mm}\\
  $\beta_{\text{ani}}$ profile &  Cst       & Cst+Osipkov        & Baes \& van Hese \vspace{0.mm}\\
 \hline
\end{tabular}\\
\vspace{1mm}
{\small 
$^\star$ \citetads{2011ApJ...733L..46W} and \citetads{2011MNRAS.418.1526C}\\
$^\circ$ \citetads{2011ApJ...742...20W} and {\em The Gaia Challenge}\\
$^\dagger$ \citetads{2009MNRAS.395.1079D} and {\em The Gaia Challenge}
}
\end{center}  
%\vspace{-3mm} 
\end{table}
The third and final suite of mock data sets is also available and
described in detail as part of \textit{The Gaia Challenge}, but in
this case the underlying models are triaxial and therefore violate the
common assumption of spherical symmetry.  These
samples are generated using the `Made-to-Measure' N-body code of
\citetads{2009MNRAS.395.1079D}.  There are two unique models in this suite, and both
 have axis ratios (for both
halo and tracer components) $b/a=0.8$ and $c/a=0.6$, with
spherically-averaged profiles described by Eq.~(\ref{eq:rho_dm_zhao}), with $\alpha_{\rm DM}=1$, $\beta_{\rm DM}=4$,
$r_s=1.5$ kpc, $\alpha_*=2.9$, $\beta_*=5.92$, $\gamma_*=0.23$ and
$r_*=0.81$ kpc.  Parameters that vary from one case to the other are
$\gamma_{\rm DM}$, which takes values of either 0.23 or 1.0, and
$\rho_{\rm DM}$, which takes values of $5.5\times 10^7M_{\odot}$ kpc$^{-3}$ (for cases with
$\gamma_{\rm DM}=1$) or $1.2\times
10^8M_{\odot}$ kpc$^{-3}$ (for cases with $\gamma_{\rm DM}=0.23$).  

Table \ref{table:mock_data} summarises the three suites of synthetic data. For each model we draw samples of $N=30$ (small), 1000 (medium) and 10000 (large)
stars in order to encompass the range of stellar-kinematic data sets
currently available for ultra-faint and classical dSphs, as well as for `ideally observed' dSphs.  
For each sample, we estimate the `observed' velocity dispersion
profile by projecting mock data along one of the principle axes,
parsing the sample into $\sqrt{N}$ bins that each contain $\sqrt{N}$
mock observations (except for the `small' samples of $N=30$, for which
we take two bins, each with 15 stars), and then computing the projected velocity
dispersion (and its variance) using the maximum-likelihood technique
as discussed by \citetads{2006AJ....131.2114W}. All our mock data are
free of background and foreground contaminations, which represent additional,
though independent, sources of systematic effects for the quantities
reconstructed (DM profile, $J$ and $D$-factors). The star contaminations will
be discussed in a separate study in the context of the analysis of real dSph data.
Figure \ref{fig:dispprofiles} shows examples of velocity dispersion profiles
calculated for small, medium and large mock samples.  
For the calculation of the $J$ and $D$-factors, all the mock dSphs
are assumed to correspond to objects at a fixed distance $d = 100~\text{kpc}$.

%________________________________________________________________________
\section{Preliminary discussion}
\label{sec:starter}

\begin{figure}
\includegraphics[width=0.51\linewidth]{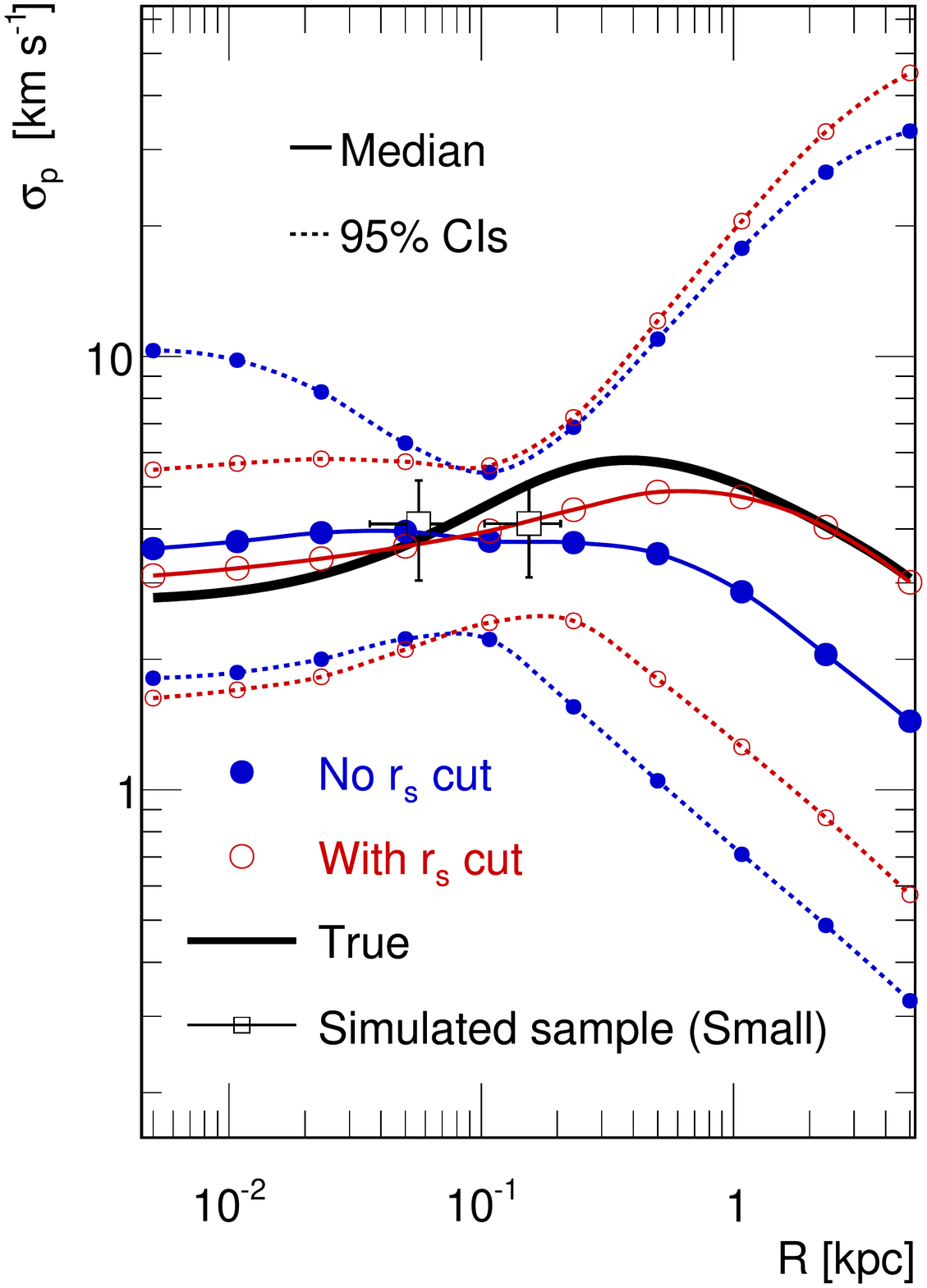}
\includegraphics[width=0.51\linewidth]{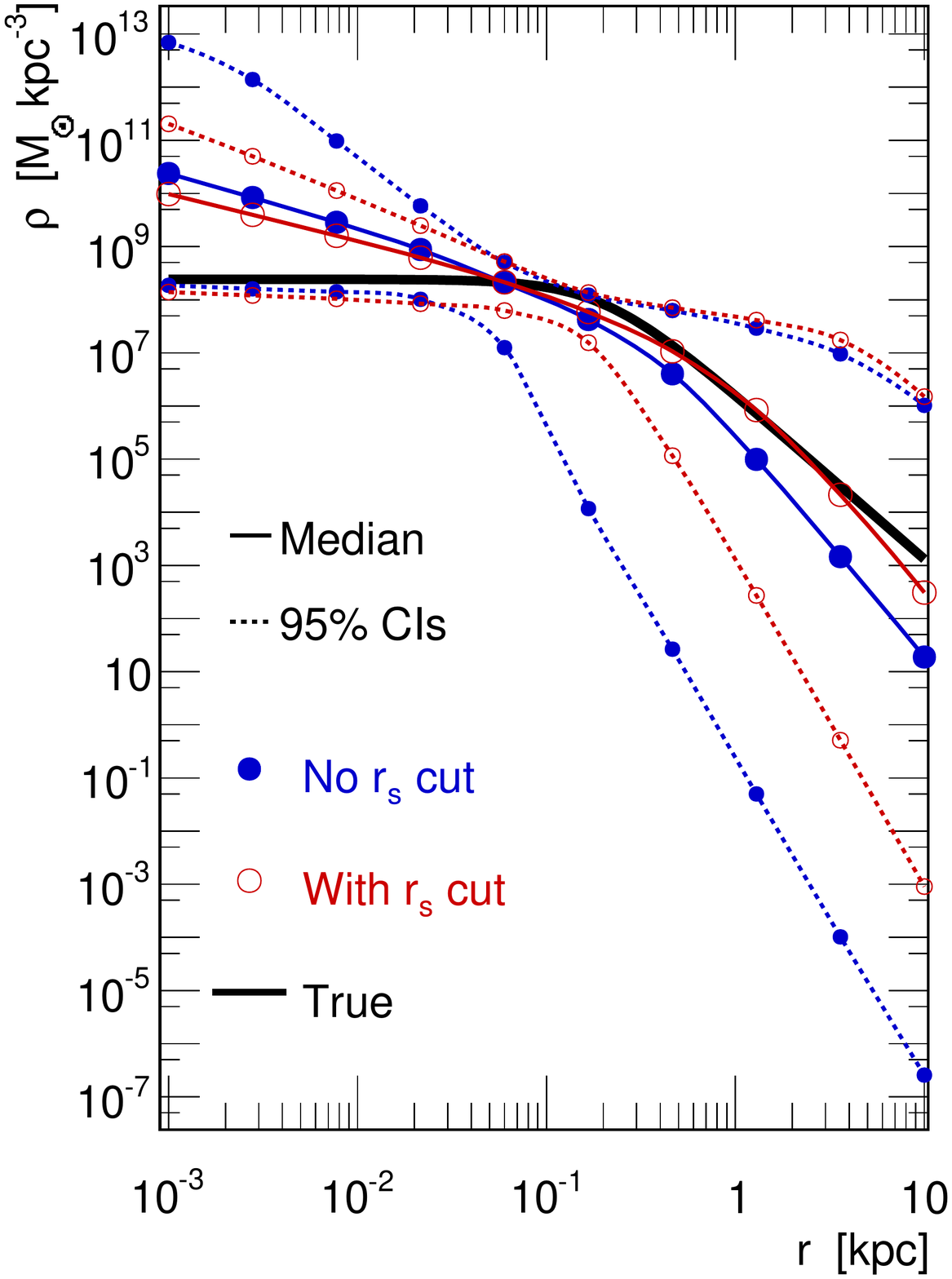}
\includegraphics[width=0.51\linewidth]{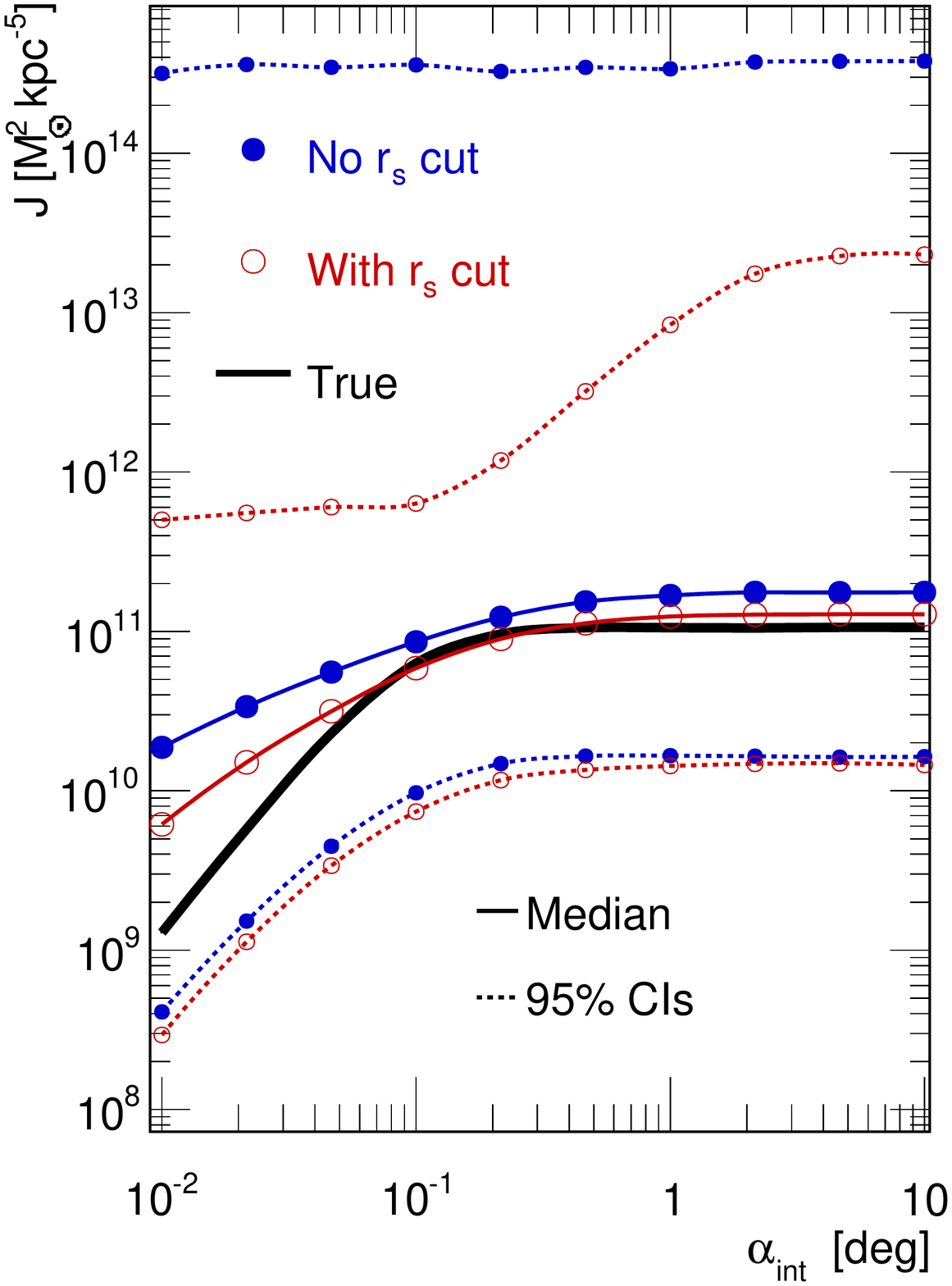}
\includegraphics[width=0.51\linewidth]{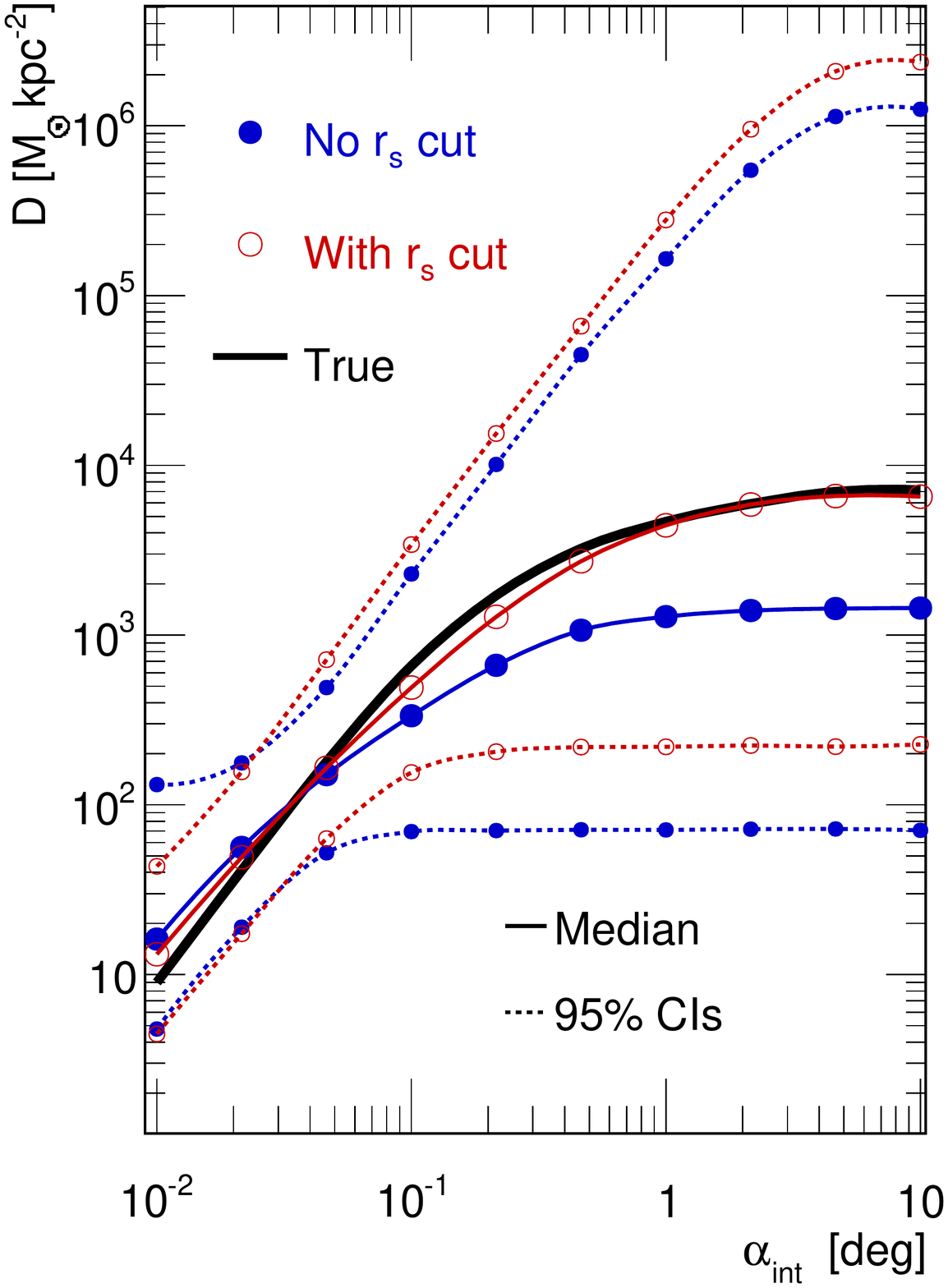}
\caption{Median values (solid lines with symbols) and 95\% CIs (thin dashed lines with symbols) for the velocity dispersion profile (top left), the DM density profile (top right),  the $J$-factor (bottom left) and the $D$-factor (bottom right). Enforcing the condition $r_s \ge r_s^\star$ (red empty circles compared to blue filled circles) in the prior of the MCMC analysis (see Table~\ref{table:priors_DM}) lead to better results, i.e. with the median value closer to the true value with smaller uncertainties (less biased and better estimator). The model shown corresponds to a mock ultra-faint dSph galaxy with $\gamma = 0$, $r_{s} = 0.2~\text{kpc}$ and $r_{s}^{*} = 0.1~\text{kpc}$.}
\label{fig:cut_rs}
\end{figure}

Figure~\ref{fig:cut_rs} illustrates (for one of the models and small
sample size) the main functions of interest in this study: the projected
velocity dispersion profile, $\sigma_p(R)$ (top left), the DM density profile
$\rho_{\rm DM}(r)$ (top right), and the $J$- and $D$-factors calculated (using Eq.~\ref{eq:J}) as functions of the integration angle $\alpha_{\text{int}}$ (bottom panels). In each panel, and in almost all plots shown in this paper, the thick black lines are reference curves, calculated with the {\em true} (i.e. known) parameters, to which the MCMC results are compared. 

On the top left panel, the empty squares correspond to the velocity dispersion data used in the
Jeans/MCMC analysis (small sample in this case). The
median (solid lines with symbols) and the 95\% lower and upper CIs
(dotted lines) are computed from \eq{eq:cl}. The two sets
of blue and red curves, discussed in \S\ref{sec:rs_cut}, correspond to
different priors, and we focus on the red curves only (filled circles)
for this preliminary discussion. 

Before moving to our new results, it is useful to underline the
similarities and differences with the analysis performed on the same
mock data considered by \citetads{2011ApJ...733L..46W} and
\citetads{2011MNRAS.418.1526C}. In these papers, only medium-size
samples were used (to be representative of classical dSphs data), with surface brightness profiles fitted with Plummer 
profiles, and with anisotropy assumed to be constant. The
more thorough analysis of this paper allows us to get more insight
into the modelling uncertainties, while reinforcing and extending the
conclusions drawn from the previous analyses
\citepads{2011MNRAS.418.1526C,2011ApJ...733L..46W}. We briefly
summarise below what conclusions these previous studies reached and
underline what this new analysis brings.
  \begin{itemize}

    \item {\em Determination of the DM inner slope:} for
      mock classical dSph galaxies $(N\sim 1000$ stars), the DM profile inner slope $\gamma$ was
      found not to be constrained by the Jeans analysis
      \citepads{2011MNRAS.418.1526C}. This is even more true for
      mock ultra-faint dSphs (see, e.g., the top right panel in
      Fig.~\ref{fig:cut_rs}, where all slopes $\gamma\in[0,1]$ are
      allowed at the 95\% level). The difficulty to constrain this
      slope is generally attributed to the degeneracy of the DM parameters
      with the anisotropy parameter $\beta_{\rm ani}(r)$, and methods
      relying on higher order of the Jeans analysis have been proposed
      to circumvent this issue
      \citepads{2013MNRAS.432.3361R,2014MNRAS.441.1584R,2014MNRAS.440.1680R}.
      See also the {\tt MAMPOSSt} approach in \citetads{2013MNRAS.429.3079M}.
      We find in Sect.~\ref{sec:jeans_ref} that this difficulty
      remains even with a perfect knowledge of the light profile and
      anisotropy, for any sample size. This indicates that constraining the
      inner slope in the standard Jeans modelling is limited by
      degeneracies among the DM profile parameters themselves. These
      degeneracies appear because of the poor sampling of the
      inner parts of dSph galaxies which remain difficult to measure.
      A better approach to address this issue may be to use different
      population tracers (e.g., \citealtads{2011ApJ...742...20W,2011MNRAS.411.2118A,2012ApJ...754L..39A}).

    \item {\em Determination of $J$ independently of the exact
      $\gamma$ value:} while $\gamma$ cannot be determined, it does
      not, however, prevent constraining the $J$-factor (\citealtads{2011MNRAS.418.1526C}, and bottom left panel of Fig.~\ref{fig:cut_rs}, where the true $J$-factor is encompassed within the CIs). A similar behaviour is found for decaying DM (illustrated by the bottom right panel of Fig.~\ref{fig:cut_rs}).      

    \item {\em Optimal integration angle:} one major finding of
      \citetads{2011ApJ...733L..46W} and
      \citetads{2011MNRAS.418.1526C} is the existence of an optimal
      integration angle for which the uncertainty on the $J$-factor is
      minimal. This is illustrated by the pinch visible in the red
      curves (with empty circles) in the bottom left panel of
      Fig.~\ref{fig:cut_rs}. This pinch is observed for all the
      quantities displayed---$\sigma_p(R)$, $\rho(r)$, $J(\alpha_{\rm
        int})$, and $D(\alpha_{\rm int})$.  For the velocity
      dispersion profile, the pinch occurs where most of the data lie,
      which is near the tracer scale radius $r_s^\star$ (0.1~kpc for
      the model displayed).  Because this is also the radius where the
      mass-anisotropy degeneracy is minimised
      \citepads{2009ApJ...704.1274W,2010MNRAS.406.1220W}, $\rho(r)$ is
      also relatively well-constrained at this radius.  In terms of
      the $J$-factor, the tightest constraint occurs when the signal
      is integrated over an angle  
       $\alpha_c^J\approx2r_s^\star/d$ \citepads{2011ApJ...733L..46W}. We recall
      that $d$ is taken to be 100~kpc throughout the paper, so that it
      corresponds to $\alpha_c^J\sim0.1^\circ$ in the bottom left
      panel of Fig.~\ref{fig:cut_rs}. As discussed in
      Appendix~\ref{app:d_crit}, we find that for DM decay, the
      critical angle $\alpha_c^D$ is half the one for annihilation DM: $\alpha_c^D\approx \alpha_c^J/2\approx r_s^\star/d$. For the model shown in the bottom right panel, $\alpha_c^D\sim0.05^\circ$.
  \end{itemize}

These results will not be further discussed in the paper, which from
here on focuses on effects that were not systematically (or not at all) studied in \citetads{2011MNRAS.418.1526C}.

%________________________________________________________________________
\section{Impact of the DM modelling: maximum knowledge setup}
\label{sec:jeans_ref}

In this section, we use the mock data described in \S\ref{subsec:mock}, in the idealised case where the light and
anisotropy profiles are known and fixed to their true values (i.e. that were used to generate the mock data).
This configuration, dubbed {\em maximum knowledge}, allows us to
investigate the direct impact of the DM profile parametrisation ({\em Zhao} or
{\em Einasto}) and of its priors. The uncertainties obtained on $\rho(r)$,
$J(\alpha_{\rm int})$, and $D(\alpha_{\rm int})$ in this ideal case
also give a flavour, for different sample sizes, of the precision that
the Jeans modelling could reach for analyses improving on the light and anisotropy-related parameters. 

The following results are based on three different sample sizes (see
Section~\ref{subsec:mock} and Fig.~\ref{fig:dispprofiles}) of 64 spherical models (first column of
Table~\ref{table:mock_data}). For each MCMC analysis, the only free
parameters are the DM profile ones.

%__________
\subsection{An optimal cut for the DM scale radius: $\boldsymbol{r_s\geq r_s^\star}$}
\label{sec:rs_cut}

Figure~\ref{fig:cut_rs} shows the results of the Jeans analysis for a typical mock ultra-faint
dSph galaxy, where blue curves (with filled circles) are
obtained using the prior $\log_{10}(r_s/\text{kpc})\in[$-$3,1]$, while
red curves (with empty circles) are obtained using the range
$[$-$1,1]$ (see also Table~\ref{table:priors_DM}). The value -1 comes
from the condition $r_s\geq r_s^\star$ (with $r_s^\star=0.1$~kpc for
the model shown), i.e. demanding the DM scale radius to be larger
than the light scale radius. Even knowing the value of the light and velocity anisotropy parameters, very large uncertainties appear on all quantities displayed in Fig.~\ref{fig:cut_rs} (especially the $J$-factor). However, they are significantly reduced using the above cut on $r_s$ values (blue vs red curves). 

\begin{figure}
\includegraphics[width=\linewidth]{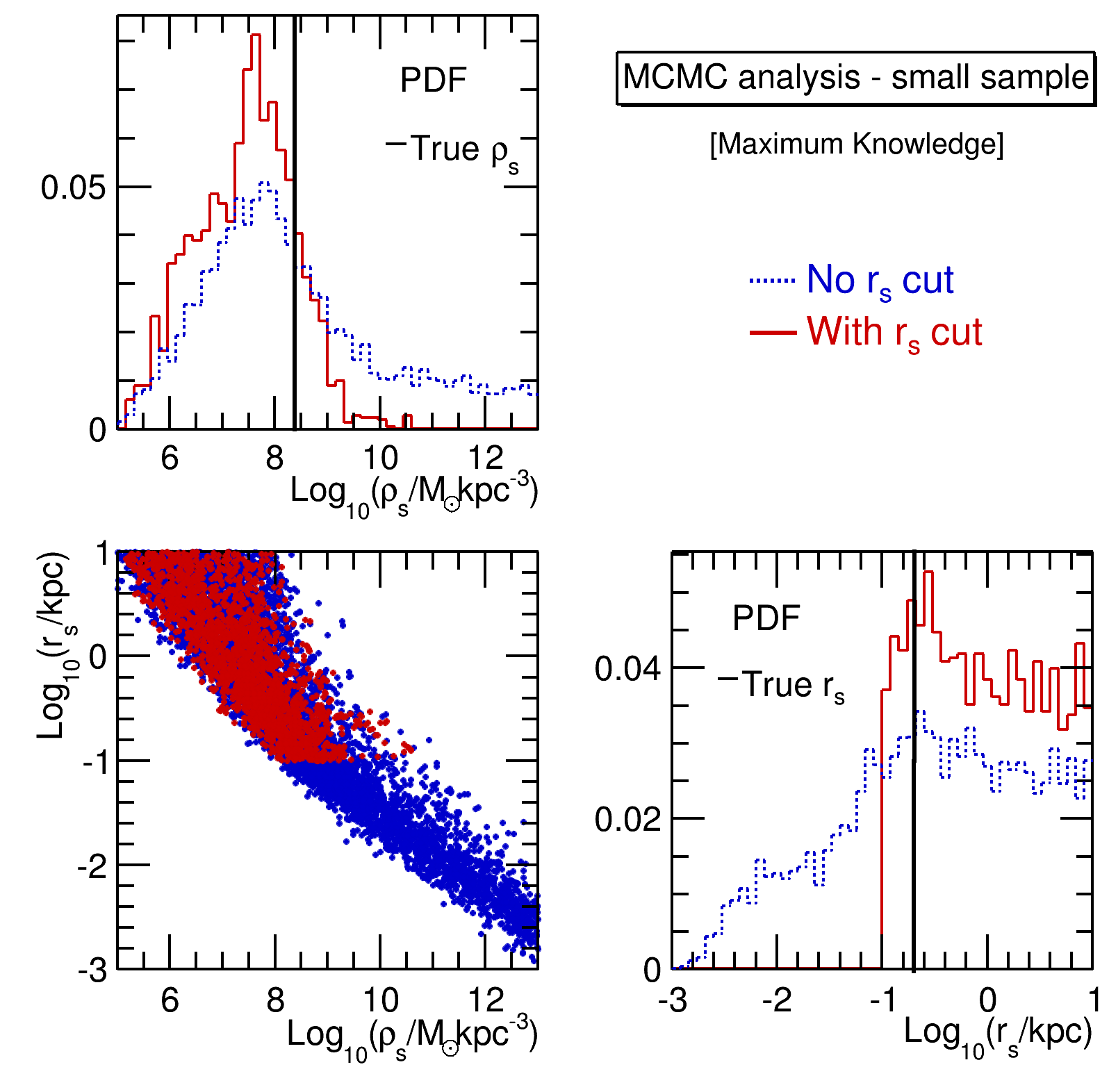}
\caption{PDFs (diagonal) and correlation (off diagonal) of $\rho_{s}$ and $r_{s}$ obtained from the \textit{maximum knowledge} MCMC analysis (same mock dSph as in Fig.~\ref{fig:cut_rs}, and same colour code). With no cut on the $r_s$ prior (blue dashed line), the anti-correlation between the two gives rise to nonphysical models (with very low DM scale radii and very large densities). Applying the condition $r_s\geq r_s^\star$ (equals to $0.1~\text{kpc}$ here), the two parameters are still poorly constrained (red solid lines), but nonphysical models are removed.}
\label{fig:pdf_cut_rs}
\end{figure}

A more detailed view of this effect is provided in Fig.~\ref{fig:pdf_cut_rs} using the same colour code for the two different priors: PDFs of $\rho_{s}$ and $r_{s}$ and their correlation are shown for the same model as in Fig. \ref{fig:cut_rs} (see \citealtads{2011MNRAS.418.1526C} for a thorough discussion of DM parameter correlations).
These plots illustrate that too-small $r_s$ lead to very high
values of the DM density in the inner parts (see top left panel), which
will propagate to large $J$ factors.
Although unrealistic, these models fit well the
poorly constrained velocity dispersion profile (top left panel of
Fig. \ref{fig:cut_rs}). Note that whether the cut on $r_s$ values is
applied or not, the Jeans analysis is unable to recover the correct
values of the DM parameters (because of the degeneracy between them), even in this {\em maximum knowledge} configuration.

\begin{figure}
\includegraphics[width=\linewidth]{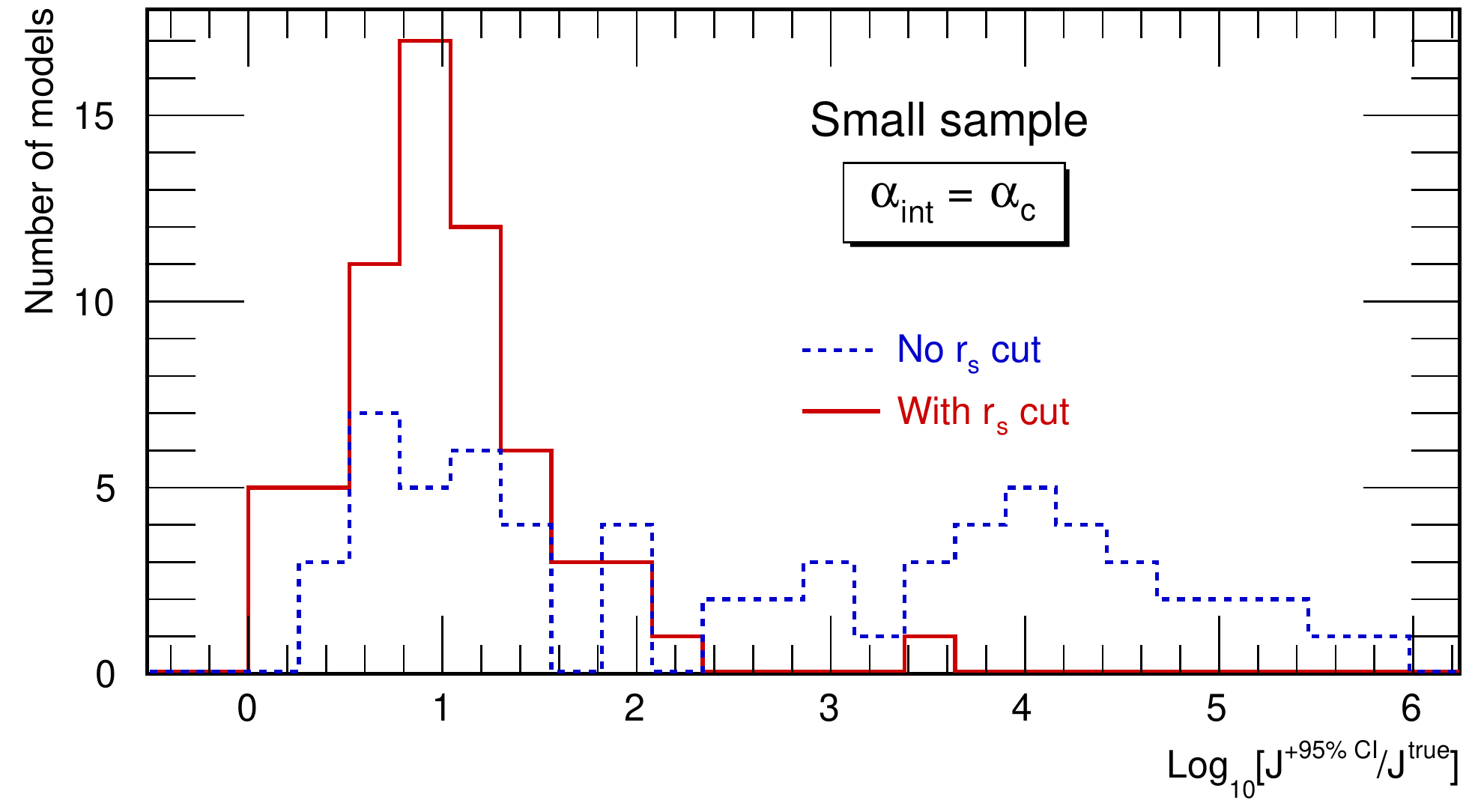}
\caption{Distribution of $J^{+95\%\rm CI}/J^{\rm true}$  for the 64 mock
  ultra-faint dSph galaxies, calculated at the critical angle
  $\alpha_c= 2r_{s}^{*}/d$. The condition $r_s\geq r_s^\star$ (red solid
  line) produced smaller uncertainties compared to the case
  where no cut is applied (blue dashed line).}
\label{fig:histo_j_cut_rs}
\end{figure}

Following \citetads{2011MNRAS.418.1526C}, we wish to
rely on as ``data-driven'' an approach as possible and therefore, do not
want to adopt priors based on DM halos produced in
cosmological N-body simulations (cf. \citealtads{2009JCAP...06..014M}). The condition $r_s\geq r_s^\star$ 
assumes that a DM halo must be at least as large as the stellar
population it hosts, which seems inevitable so long as the stars form
from gas that sinks to the centre of a DM halo as it cools \citepads{1978MNRAS.183..341W}.  This prior leads to less biased
reconstructed values with smaller uncertainties for mock ultra-faint dSphs (see all panels in
Fig. \ref{fig:cut_rs}). The
analysis has been repeated on the 64 mock dSph galaxies and for each, it always led to the same
conclusion. Figure~\ref{fig:histo_j_cut_rs} shows the distribution of
$J^{+95\%\rm CI}/J^{\rm true}$ values among the 64 models, before (blue
dashed line) and after (red solid line) applying the condition
$r_s\geq r_s^\star$. Before the cut, more than half of the models had $10^2 J^{\rm true}
\lesssim J^{+95\%\rm
  CI}\lesssim 10^6 J^{\rm true}$. The improvement is significant once the cut
is applied, with $J^{+95\%\rm CI}\lesssim 100~J^{\rm true}$ for all models (but one). This result is obtained for an integration angle $\alpha_{\rm int}=\alpha_c$, but similar improvements are observed at other angles\footnote{Caution is required when interpreting the histogram in Fig.~\ref{fig:histo_j_cut_rs}. It should not be used to infer what the mean
upper 95\% CI should be, as the distribution of the
profiles in the mock dSph sample may not be representative of the (unknown)
Milky Way's dSph galaxies profile distribution. However, it clearly shows the
benefit of the cut $r_s \ge r_s^\star$ on a sample presenting a
large variety of DM profiles, and that is expected to encompass the
range in which actual Milky Way's dSph profiles lie.}. This cut is less crucial for the mock classical samples (not shown), and has no effect for dSphs from the large sample. Note that the $J^{-95\%\rm CI}$ values are not affected by this cut (bottom left panel of Fig.~\ref{fig:cut_rs}).

In the remainder of the paper, the cut on $r_s$ will always be
applied. For the sake of legibility, only results related to the
$J$-factors are discussed below: similar effects are generally observed
for the other quantities ($\sigma_p$, $\rho$, and $D$), and we refer
the interested reader to Appendix~\ref{app:Dplots} for results on $D$-factors.

%__________
\subsection{Impact of the DM profile: Zhao vs Einasto}
\label{sec:zhao_einasto}

\begin{figure}
\includegraphics[width=\linewidth]{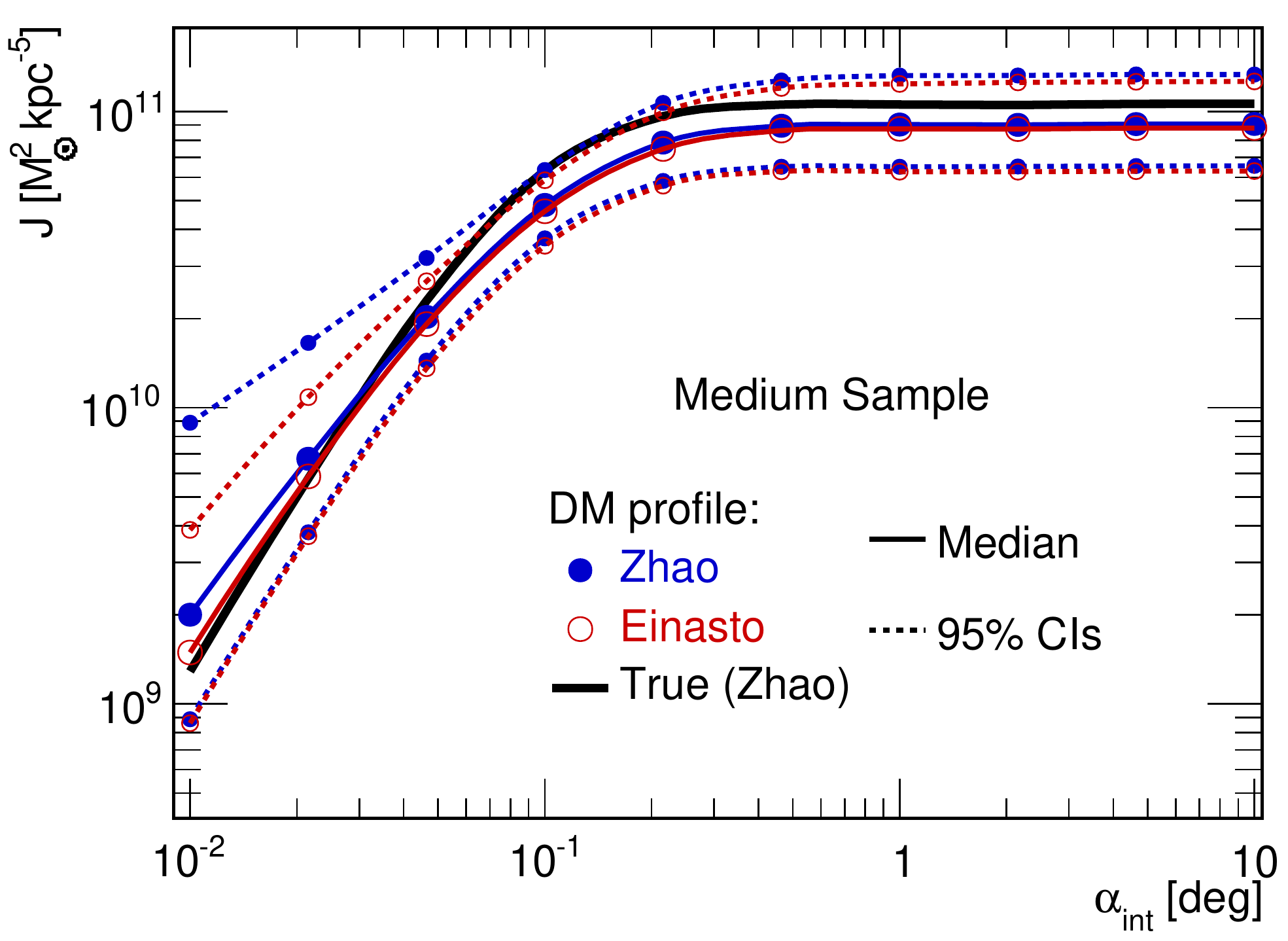}
\caption{$J$-factor median values (solid lines with symbols) and
  95\% CIs (thin dashed lines with symbols) for a mock classical
  dSph galaxy for which the light and anisotropy profiles are perfectly known. MCMC/Jeans analyses with a Zhao (blue filled circles)
  or a Einasto (red empty circles) DM profile give similar results,
  the latter being much faster to run (three free parameters instead of
  five). The true value (thick black solid line) is encompassed within the 95\% CIs.}
\label{fig:j_einasto_zhao}
\end{figure}
\begin{figure}
\begin{center}
\includegraphics[width=\linewidth]{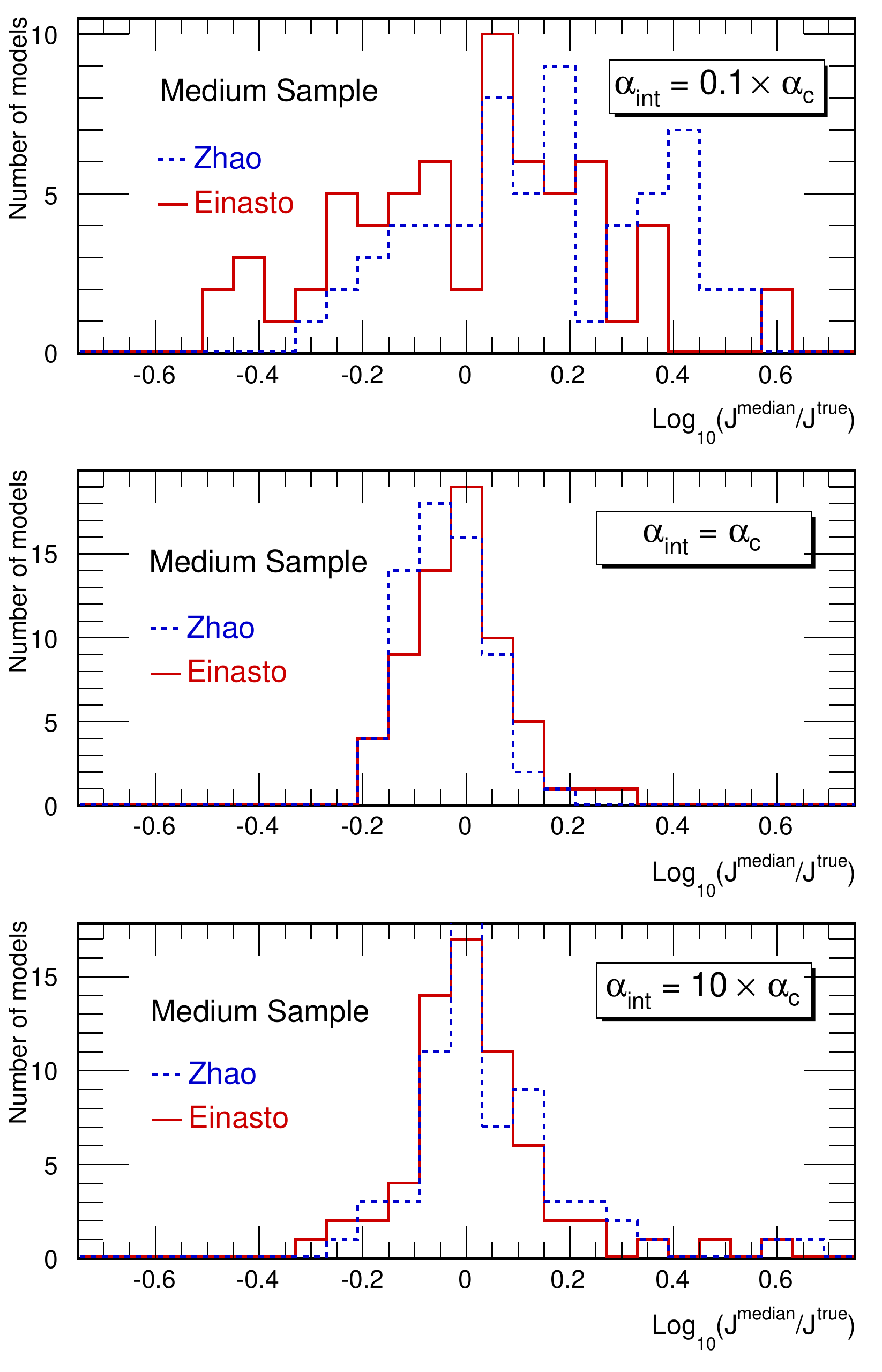}
\caption{Distribution of $J^{\rm median}/J^{\rm true}$ values for the 64
  models in the medium-size configuration, for three integrations
  angles: $0.1~\alpha_{c}$ ({\em top}), $\alpha_{c}$ ({\em middle}) and
  $10~\alpha_{c}$ ({\em bottom}). Using a Zhao (dashed blue lines) or an Einasto (red solid lines) parametrisation gives similar results.}
\label{fig:histo_j_einasto_zhao}
\end{center}
\end{figure}

Jeans analyses of dSph galaxies have typically used NFW or core profiles
(i.e., Zhao profiles with fixed slope parameters, e.g. \citealtads{2004PhRvD..69l3501E,2007PhRvD..75h3526S}), while
\citetads{2011ApJ...733L..46W} and \citetads{2011MNRAS.418.1526C} have recently
extended this approach to more generic Zhao halos. Einasto profiles
are also becoming a popular
choice (e.g., \citealtads{2009JCAP...06..014M,2010PhRvD..82l3503E}). This section aims at checking if the choice of the DM profile parametrisation has an
impact on the $J$ values. To do so, the results for the 64 models (and the three sample sizes) are compared, when processed using the 5-parameter Zhao profile or the 3-parameter Einasto profile.

The very good agreement (on a wide range
of integration angles) between the two parametrisations 
is illustrated for one medium-size model
(i.e., mock classical dSph) in Fig.~\ref{fig:j_einasto_zhao}, where
$J(\alpha_{\rm int})$ is plotted for both Zhao (blue filled circles)
and Einasto (red empty circles) DM profiles. Median values and CIs are
similar in both cases, with a small deviation occurring for very small
integration angles, $\alpha_{\rm
int} \lesssim$~a~few~$10^{-2}$~deg.
Figure~\ref{fig:histo_j_einasto_zhao} compares the distribution of $J^{\rm median}/J^{\rm
  true}$ values obtained using all the 64 models, at three integration angles (from top to bottom, $0.1~\alpha_c$, $\alpha_c$, and
$10~\alpha_c$). The distributions for the Zhao (dashed blue) and
Einasto (solid red) cases are in very good agreement (the same conclusion holds for $\rho$ and $D$ calculations). There is an indication that the Einasto profile provides a slightly better fit for
$\alpha_{\rm int}=0.1~\alpha_c$ (top panel) as the distribution appears
more centred around zero. However, the effect is small and we only mention it as a possibility.

The Einasto DM profile has less free
parameters than the Zhao parametrisation, which allows
for faster runs of the MCMC chains (fewer points required to reach
convergence). Therefore, it is used in the remainder of the paper.

%__________
\subsection{Importance of the sample size}
\label{sec:sample_size}

Knowing the light and velocity anisotropy parameters, we can establish the best limits that could be reached (but not
overcome) using the data-driven Jeans analysis (i.e. without strong
cosmological priors), for classical and ultra-faint dSphs. 

\begin{figure*}
\includegraphics[width=\linewidth]{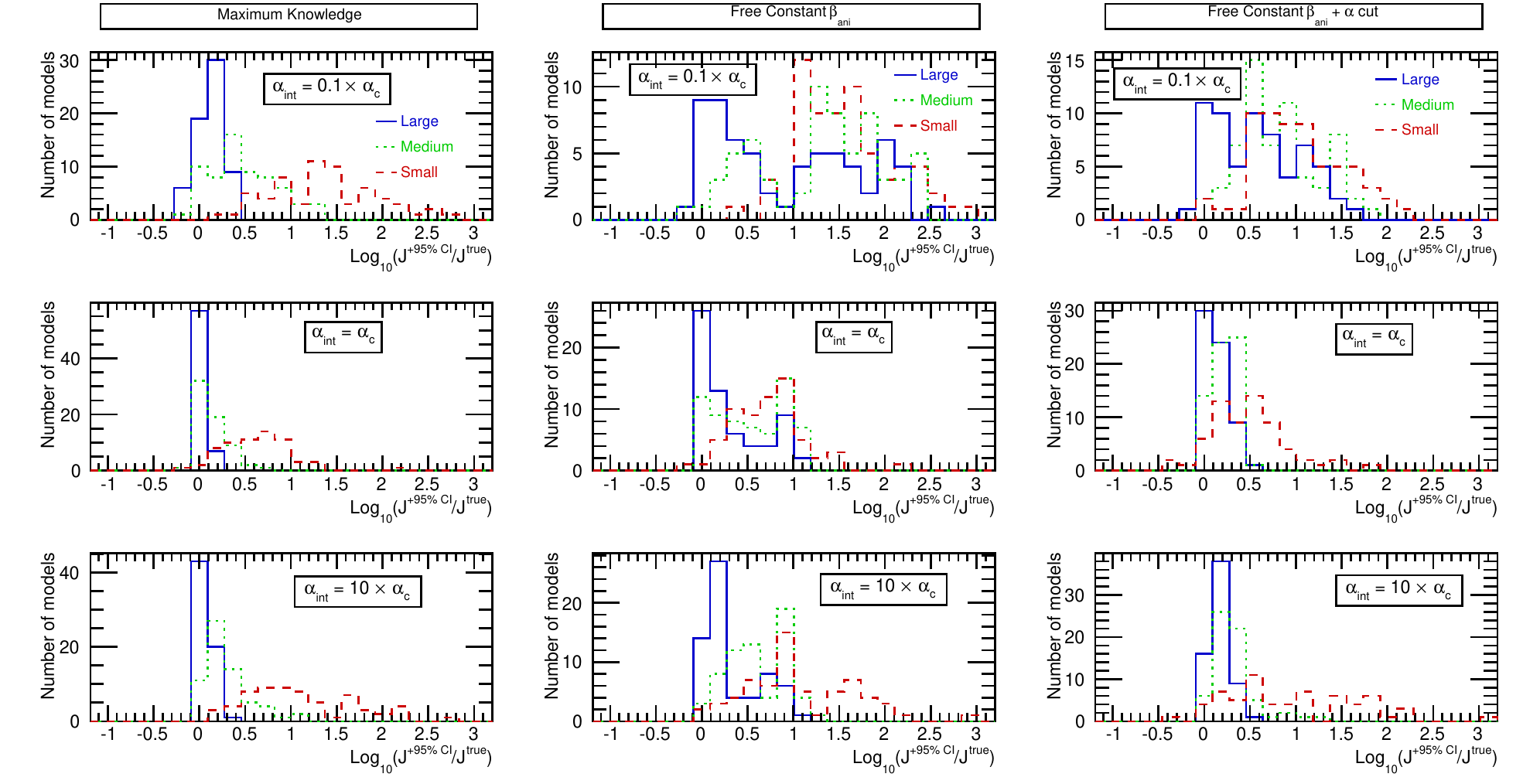}
\caption{Distribution of $J^{+95\%\rm CI}/J^{\rm true}$ values for the
  64 dSph models. These plots show the impact of the data sample
  size|large sample in solid blue lines, mock classical dSph galaxies in
  dotted green lines, and mock ultra-faints in dashed red lines|on the
  $J$-factor uncertainties for `different' Jeans analysis
  configurations (with Einasto DM profile priors set to
  Table~\ref{table:priors_DM} values and using $r_{-2}\geq
  r_s^\star$). Rows are different integration angles ($0.1~\alpha_{c}$:
  top; $\alpha_{c}$: middle; $10~\alpha_{c}$: bottom), and columns are
  the different configurations. \textit{Left panels}: the setup is
  {\em Maximum Knowledge} (\S\ref{sec:sample_size}), i.e. known light and anisotropy
  parameters. \textit{Middle panels}: as for the previous setup (free
  DM parameters), but with free constant anisotropy
  $\beta_0$ (\S\ref{subs:freebeta}). \textit{Right panels}: as for the previous setup, but
  adding the cut $\alpha \geq 0.12$ on the Einasto slope (\S\ref{subs:freebeta}). See text for a discussion of the plots. Note that the distributions are almost always positive, indicating that the reconstructed $J$-factors are not strongly biased w.r.t. the true values.}
\label{fig:histo_j_beta}
\end{figure*}

In figure~\ref{fig:histo_j_beta}, the left column shows the
distributions of the $J^{+95\%\rm CI}/J^{\rm true}$ values of the 64
models in the {\em maximum knowledge} analysis: top to bottom panels
correspond to different integration angles and the different
colours/linestyles to the three sample sizes. As already underlined,
the uncertainty on the $J$-factor is smaller for $\alpha_{\rm
  int}=\alpha_{c}$ (middle panel). At this optimal integration angle,
the best limit expected to be set on the $J$-factor is uncertain up to
a factor $\sim 3$ for mock classical dSph galaxies (green dotted lines), and
up to a factor $\sim 25$ for mock ultra-faint ones (red dashed
lines). For smaller ($\alpha_{\rm int}=0.1~\alpha_{c}$, top panel) or
larger ($\alpha_{\rm int}=10~\alpha_{c}$, bottom panel) integration
angles, these uncertainties are an order of magnitude larger. 
The $J^{-95\%\rm CI}/J^{\rm true}$ distributions (not shown) have the same behaviour, with comparable widths.

Note that smaller uncertainties are quoted in studies relying
on analyses using ``cosmological priors"
(e.g., \citealtads{2013arXiv1309.2641M}); however, our data-driven
approach mitigates biases that would arise in the case that real dSphs
are hosted by DM subhalos that differ structurally from
simulated ones.  This would be the case if, for example, the DM differs
from CDM, as in `warm' \citepads{2001ApJ...556...93B} or `self-interacting' \citepads{2000PhRvL..84.3760S} DM models, or if
feedback from star formation significantly alters the internal
structure of low-mass subhalos \citepads{2012MNRAS.421.3464P} with respect to the CDM-only
simulations from which otherwise-cosmological priors have been derived.  

%________________________________________________________________________
\section{Impact of the anisotropy profile}
\label{sec:anisotropy}

The velocity anisotropy profile $\beta_{\rm ani}(r)$
needed in the Jeans modelling,
Eqs.~(\ref{eq:jeans_sol}) and (\ref{eq:fr}), is degenerate with
the mass profile and cannot be directly measured
from stellar velocities.  Instead, the anisotropy profile is often parameterized and treated in the Jeans modelling as another
unknown of the problem, along with the DM profile. Many Jeans analyses
employed to study new physics in the context of DM indirect detection
rely on a constant anisotropy (one free parameter, see
Eq.~\ref{eq:beta_constant}).

In this section, we test whether lifting this strong assumption
and using instead an {\em Osipkov - Merritt} (1 free parameter, see Eq.~\ref{eq:beta_osipkov}) or the more generic {\em
  Baes \& van Hese} anisotropy profile (4 parameters, see
Eq.~\ref{eq:beta_baes}) significantly changes the results with respect
to the constant anisotropy case. To this end, we use the 64 constant
anisotropy models already used in the previous sections, as well as the 16
constant and 16 non-constant anisotropy mock dSph galaxies generated for {\em The Gaia
Challenge}. A summary of the models properties are given in Table~\ref{table:mock_data}. All analyses below rely on the {\em Einasto} DM profile, and mock data light profile parameters are set to their true values. Uniform priors are used for the anisotropy profile parameters, see Table~\ref{table:priors_ani}.

%__________
\subsection{Priors and optimal cuts}
\label{subs:beta_prior_cuts}

The interplay between stellar parameters on the one
hand, and the degeneracy between the anisotropy and mass profiles on
the other hand set specific constraints on the anisotropy and DM profile parameters.

\begin{table}
\begin{center}
\caption{Range of uniform priors used for the velocity anisotropy profile parameters. Note that all models must satisfy the \textit{Global Density-Slope Anisotropy Inequality} (see Section~\ref{sec:anisotropy}): for instance, this reduces $\beta_0$ to the range [$-$9,0] for a Plummer light profile}.
\label{table:priors_ani}
\begin{tabular}{ccc}\hline\hline
 Anisotropy profile & Parameter & Prior \\
 \hline
``Cst"                      & \multirow{2}{*}{$\beta_{0}$}               &  \multirow{2}{*}{$[$-$9,1$]}   \\\vspace{5mm}
Eq.~(\ref{eq:beta_constant})  &                                            &                              \\%\cline{2-5}
``Osipkov-Merritt"                       & \multirow{2}{*}{$\log_{10}(r_{a})$}  & \multirow{2}{*}{$[$-$3,1$]} \\\vspace{5mm}
Eq.~(\ref{eq:beta_osipkov})   &                                            &                     \\%\cline{2-5}
\multirow{3}{*}{``Baes \& van Hese"}         & $\beta_{0}$                                & $[$-$9,1]$        \\
\multirow{3}{*}{Eq.~(\ref{eq:beta_baes})}        & $\beta_{\infty}$        & $[$-$9,1]$        \\
                              & $\log_{10}(r_{a})$                         & $[$-$3,1]$        \\
                              & $\eta$                                     & $[0.1,4]$       \\\hline                    
\end{tabular}
\end{center}
\end{table}

\paragraph*{Interplay between $\beta_{\rm ani}(r)$ and $\nu(r)$: nonphysical models}

An inappropriate choice of anisotropy parameters can lead to nonphysical profiles. The so-called \textit{Global Density-Slope Anisotropy Inequality} \citepads{2010MNRAS.408.1070C} ensures that solutions to the Jeans equation correspond to physical models where the phase-space distribution function is positive. This condition reads
\begin{equation}
\label{eq:beta_condition}
\beta_{\text{ani}}(r) \leq -\frac{1}{2}\frac{\text{d}\log\nu(r)}{\text{d}\log(r)},
\end{equation}
and is applied to all the dynamical models used in this study. Note that this inequality is a generalisation to all radii of the results of \citetads{2006ApJ...642..752A}.

\paragraph*{Degeneracy between $\beta_{\rm ani}(r)$ and $\rho_{\rm DM}(r)$: optimal cut $\alpha\!\geq\!0.12$}
The differences between the plots in the left and middle panels of
Fig.~\ref{fig:histo_j_beta} illustrate (for the 64 constant anisotropy
models used in previous sections) the impact of the degeneracy between
the anisotropy and the DM profiles on the $J$ values
(width of the uncertainty distributions). Moving from the {\em maximum knowledge}
setup (known anisotropy, left panels) to a configuration with a free
constant anisotropy $\beta_0$ (middle panels) leads to a significant
increase of the width of the upper CI distribution for the large
and medium size samples. The effect is less pronounced for the \new{lower CI} distributions (not shown). We further discuss the sample sizes in \S\ref{subs:freebeta}.

Because of this degeneracy, the slope of the DM profile ($\gamma$ for
a Zhao, or $\alpha$ for an Einasto) becomes a crucial parameter. As
pointed out by \citetads{2011MNRAS.418.1526C} using medium size
samples (classical dSphs), restricting the range of the inner slope
$\gamma$ to $[0,1]$ drastically reduces the
uncertainties on the $J$-factors. In a similar fashion for the Einasto
profile, we restrict $\alpha$ from $[0.05,1]$ to $[0.12,1]$,
i.e. excluding the steepest slopes\footnote{There is no direct equivalency
  between the inner slope of the Zhao profile and the logarithmic
  slope of the Einasto profile $\textrm{d}\!\log\rho/\textrm{d}\!\log
  r=-2(r/r_{-2})^{\alpha}$. The lower limit on $\alpha$ is
  chosen so that the logarithmic slope is equal to -1 for $r/r_{-2} =
  1/300$, leading to $\alpha\sim0.12$. The value $\alpha = 0.05$ used
  for the base prior of Table~\ref{table:priors_DM} corresponds to a
  logarithmic slope of -1.5 for the same value of $r/r_{-2}$.}. As illustrated in the right panels of
Fig.~\ref{fig:histo_j_beta}, this cut is very efficient in reducing
the range of the upper CIs of the large and medium samples, and
gives a more robust (less biased and more precise) estimation of the
$J$-factors. It has nevertheless no impact on the lower CIs (not shown). 
This cut is always applied in the remainder of the paper. Note that this range 
of slopes is consistent with the observations of two Milky Way's dSphs, Fornax and Sculptor, for which analyses using either Schwarzchild modeling \citepads{2012ApJ...746...89J,2013MNRAS.433.3173B} or multiple stellar populations \citepads{2011ApJ...742...20W,2012MNRAS.419..184A,2012ApJ...754L..39A} seem to disfavour cuspy DM profiles. 
It is also in agreement with the results of recent cosmological simulations 
of spiral and dSph galaxies including both baryons and DM, which seem to favour 
flat over very steep
DM density profiles \citepads{2010Natur.463..203G,2014arXiv1405.4318M,2014ApJ...786...87B}.

\subsection{$\beta_{\rm ani}^{\rm Cst}$ analysis: uncertainties for different sample sizes}
\label{subs:freebeta}

The right column of Fig.~\ref{fig:histo_j_beta} shows the impact of
restricting the prior on $\alpha$ (to the range $[0.12,1]$) on the
upper 95\% $J$-factor CIs for all sample sizes. For the medium-size samples
(mock classical dSph galaxies, green dotted lines), 
the 95\% upper CIs decrease from a factor $10$ to a factor
$3$ at the critical angle $\alpha_{c}$. However, this cut has almost no
effect on the small-size samples (red dashed lines), for which the statistical
uncertainties completely dominate the error budget. We find no significant effect on the 95\% lower CIs.

The comparison between the {\em maximum knowledge}
setup (left panels) and the $\beta_{\rm ani}^{\rm Cst}$ analysis
(right panels) is also interesting. The latter is very often used in the literature
(e.g., \citealtads{2007PhRvD..75h3526S,2008ApJ...678..614S,2011MNRAS.418.1526C}),
allowing for a free anisotropy in the simplest way. The impact of the
anisotropy-mass degeneracy is significant for large sample sizes
(ideal case) for which the upper 95\% CIs are twice as large as in the
{\em maximum knowledge} setup. The difference is less pronounced for
medium-size samples (mock classical dSphs) and there is no difference at all for
small-size samples (mock ultra-faint dSphs). In the latter two cases,
the velocity dispersion data are simply too sparse to strongly
constrain the $J$-factors, even when avoiding the degeneracy built in the
Jeans equation by forcing the anisotropy to its real value.

%__________
\subsection{$\beta_{\rm ani}^{\rm Cst}$ vs $\beta_{\rm ani}^{\rm Osipkov}$: wrong assumption leads to wrong result}
\label{sub:beta_r}

To further explore the effects related to the velocity anisotropy
prescription, we now use the 32 spherical mock dSph galaxies
generated for {\em The Gaia Challenge} (second column of Table \ref{table:mock_data}). They are divided in 16 pairs of models with the same DM profiles, but with either a constant or an Osipkov-Merritt velocity anisotropy.

\begin{figure}
\includegraphics[width=\linewidth]{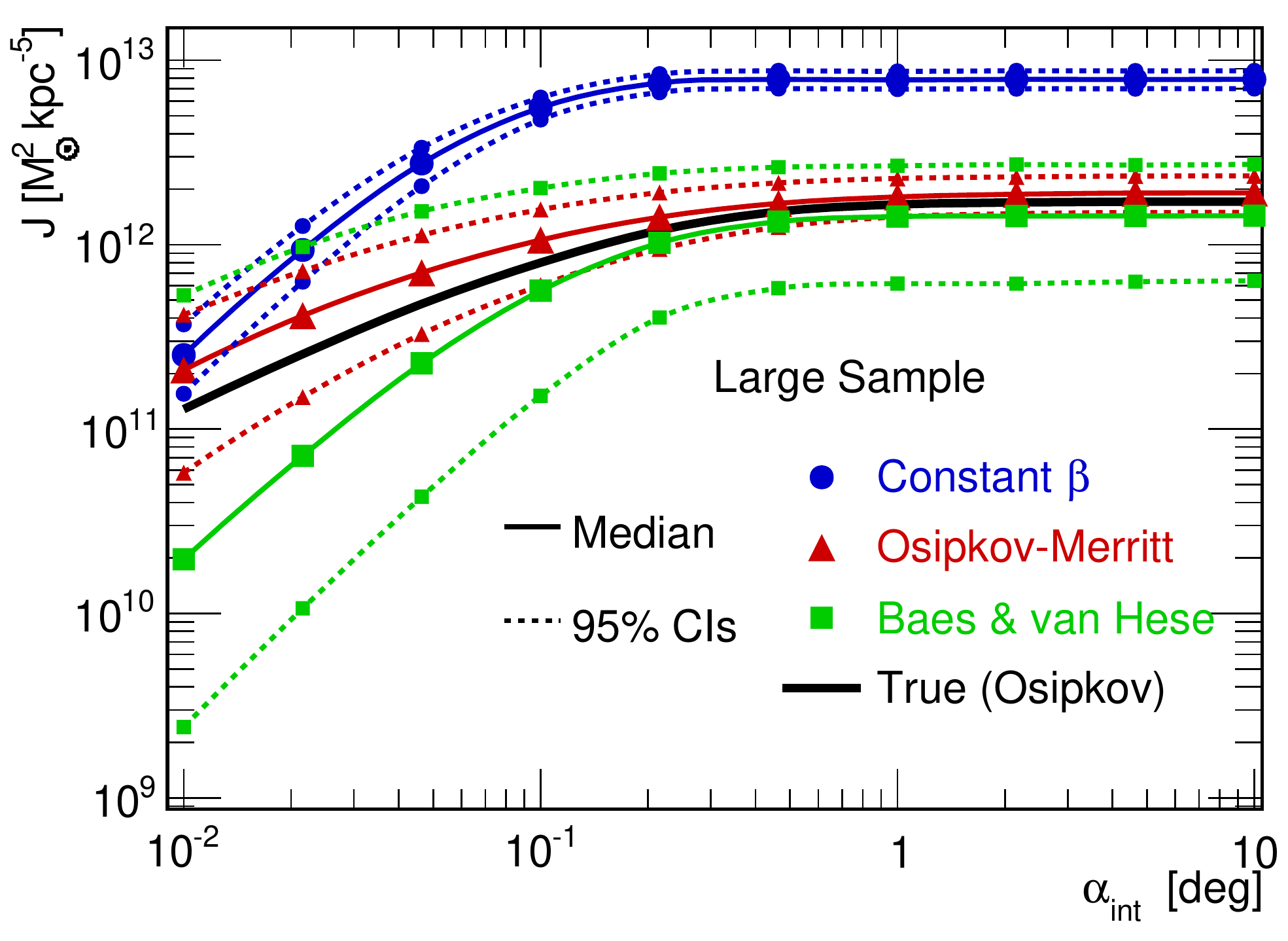}
\caption{Median values (solid lines with symbols) and $95 \%$ CIs
  (dotted lines with symbols) of $J(\alpha_{\rm int})$ for a mock dSph
  (large sample) generated with an Osipkov-Merritt velocity
  anisotropy. The true $J(\alpha_{\rm int})$ is given in solid black. The $J$-factor has been reconstructed using different
  anisotropy prescriptions: i) constant (blue circles), ii)
  Osipkov-Merritt (i.e., the correct parametrisation, red triangles)
  and iii) Baes \& van Hese (more general than Osipkov-Merritt, green
  squares). See \S\ref{sub:beta_r} and \S\ref{subsec:baes}. }
\label{fig:j_baes}
\end{figure}

We find that using the wrong anisotropy parametrisation can have
dramatic effects on the reconstruction of the $J$-factor. This is
exemplified in Fig.~\ref{fig:j_baes}, where $J(\alpha_{\rm int})$ is
computed for a mock dSph (large sample) generated with an Osipkov-Merritt anisotropy,
using either a constant (blue circles) or an Osipkov-Merritt (red
triangles) parametrisation in the Jeans analysis. The $J$-factor
obtained using a constant anisotropy profile lies one order of
magnitude above the true value. When assuming the correct
parametrisation, i.e. Osipkov-Merritt, the estimated $J$-factor becomes
compatible with the true value. This effect can also be important for
mock classical dSphs (medium sample), but not for mock ultra-faint ones
(small sample), for which the statistical uncertainties are
dominant. For completeness, Appendix~\ref{app:bimodality} extends the
discussion to the DM density profile and mass of dSph galaxies.

\subsection{Recommended option: $\beta_{\rm ani}^{\rm Baes}$ analysis}
\label{subsec:baes}
In light of the previous result, when kinematic samples are large, it is important to have an as
general as possible model for the anisotropy profile parametrisation. Indeed,
for real data, the true model for the anisotropy profile is obviously unknown. The Baes \& van Hese model (4 free parameters) is a good option since it encompasses both the constant and Osipkov-Merritt anisotropy profiles. 

\begin{figure}
\includegraphics[width=\linewidth]{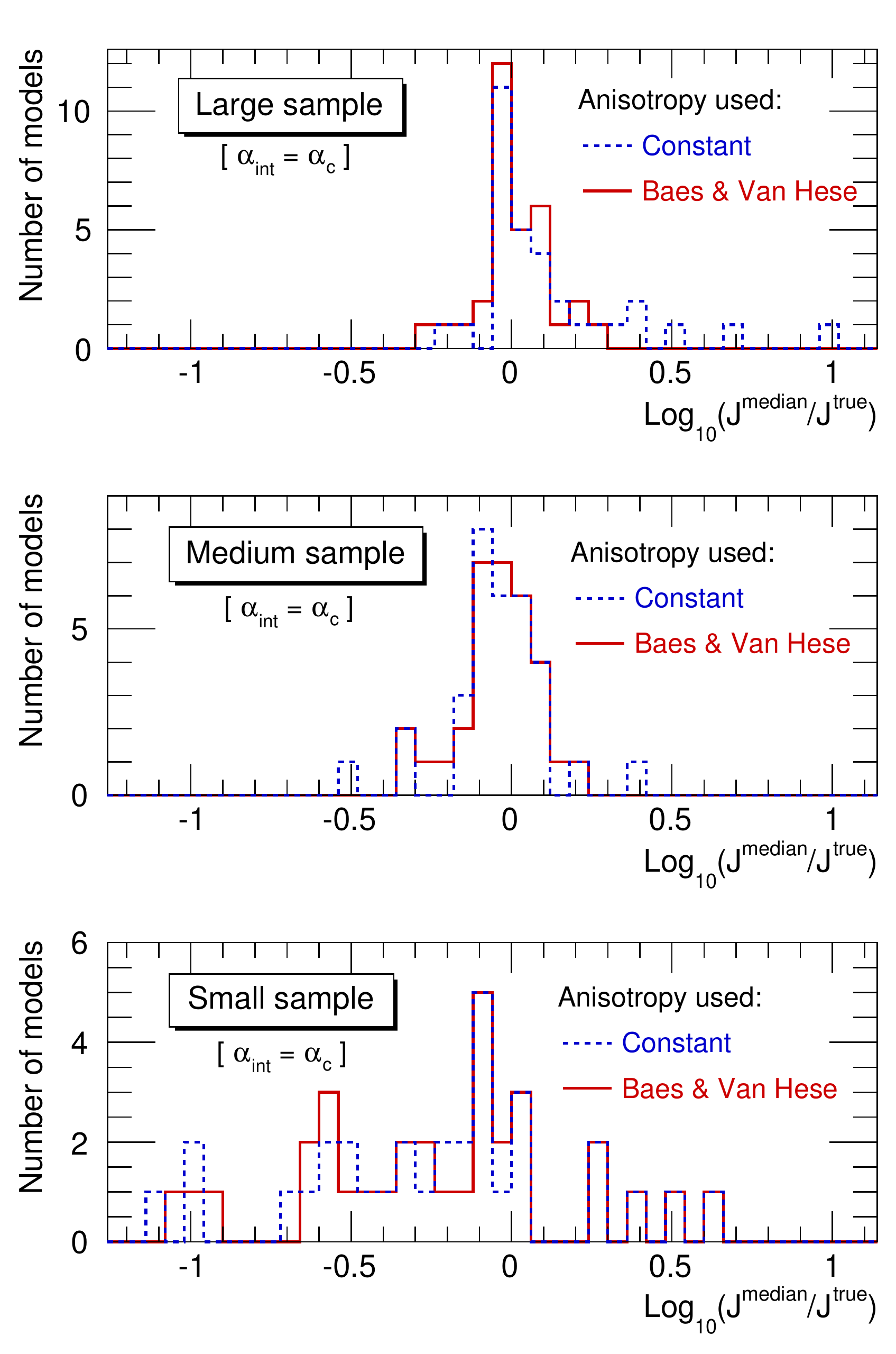}
\caption{Distribution of the $J^{\rm median}/J^{\rm true}$ at the critical integration angle $\alpha_{c}$ using either a constant (blue dashed line) or a Baes \& van Hese (red solid line) anisotropy profile, for the three sample sizes of the 32 Gaia Challenge models. The analysis of the mock ultra-faint samples (bottom panel) is dominated by the uncertainties on the $\sigma_p$ data, and hence are not sensitive to a wrong choice of the anisotropy profile. This is no longer the case for mock classical samples (small effect, middle panel), and crucial for the ideal case of a large sample (top panel).}
\label{fig:histo_j_baes}
\end{figure}

As shown in Fig.~\ref{fig:j_baes}, the CIs of the
$J$-factors are then larger using $\beta_{\rm ani}^{\rm Baes}$
(green squares) because of the extra degrees of freedom in the parametrisation. The analysis on the 32 models in Fig.~\ref{fig:histo_j_baes} (distribution of $J^{\rm
  median}/J^{\rm true}$ at $\alpha_c$) also shows that the use of Baes
\& van Hese profile (red solid line) can reduce biases coming from the wrong anisotropy parametrisation, except for small samples where once again the statistical errors on $\sigma_p$ dominate the error budget.

At worst, using $\beta_{\rm ani}^{\rm Baes}$ increases the $J$-factor uncertainty
by $\sim 2$ for mock classical dSphs compared to using a simpler anisotropy model,
but it can avoid biases for some models. We therefore recommend the use of the Baes \& Van Hese
parametrisation for the velocity anisotropy.

%________________________________________________________________________
\section{Impact of the light profile}
\label{sec:light}

Another key ingredient of the Jeans analysis is the light profile,
which appears both in its projected $I(R)$ and deprojected form
$\nu(r)$ in the computation of the velocity dispersion (Eq. \ref{eq:jeansproject}). A parametric model is usually fitted to
the observed projected light profile, and then deprojected using the
inverse Abel transform (see Sect.~\ref{sec:jeans}). In most dSph studies,
Plummer or King profiles are fitted to the surface brightnesses
\citepads{2008Natur.454.1096S,2009JCAP...06..014M,2011ApJ...733L..46W}, but exponential and S\'ersic profiles (Sect.~\ref{subsubsec:light_prof}) are also often used  (e.g., \citealtads{1995MNRAS.277.1354I,2001MNRAS.327L..21L}). We now investigate the impact of such parametrisations on the reconstruction of the $J$-factor. A preliminary (and less systematic) study of this effect was performed in \citetads{2011MNRAS.418.1526C}|see their Appendix~H.

The free parameters of the analysis are the Einasto DM profile parameters, using the optimal priors of Table~\ref{table:priors_DM}. The light profile parameters are fitted separately, as described below, and the anisotropy parameters are fixed to their true values in order to be more sensitive to the effects of the light profile.

\subsection{Subset of models and fit of the light profile}
\label{sec:diff_light_prof}
To avoid heavy computation, we select a subset of 3 `representative' spherical models from {\em The
Gaia Challenge} (second column of Table \ref{table:mock_data}), chosen according to their $J^{\text{median}}/J^{\text{true}}(\alpha_{c})$ value obtained in the previous section: one close to 1, and two extreme models (maximal and minimal value).

For each model and each sample size (small, medium, large), binned surface brightness profiles are generated from
the positions of the stars. For a given sample size, we use ten times more stars
to create the light profile than what was used for creating the velocity
dispersion profile; this aims at mimicking the observational data of
real dSphs, i.e. where there are less velocity measurements than
stars detected in the object (for instance for a mock classical dSph galaxy, 10000 stars are used for $I(R)$
and $1000$ for $\sigma_p(R))$. Light profile parameters are fitted using a likelihood function similar to Eq.~(\ref{eq:likelihood}):
\begin{equation}
  \mathcal{L}\!=\! \prod_{i=1}^N \frac{1}{\sqrt{2\pi}\,\Delta I(R_i)}\exp\biggl [\!-\frac{1}{2}\biggl (\!\frac{I_{\rm obs}(R_{i}) \!-\!I(R_i)}{\Delta I(R_i)}\!\biggr )^{2}\biggr ].
  \label{eq:likelihood_light}
\end{equation}
Because of the sharp decrease of the light profile (see, e.g., top
left panel of Fig.~\ref{fig:I_J}), the fit is very sensitive to the
sparse data points lying at large radii. It is possible to
perform an unbinned analysis on the light profile (as described, e.g.,
by \citealtads{2008ApJ...684.1075M}). We have checked that the binned and unbinned analyses give the same result if the error on both the $x$ and $y$ axes are taken into account. Namely, the error $\Delta^2 y\equiv[\Delta I(R_i)]^2$ in Eq.~(\ref{eq:likelihood_light}) is replaced by
\[
 \Delta^2 y \rightarrow \Delta^2 y + \left(\frac{1}{2} [f(x+\Delta x) - f(x-\Delta x)]\right)^2,
\]
where $\Delta x$ corresponds to the dispersion around the mean position $R_i$ in the bin $i$, and where $f(x)\equiv I(R)$.

\subsection{Impact of different light profile assumptions}
\label{subsec:different_light}

\begin{figure}
\includegraphics[width=0.51\linewidth]{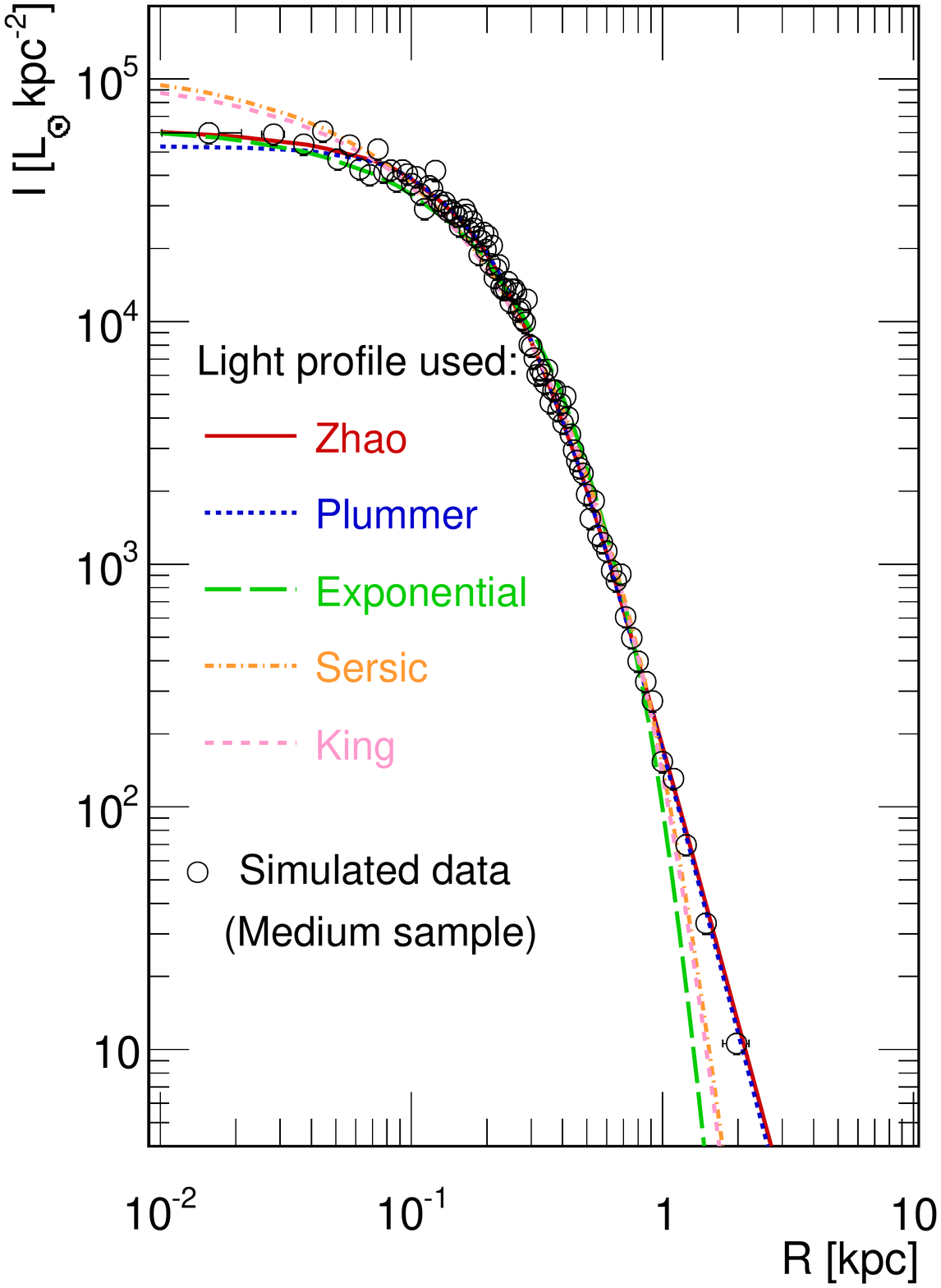}
\includegraphics[width=0.51\linewidth]{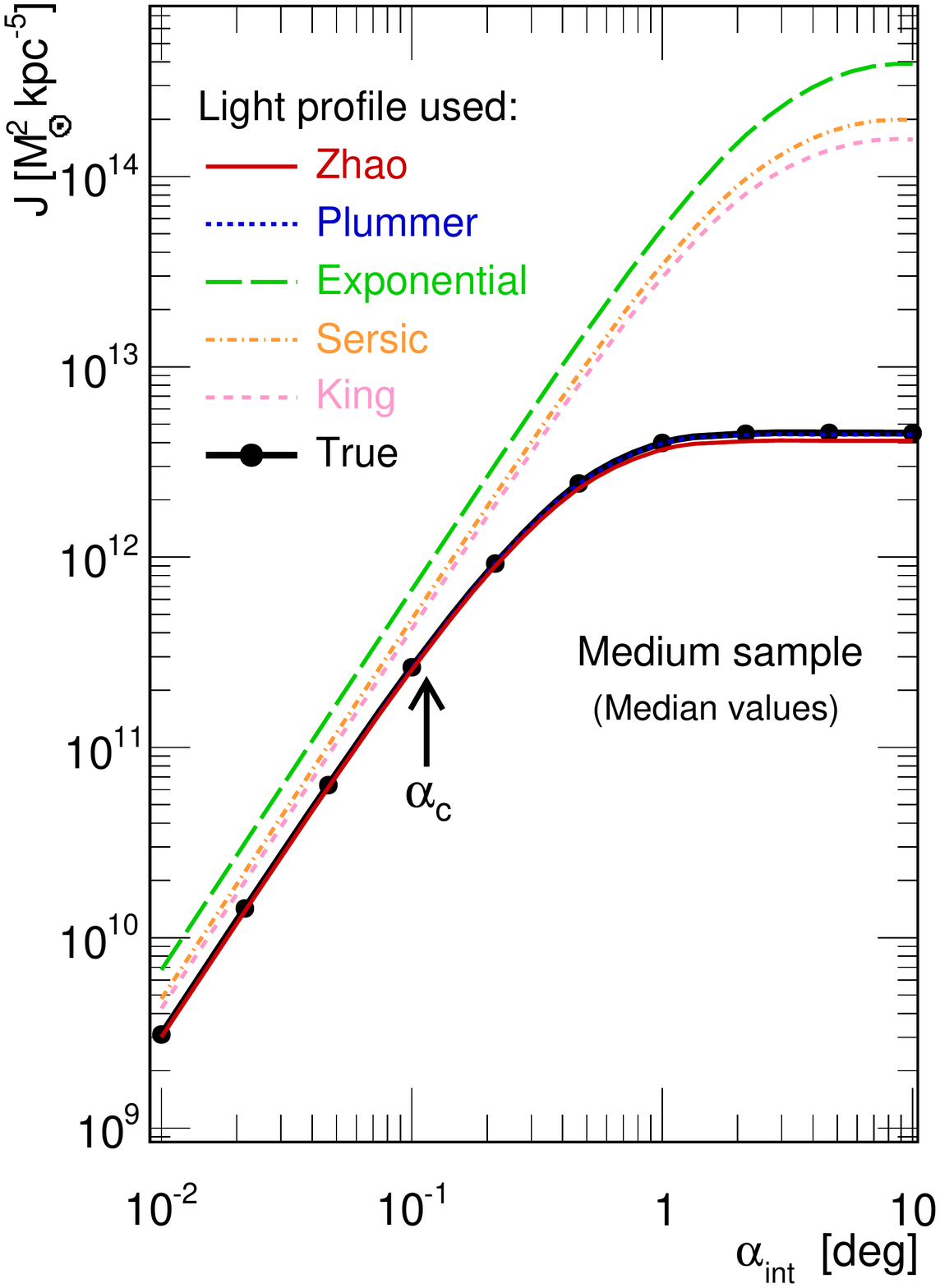}
\includegraphics[width=0.51\linewidth]{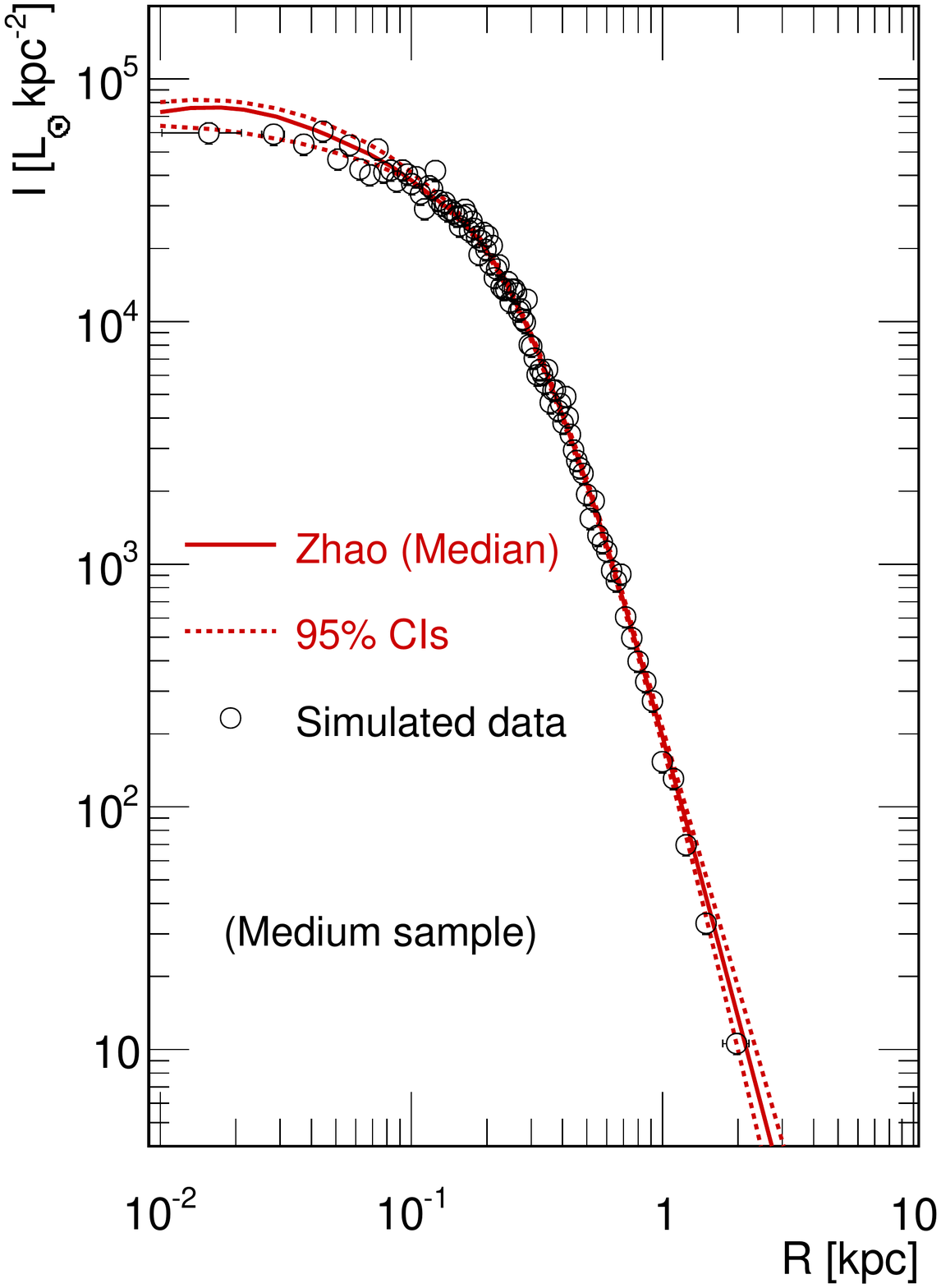}
\includegraphics[width=0.51\linewidth]{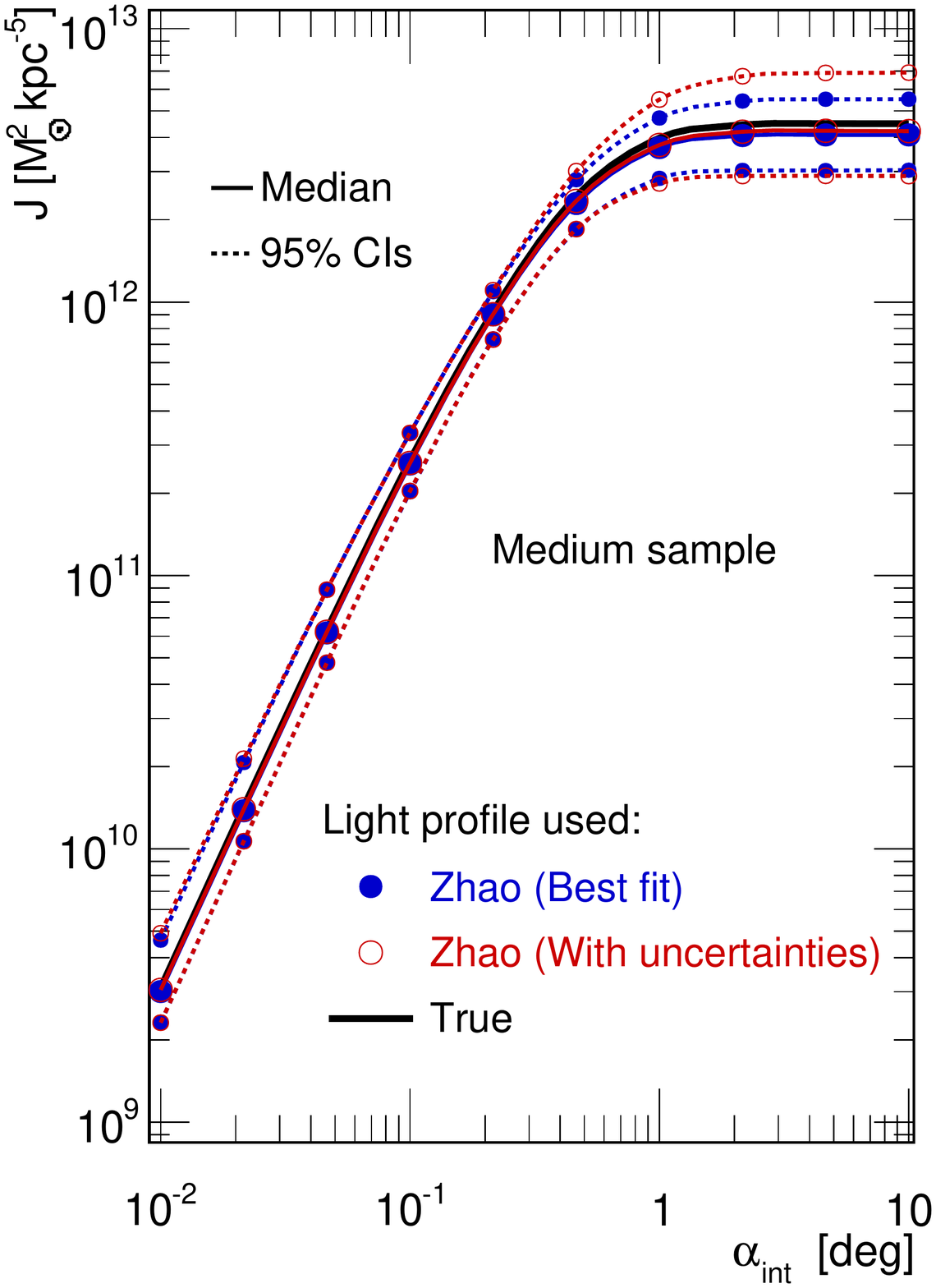}
\caption{{\em Top panels}: best-fit models of the surface brightness
  $I(R)$ for the five light profile parametrisations given in
  \S\ref{subsubsec:light_prof} (left); $J(\alpha_{\rm int})$ obtained when
  using each of the five best-fit light profile in the Jeans modelling
  (right)|see \S\ref{sec:diff_light_prof}. {\em Bottom panels}: Zhao
  median surface brightness and its uncertainties (left) and
  propagation of these error bars to $J(\alpha_{\rm int})$ (right). The
  same mock dSph galaxy (with medium-size velocity dispersion sample) has been used in all panels.}
\label{fig:I_J}
\end{figure}

For the three models and three sample sizes, we fit five different light profiles (Plummer, Zhao, exponential, S{\'e}rsic and King, see Section~\ref{subsubsec:light_prof}) to the surface brightness, using the likelihood function of Eq.~(\ref{eq:likelihood_light}). For one of the models, we show in the top panel of Fig. \ref{fig:I_J} the best fits obtained for the medium-size sample. For this
model, the true light profile (Zhao) is very close to a Plummer
profile ($\gamma^{*} = 0.1$ and $\beta^{*} = 5$), and Zhao (solid red)
and Plummer (dotted blue) parametrisations provide an excellent fit
to the data. The three other parametrisations significantly undershoot
the data at large radii (top left panel), with both King and S\'ersic also overshooting in the inner parts. 

We then run our Jeans/MCMC analysis using each light profile best-fit. The top right panel of Fig.~\ref{fig:I_J} shows the median $J$-factor
obtained with the Jeans/MCMC analysis done with each of these best-fit light
profiles, compared to the one obtained with an analysis run with the true light profile (black solid line). Models that fit well the light profile (Zhao and Plummer) lead to a reconstruction of the $J$-factor as good as the one obtained when using the true light profile. With the three others, the
$J$-factors systematically overshoot the latter: the CIs are not
shown (for legibility purposes), but they encompass this reference $J$-value
only at small integration angles $(\alpha_{\rm int} < 0.1$, which
corresponds to the optimal integration angle $\alpha_c$ of that
particular model). There is up to a factor $\sim 3$ systematic bias
below $\alpha_c$, which increases to $\sim 100$ at large
angles. Here we are showing the most pathological model of the
three that have been studied, but the effect is always present, though
less pronounced, for the two other models. A similar bias is obtained
for mock ultra-faint dSph galaxies, but their $J$-factor CIs (not shown) are
larger and therefore encompass this bias. 
 
This shows that the light profile parametrisation plays a
significant role in the $J$-factor reconstruction. It is therefore of particular importance to both measure
and fit precisely the surface brightness profiles of dSph galaxies. We
advocate the use of models with large degrees of freedom (e.g., Zhao), in
order to obtain the best possible fits to the data and reduce biases
in the derived $J$ values.

\subsection{Propagation of the light profile uncertainties on $J$}
\label{sec:propag_light_error}

Once a flexible-enough parametrisation is selected (here a Zhao profile) for
fitting the surface brightness, the uncertainties on the fit must be
propagated to the $J$-factor. 

We use our MCMC engine to recover both the median and CIs of the light
profile, as illustrated in
the bottom left panel of Fig.~\ref{fig:I_J}. Once done, we perform the
standard Jeans/MCMC analysis where, for each new step, a random point
of the previously-built light profile chains is chosen: this effectively propagates the surface brightness profile uncertainties to the posterior distributions of the DM and anisotropy parameters.

For any sample size, we find that the small light profile
uncertainties only weakly affect the $J$-profile reconstruction, as
shown in the bottom right panel of Fig.~\ref{fig:I_J}. This is
emphasised by the comparison of the median value (solid lines) and CIs
(dotted lines) obtained from the best-fit light profile only (blue
filled circles), or including the propagation of the error on the latter (red empty circles). Since the implementation of these errors is quite straightforward in the MCMC analysis, we nonetheless encourage their inclusion in the analysis when dealing with real data.

%________________________________________________________________________
\section{Geometrical effects: DM halo triaxiality}
\label{sec:triaxiality}

In the spherical Jeans analysis, it is assumed that the stellar
component is spherical, while it is known that dSphs have non-zero flattening \citepads{1995MNRAS.277.1354I,2013pss5.book.1039W}. The DM halo is also considered spherical, but cosmological N-body simulations have shown that both isolated DM halos and their substructures have triaxial shapes \citepads{1988ApJ...327..507F,2005ApJ...627..647B,2007MNRAS.376..215B,2011MNRAS.411..584M}. 

\citetads{2012ApJ...755..145H} have used an axisymmetric version of the Jeans equation in order to assess the impact of non-sphericity on the mass reconstruction. However, most studies rely on the spherical Jeans equation. In this section, we quantify the biases introduced by using a spherical Jeans analysis on triaxial DM halos.

\subsection{Triaxial halos: description and analysis}

The geometry of triaxial halos is described by three principal axes
$a$, $b$ and $c$, with $a \geq b \geq c$ and $abc =
1$. \citetads{2007ApJ...671.1135K} found using the Via Lactea
simulation \citepads{2007ApJ...657..262D} that for dSphs-like
sub-halos, the ratios $b/a$ and $c/a$ are in average equal to $0.83$
and $0.68$ respectively, and using the Aquarius simulation
\citepads{2008MNRAS.391.1685S}, \citetads{2014MNRAS.439.2863V} found
ratios close to $0.75$ and $0.6$. These objects appear therefore to be mildly
triaxial, and simulations predict more triaxiality for more massive halos \citepads{2012JCAP...05..030S}. 

\paragraph*{Mock data}
For the analysis, we use the two triaxial mock dSphs made available by
{\em The Gaia Challenge} (third column of Table \ref{table:mock_data}, see
also section \ref{subsec:mock}). We recall that each model consists of a triaxial
stellar distribution embedded in a triaxial DM halo, with $b/a$ and
$c/a$ ratios of $0.8$ and $0.6$ respectively for both stellar and DM
distributions. The velocity anisotropy profile of the stars is Baes
\& van Hese for both models, and the light and DM profiles have Zhao parametrisations. The only difference between the two models is the DM profile, which is cusped for one ($\gamma = 1$) and cored for the other ($\gamma = 0.23$).

\paragraph*{Analysis steps}
As described in section \ref{subsec:mock}, we once again create three
sample sizes for each model, mimicking ultra-faint, classical and ideal
dSphs. For each sample size, we build binned velocity dispersion and
binned light profiles for three different line of sights (l.o.s.),
chosen along the three principal axes of the object. This allows us to
investigate the effect of the different orientations in the reconstruction of the $J$-factor.

\paragraph*{$J$-factor calculation}
To compute the true $J$-factors for each l.o.s., we replace the spherical radius $r$ in the expression of the Zhao DM profile (Eq. \ref{eq:rho_dm_zhao}) by its ellipsoidal counterpart:
\begin{equation}
r_{e} = \sqrt{\frac{X^2}{a^2} + \frac{Y^2}{b^2} + \frac{Z^2}{c^2}},
\end{equation}
with $X$, $Y$ and $Z$ the Cartesian coordinates in the frame aligned
with the three principal axes of the object. To obtain the $J$
(resp. $D$)-factors, we perform the l.o.s. integration of the squared density profile (resp. density profile). The full 3D integration for any halo orientation is a new feature added in {\tt CLUMPY}.

\subsection{Projection effects}
Triaxiality implies projection effects on both the observed stellar component and the DM profile. This depends on the unknown orientation of the halos w.r.t. the observer line of sight.

\begin{figure}
\includegraphics[width=\linewidth]{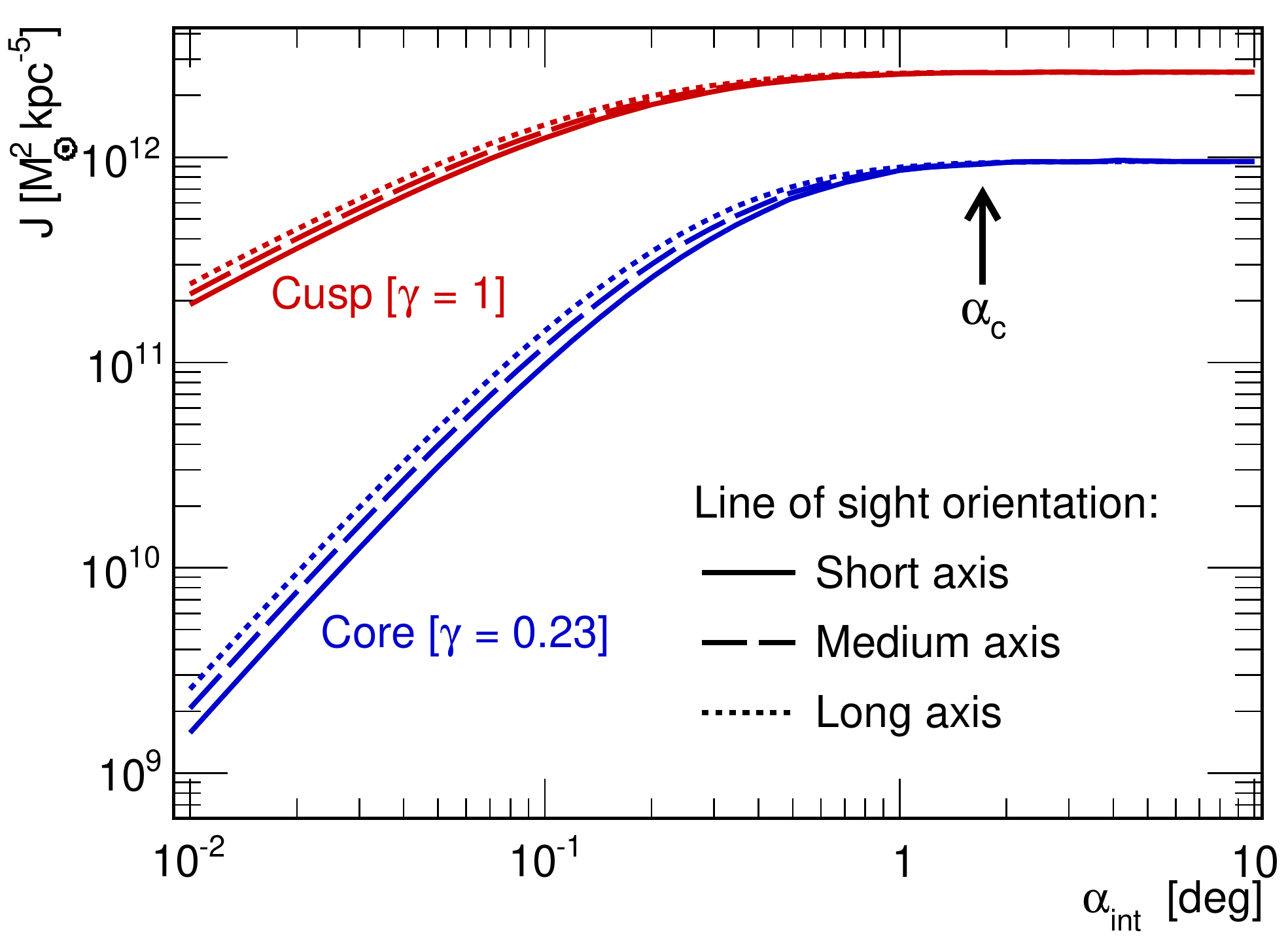}
\caption{$J$-factor true values for the cusp (red) and core (blue)
  mock triaxial dSph galaxies. The three curves correspond to the l.o.s. aligned with the short (solid), medium (dashed) or long (dotted) axes of the halo.}
\label{fig:true_j_triax}
\end{figure}

\paragraph*{Impact on $J$-factor true values}
Figure~\ref{fig:true_j_triax} shows the $J$-factor true values
obtained for the cusp (red) or the core (blue) dSph for three
l.o.s. orientations: along the short (solid), medium (dashed), and
long (dotted) axes. First and as expected, the cuspy DM profile gives larger $J$ values than the core. The projection effect on $J$ reaches at most a $30 \%$ difference at very small integration angles, for these mildly triaxial mock data. 

\paragraph*{Impact on the velocity dispersion profile}

\begin{figure}
\includegraphics[width=0.5\linewidth]{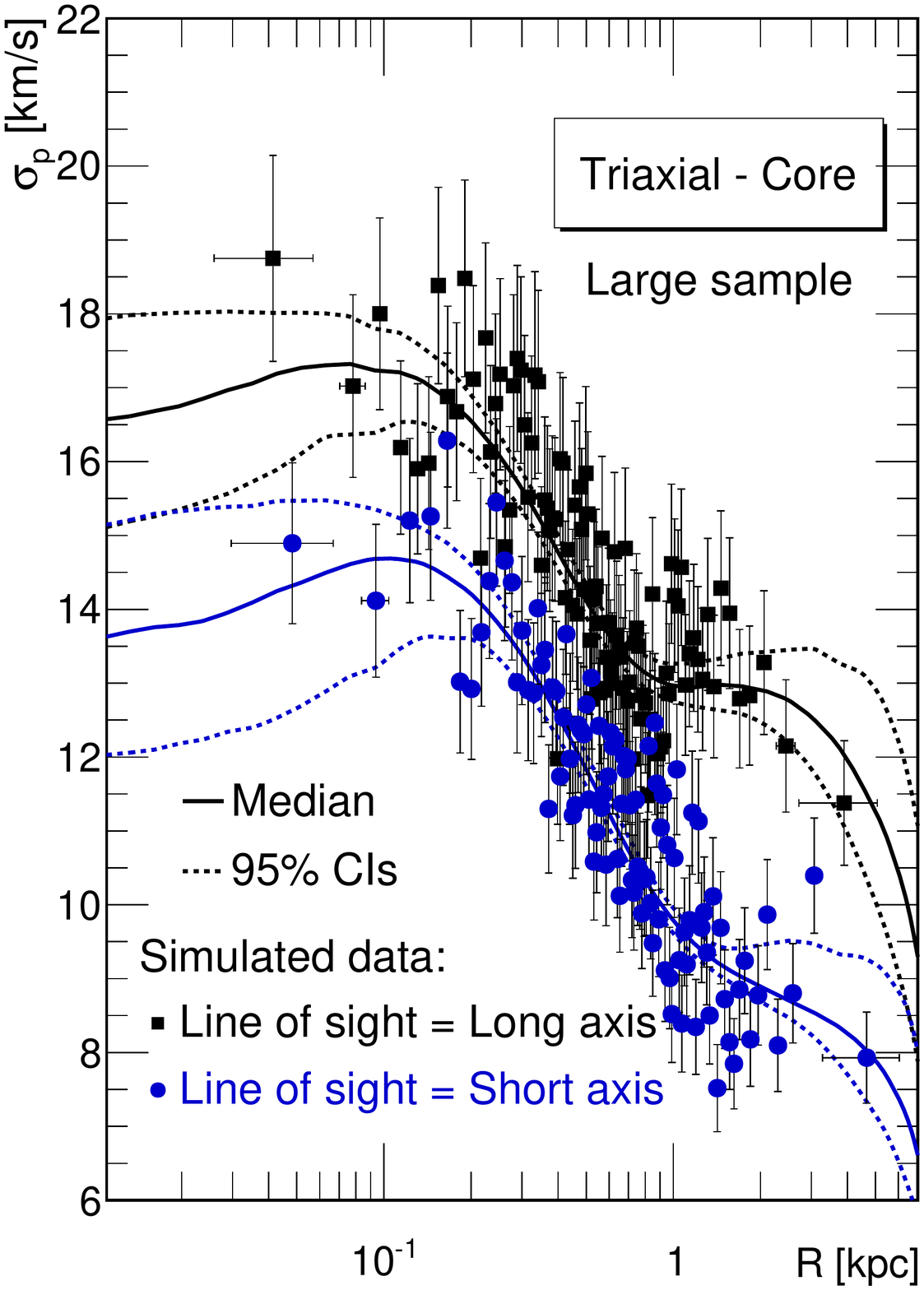}
\includegraphics[width=0.5\linewidth]{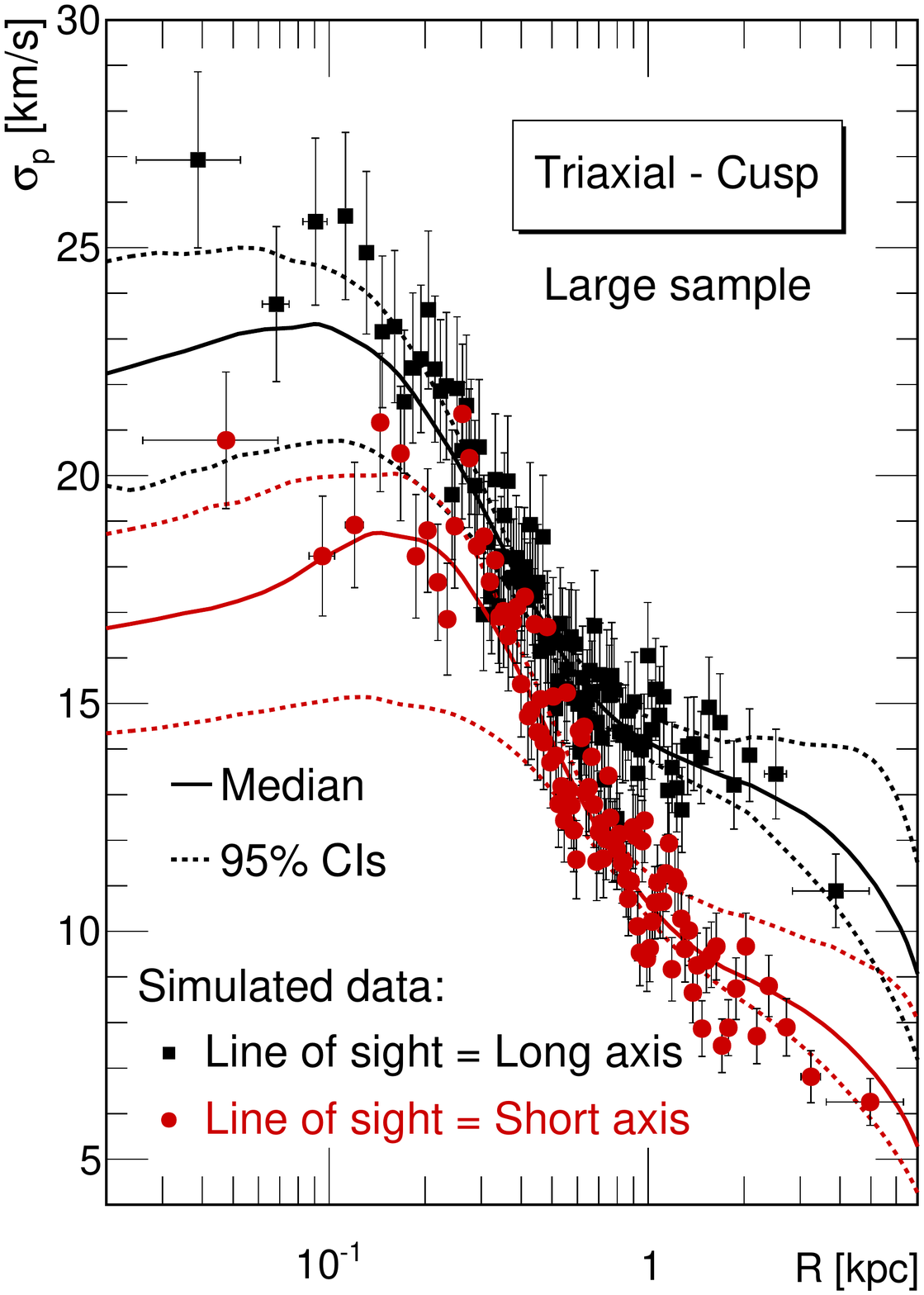}
\caption{Projection effects (along short and long axis) on the reconstructed velocity dispersion profiles (median and CIs in solid and dashed lines). The triaxial models shown are a core ({\em left panel}) and cusp ({\em right panel}), for a large-size sample.}
\label{fig:dispproftriax}
\end{figure}

Figure~\ref{fig:dispproftriax} shows the velocity dispersion profiles
obtained for the large samples (in order to emphasise the effect) of the two models (left panel for the
core, right for the cusp), when looking either along the long axis $a$
(black squares) or the short axis $c$ (circles). The projection
effects have a strong impact on the velocity dispersion: while the global shape of the profile is preserved, it is shifted to larger values when the l.o.s. alignment moves from the short to the long axis. This is expected to have a significant effect on the reconstruction of the $J$ and $D$-factors.

\subsection{Triaxiality-induced bias on $J$ for a spherical Jeans analysis}
\label{subsec:j_triax}

We run our Jeans analysis on the two triaxial models, with all the findings of the previous sections (for the DM, anisotropy, and light profiles, i.e. using the Einasto DM profile, Baes \& van Hese anisotropy and Zhao light profile). It is performed for the three sample sizes and three orientations (the light profile is fitted separately for each orientation).

\begin{figure}
\includegraphics[width=\linewidth]{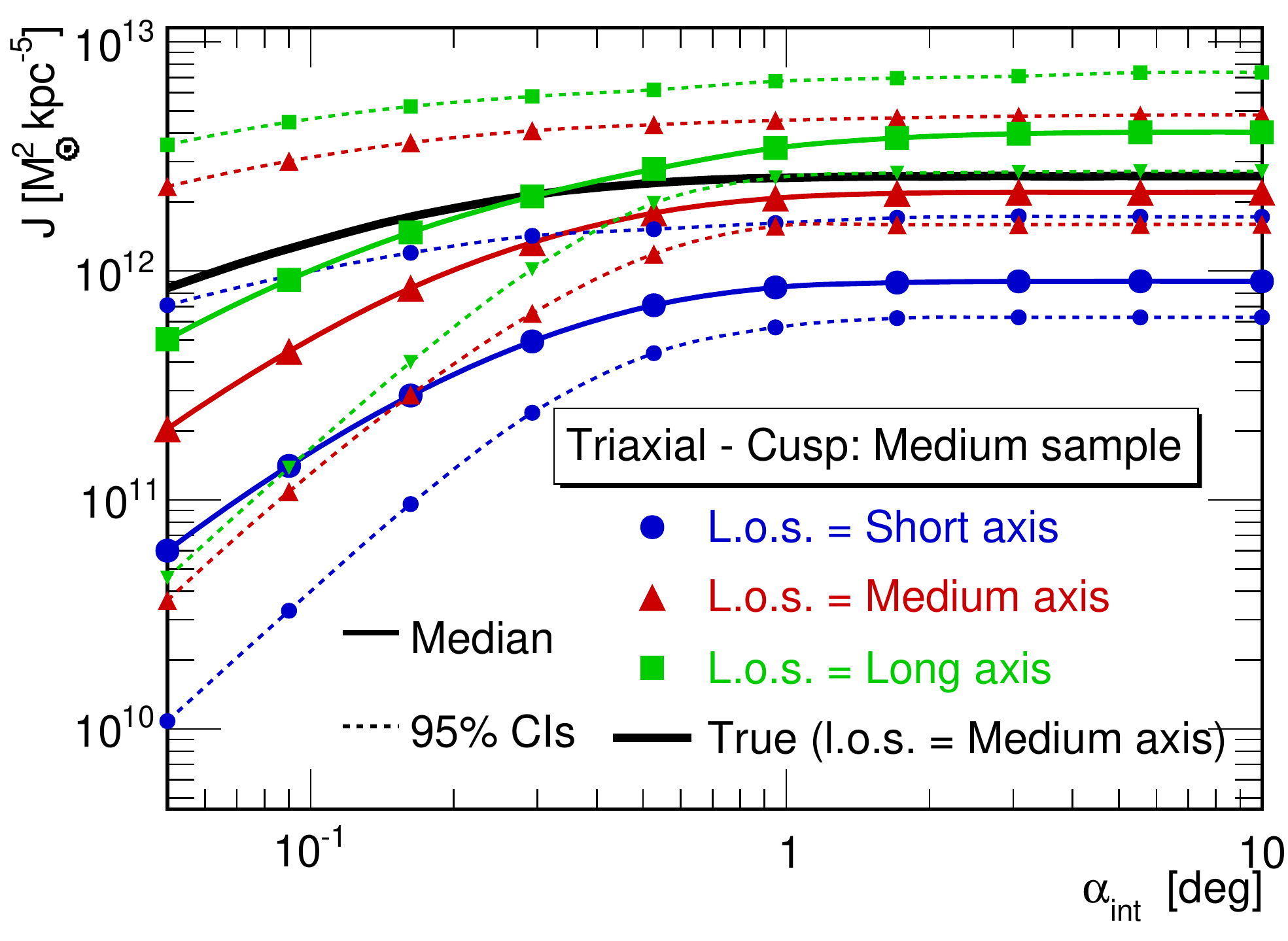}
\caption{Median (solid lines) and $95\%$ CIs (dotted lines) $J$ values reconstructed with the spherical Jeans analysis on a mock classical triaxial dSph (cuspy profile), for three l.o.s. orientations. The critical angle $\alpha_{c}$ is the same as in Fig. \ref{fig:true_j_triax}.}
\label{fig:j_triaxial}
\end{figure}

Figure~\ref{fig:j_triaxial} shows the $J$-factors obtained for the
mock classical cuspy dSph (similar results are obtained for the core
profile), for l.o.s. oriented along the short (blue circles), medium (red triangles), and long
axes (green squares). They are compared to the true value in black
solid line (only the orientation along the intermediate axis is
shown). A systematic shift appears between the three orientations,
with the $J$-factor being maximum for the l.o.s. along the long
axis. This is caused by the orientation-dependent velocity dispersion
profiles (shown in Fig.~\ref{fig:dispproftriax} for the large-size
sample). At the optimal integration angle $\alpha_{c} \simeq
1.7^{\circ}$, the reconstructed $J$-factors overshoot (or undershoot)
the true values by a factor $\lesssim 2.5$. Some of the $95 \%$ CIs
(dotted lines) do not even encompass the true value for any integration angle.

Applying the spherical Jeans analysis to triaxial halos therefore biases the $J$-factor reconstruction. The bias must be accounted for in the $J$-factor determination. We emphasise the fact that all observed dSphs display elliptical isophotes \citepads{1995MNRAS.277.1354I,2008ApJ...684.1075M} and thereby violate the common assumption of spherical symmetry. This assumption can be relaxed, for example, in the axisymmetric Jeans models of \citetads{2008MNRAS.390...71C}, or in alternative techniques that rely on orbit-based modelling \citepads{1979ApJ...232..236S} or made-to-measure models \citepads{1996MNRAS.282..223S,2010MNRAS.405..301L}.  One disadvantage of these models is that their greater computational expense inhibits the range of systematic tests that can be performed in a reasonable amount of time.  Nevertheless, we expect that the greater flexibility of such models can reduce the biases we found in tests against mock data drawn from triaxial models.

%________________________________________________________________________
\section{Conclusions}
\label{sec:conclusions}

\begin{table*}
\begin{center}
\caption{Summary of all effects discussed in the paper for annihilation and decay ($J$ and $D$-factors).
The upper block corresponds to biases induced by the choices of parametrisation and halo triaxiality. The
lower block gives the minimum (\textit{maximum knowledge}) and typical ($\rho_{\rm DM}^{\rm Einasto}$+
$\beta_{\rm ani}^{\rm Baes}$ modelling) uncertainties expected in a data-driven Jeans analysis. Note that
we show the quantity $J^{\pm 95\% \rm CI}/J^{\rm median}$ instead of $J^{\pm 95\% \rm CI}/J^{\rm true}$
(presented in Figs. \ref{fig:histo_j_cut_rs} and \ref{fig:histo_j_beta}), in order to be comparable
to the expectations of real data analyses.}
\label{table:recap}
\begin{tabular}{lcccccccccl}\hline\hline
             &        \textbf{Section} &  ~~~ &                          \multicolumn{3}{c}{\textbf{Annihilation}} & ~~~ & \multicolumn{3}{c}{\textbf{Decay}} & \textbf{Comments}  \\
             &         &  ~~~ &                          {\em Ultra-faint} & {\em Classical} & {\em Ideal} & ~~~ & {\em Ultra-faint} & {\em Classical} & {\em Ideal} &   \\
  \hline
  \textbf{Bias from:} & & & \multicolumn{3}{c}{$J^{\rm median}/J^{\rm true} (\alpha_{c}^{J})$} & & \multicolumn{3}{c}{$D^{\rm median}/D^{\rm true}(\alpha_{c}^{D})$} & \vspace{1mm}\\
  ~~~~Einasto vs Zhao & \S\ref{sec:zhao_einasto} & & none & none & none & & none & none & none  & $\rightarrow$ use Einasto + cuts$^{\ddagger}$ \vspace{1mm}\\ 
  ~~~~Wrong $\beta_{\rm ani}$ & \S\ref{sub:beta_r} & & none & $\lesssim 3$ & $\lesssim 10$   & & none & $\lesssim 2.5$ & $\lesssim 2$       & $\rightarrow$ use $\beta_{\rm ani}^{\rm Baes}$ \vspace{1mm} \\
  ~~~~Wrong $I^{\rm light}$ & \S\ref{subsec:different_light} & & $\lesssim 2$ & $\lesssim 3$ & $\lesssim 3$ & & $\lesssim 1.5$ & $\lesssim 4$ & $\lesssim 4$ & $\rightarrow$ use Zhao \vspace{1mm}\\
  ~~~~Triaxiality &\S\ref{subsec:j_triax}& & $\lesssim 2.5$ & $\lesssim 2.5$ & $ \lesssim 2.5$ & & $\lesssim 2$ & $\lesssim 2$ & $ \lesssim 2$ & Systematic uncertainty \vspace{3mm}\\
    
  \textbf{Uncertainties$^{\dagger}$:} & & & \multicolumn{3}{c}{$J^{\pm 95\% \rm CI}/J^{\rm median} (\alpha_{c}^{J})$} & & \multicolumn{3}{c}{$D^{\pm 95\% \rm CI}/D^{\rm median}(\alpha_{c}^{D})$} & \vspace{1mm}\\
  ~~~~{\em Maximum knowledge}             & \S\ref{sec:sample_size}      &             & $\lesssim 20$ & $\lesssim 2$              & $\lesssim 1.5$ & &  $\lesssim 8$ & $\lesssim 1.5$              & $\lesssim 1.25$ & DM only + cuts$^{\ddagger}$ \vspace{1mm}\\
  ~~~~$\rho_{\rm DM}^{\rm Einasto}$+ $\beta_{\rm ani}^{\rm Baes}$ modelling        &     \S\ref{subsec:baes}              &    & $\lesssim 20$ & $\lesssim 4$ & $\lesssim 2.5$   & & $\lesssim 10$ & $\lesssim 2$ &    $\lesssim 2$    & $\dots$  \vspace{1mm}\\
  
 \hline
\end{tabular}\\
\vspace{1mm}
{\small 
$^{\ddagger}$ Enforce $r_{s} \geq r_{s}^{\star}$ and $\alpha \geq 0.12$ in the priors. \\
$^{\dagger}$ Light profile uncertainties have a very small effect on $J$ and $D$ at $\alpha_{c}$, and are not shown here. \\
}
\end{center}  
\end{table*}
We have studied the impact of the different ingredients of the spherical Jeans analysis on the reconstruction of the astrophysical factors
$J$ and $D$ (respectively for annihilating and decaying DM) of dSph galaxies. 
We find that the assumptions made regarding the ingredients of the
analysis (dark matter, velocity anisotropy and light profiles; spherical symmetry) may
significantly impact those quantities.
Coupling
the Jeans analysis to an MCMC engine and relying on a set of mock dSph
galaxy data, we were able to quantify the biases (seen as trends for systematic offsets of the reconstructed median values from the true ones) and uncertainties (assessed by the width of the 95 \% credibility intervals) associated to each assumption, using three sizes of mock samples to mimic datasets of ultra-faint (small sample), classical
(median sample), and `ideally observed' (large sample) dSph galaxies. Table~\ref{table:recap} summarises the main findings of this study.

\paragraph*{Impact of the various ingredients on $J$- and $D$-factors}
\begin{itemize}

  \item {\em DM profile:} given the precision and the small spatial range
    covered by velocity dispersion measurements, we find that it is equivalent
    to use a Zhao or an Einasto DM parametrisation in the Jeans
    analysis. We recommend the use of the Einasto profile as the
    smaller number of parameters produces less degeneracies
    and faster MCMC analyses than for the Zhao case. To avoid
    extremely large upper limits (up to factors $\sim 10^{6}$ for $J^{+95\%\rm CI}/J^{\rm median}$ of mock
    ultra-faints), coming from the sampling of unrealistic models, it
    is necessary to put weak priors on the scale radius and slope
    of the Einasto profile, namely  $r_s\geq r_s^\star$ and $\alpha \geq 0.12$;

  \item {\em Velocity anisotropy profile:} making the wrong assumption
    on the anisotropy profile parametrisation can lead to strongly
    biased astrophysical factors (with median $J$ values up to factors of a few above or below the true values depending on the sample
    size, with the 95\%~CIs not encompassing the true values). Instead
    of using a constant anisotropy profile (as done in many studies),
    we recommend the use of a flexible profile such as that of Baes \& van Hese. The
    latter encompasses both the constant and Osipkov-Merritt
    profiles, and its four
    parameters allow for more flexibility in the fit.
  
  \item {\em Light profile:} using the wrong light profile
    parametrisation can lead up to a factor $\gtrsim 10$ bias (of the
    astrophysical factor) at large integration angles $\alpha_{\rm
      int}\gtrsim \alpha_c$, and $\sim 3$ below. We recommend the use
    of a Zhao profile for the light as it appears a good choice to fit the
    generally well-sampled light profiles. Propagating the errors on
    the light profile to the astrophysical factor makes no significant
    difference as these errors are always much smaller compared to
    other uncertainties, regardless of the sample size.

  \item {\em Triaxiality:} the last important effect investigated is
    the use of a spherical analysis on DM halos that are likely to be
    triaxial. First, even with a perfectly known DM profile,
    projection effects lead to a $\sim 30\%$ systematic effect, depending on
    the DM halo orientation with respect to the l.o.s. (for
    $\alpha_{\rm int}\lesssim \alpha_c$). Second, projection effects
    in the velocity dispersion data lead to $J$ and $D$ values that
    can undershoot or overshoot the true value by a factor of a
    few. This systematic effect must be accounted for separately in
    the error budget, since the orientations of the dSph galaxies
    remain unknown.

\end{itemize}
When all these effects are taken into account, we confirm that (i) the
inner slope of DM profiles for dSph galaxies is not well-constrained by the Jeans analysis, even when the well-known {\em velocity anisotropy - mass} degeneracy is broken; (ii) the astrophysical factor can nonetheless be well constrained; and (iii) the critical angle (for which the uncertainty is the smallest) found for the $J$-factor also exists for the $D$-factor, with $\alpha_{c}^{D} \approx r_{s}^{\star}/d = \alpha_{c}^J/2$.

\paragraph*{Conclusions for the different sample sizes}
\begin{itemize}
   \item {\em Ultra-faint dSph galaxies:} these objects are the most
     promising dSph galaxy targets for indirect detection as some of them
     are found very close to us ($\sim 20-40$~kpc). However, they are
     also the most uncertain because of the few kinematic data
     available. In order to get the best constraints on these objects
     in a `data-driven' analysis, a cut on the prior of the scale radius ($r_s\geq
     r_s^\star$) is mandatory. The statistical uncertainties linked to the
     sparsity of the data always dominate the error budget, regardless
     of the Jeans modelling (anisotropy, light and DM profile
     parametrisations): we typically find $J^{\pm 95\%\,{\rm CI}}/J^{\rm median}\lesssim 20$ at $\alpha_{\rm int}=\alpha_c$, and $\lesssim 100$ at  $0.1~\alpha_c$ and $10~\alpha_c$.

   \item {\em Classical dSph galaxies:} these objects have already
     been analysed in a similar framework in
     \citetads{2011MNRAS.418.1526C}. The analysis of this paper goes
     further, with several new identified sources of biases and
     uncertainties. On the one hand, the cuts on the priors were
     not all included in \citetads{2011MNRAS.418.1526C}, so that the
     CIs obtained by these authors may be slightly overestimated. On
     the other hand, the use of more generic anisotropy and light
     profiles, and the systematic effect of triaxiality may slightly
     shift and increase these errors. A re-analysis is required to get
     better estimates on real data. In any case, typical uncertainties
     on the 95\% CIs for these objects are $J^{\pm 95\%\,{\rm
         CI}}/J^{\rm median} \lesssim 4$ at $\alpha_{\rm int}=\alpha_c$, and $\lesssim 10$ at $0.1~\alpha_c$ and $10~\alpha_c$.

   \item {\em `Ideally observed' dSph galaxies:} future instruments
     may provide much more precise and numerous kinematical data,
     if the observed objects contain enough stars (which is
     the ultimate limitation). If this is the case, halo triaxiality will
     need to be handled by non-spherical Jeans analyses
     (see e.g., \citealtads{2012ApJ...755..145H}) as it will then become
     the main source of \textit{bias} in the spherical Jeans analysis. The main source of \textit{uncertainty} in this
     ideal case is the velocity anisotropy profile. Higher-order Jeans analyses will probably
     appear as very helpful tools for reducing these uncertainties
     (see e.g., \citealtads{2013MNRAS.432.3361R,2014MNRAS.441.1584R,2014MNRAS.440.1680R}). However,
     we have found that even in the context of a {\em maximum
       knowledge} analysis (i.e., with the light profile and velocity
     anisotropy perfectly known), the best that can be achieved is $J^{\pm 95\%\,{\rm
         CI}}/J^{\rm median}\lesssim 1.5$ at $\alpha_{\rm int}=\alpha_c$, and $\lesssim 2.5$ at  $0.1~\alpha_c$ and $10~\alpha_c$.
\end{itemize}

\paragraph*{Suggestions and future studies}

The use of mock data has lead to the determination of an optimal
strategy to get the least biased and best constrained results on the
astrophysical factors. In the context of `data-driven' spherical Jeans
analyses, we recommend the use of the Einasto DM profile (with
$r_s\geq r_s^\star$ and $\alpha \geq 0.12$), the Baes \& van Hese velocity
anisotropy profile, the Zhao profile for the light, and accounting for a possible systematic bias from
triaxiality. In a forthcoming article, we will analyse all dSph galaxy data available to us using this optimal setup (in prep.).

%________________________________________________________________________
\section*{Acknowledgements}

We thank the anonymous referee for his careful reading of the manuscript that led to many useful improvements.
We are very grateful to M.~I.~Wilkinson for allowing us to use the mock data from \citetads{2011MNRAS.418.1526C}. V. Bonnivard would like to warmly thank L. Derome and A. Putze for their help using and interfacing the {\tt GreAT} package and for the useful discussions about the MCMC. This work has been supported by the ``Investissements d'avenir, Labex ENIGMASS". This study used the CC-IN2P3 computation centre of Lyon.

%________________________________________________________________________
\appendix

%________________________________________________________________________
\section{MCMC optimisation}
\label{app:MCMC_opt}
\begin{figure*}
\includegraphics[width=0.498\linewidth]{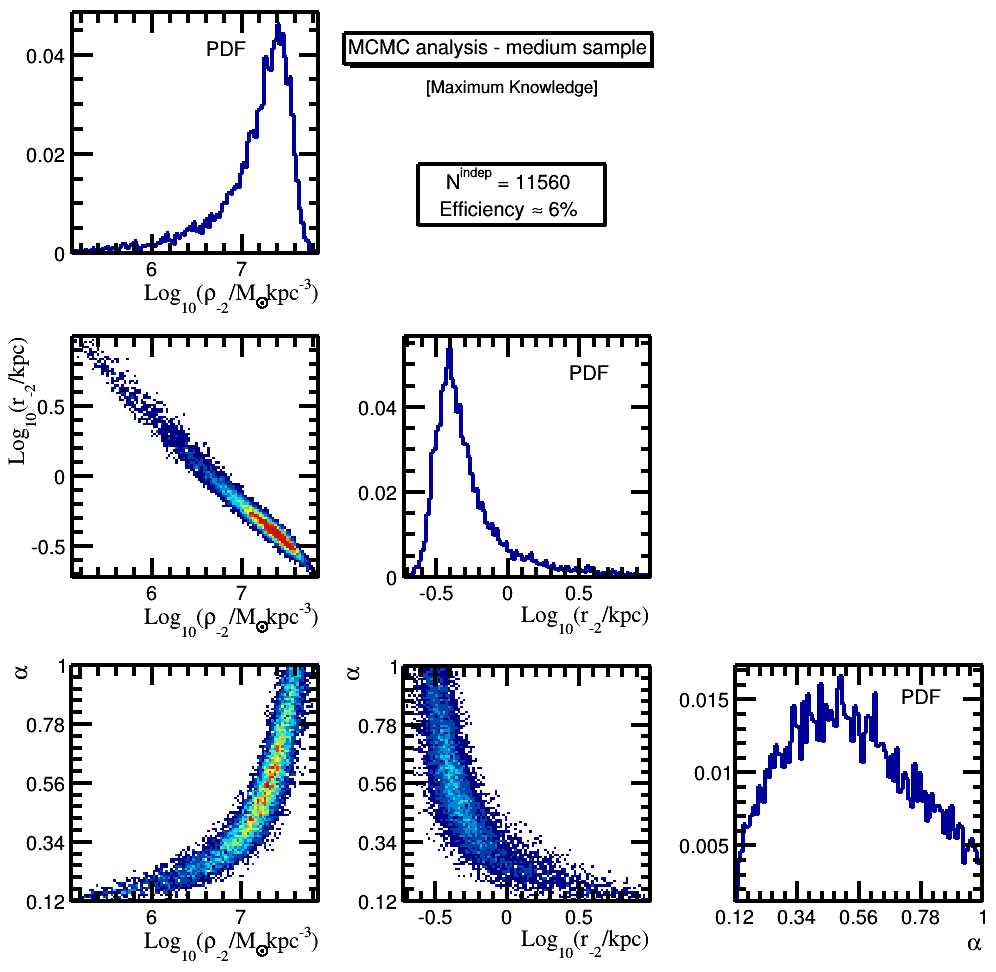}
\includegraphics[width=0.498\linewidth]{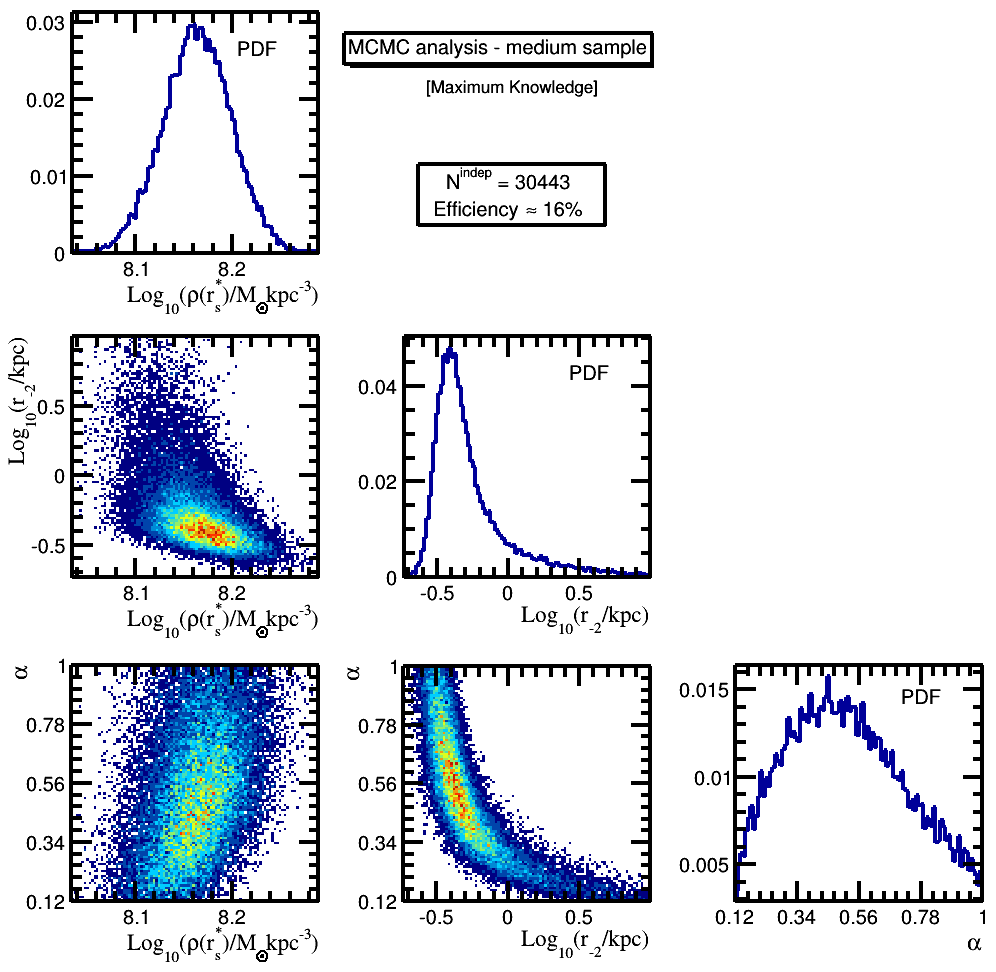}
\caption{Correlations and PDFs of three parameters describing the Einasto DM profile, for a \textit{maximum knowledge} run on a mock classical dSph. {\em Left panel}: MCMC analysis with the combination of parameters $\left\{\log_{10}(\rho_{-2}), \log_{10}(r_{-2}), \alpha \right\}$, with the priors and cuts of Table~\ref{table:priors_DM}. Because of the non-Gaussianity, the efficiency is rather poor ($\sim 6~\%$). {\em Right panel}: MCMC analysis  with the combination of parameters $\left\{\log_{10}(\rho(r_{s}^{\star})), \log_{10}(r_{-2}), \alpha \right\}$. The first parameter is now approximately Gaussian, and the correlations are much weaker. The efficiency is significantly larger ($\sim 16~\%$).}
\label{fig:mcmc}
\end{figure*} 

Our MCMC analysis relies on a multivariate Gaussian proposal function. We recall that the closer to the proposal function the target distribution, the better the MCMC efficiency. The latter is defined as the ratio between the number of accepted points (after taking into account the burn-in and correlation lengths) and the number of computed points.

Figure~\ref{fig:mcmc} shows the correlations and PDFs of the three parameters describing the Einasto DM profile, for a \textit{maximum knowledge} analysis on a mock classical dSph galaxy. The two panels correspond to the output of two MCMC runs, based on two different combinations of the Einasto parameters. Using $\left\{\log_{10}(\rho_{-2}), \log_{10}(r_{-2}), \alpha \right\}$ (left panel) leads to strongly correlated parameters, with long tails for both $\rho_{-2}$ and $r_{-2}$, hence a rather poor efficiency ($\sim6~\%$). Using $\left\{\log_{10}(\rho(r_{s}^{\star})), \log_{10}(r_{-2}), \alpha \right\}$ instead increases the efficiency almost threefold ($\sim16~\%$): the DM density is actually best constrained at $r = r_{s}^{\star}$ (see also Fig.~\ref{fig:cut_rs}), and the variable $\log_{10}(\rho(r_{s}^{\star}))$ is close to be Gaussian distributed (top right panel of Fig.~\ref{fig:mcmc}). Note that this second combination of parameters can also be made for a Zhao DM profile.

The prior used here for $\log_{10}(\rho(r_{s}^{\star}))$ is similar to the one we used on $\log_{10}(\rho_{-2})$: it is flat within the range [5, 13]. We have checked that using this prior as well as the usual priors of Table \ref{table:priors_DM} on $\log_{10}(r_{-2})$ and $\alpha$, we recover a flat distribution on $\log_{10}(\rho_{-2})$. Therefore, these two combinations of parameters lead to the same prior distributions, and give the same results.

We have focused in this study on the use of uniform priors, but there is actually no clear answer regarding the best choice of prior distribution. It can however have a strong impact on the results, particularly for ultra-faint dSphs for which the data constrain only weakly the physical parameters.

%________________________________________________________________________
\section{Optimal integration angle for $J$ and $D$}
\label{app:d_crit}

\citetads{2011ApJ...733L..46W} have shown that the integration angle
\begin{equation}
  \alpha_{c}^J \approx 2\times \frac{r_s^\star}{d}\,,
  \label{eq:alphacJ}
\end{equation}
with $r_s^\star$ the half-light radius (i.e. the scale radius of the
Plummer profile used in their analysis) and $d$ the distance to the dSph, is a good
compromise between maximising the $J$-factor and minimising its
uncertainties. We have found here that this result holds for all data
sample sizes, and for all light profiles (Plummer, Zhao, S\'ersic,
exponential and King) when using the scale radius of each profile accordingly.

\begin{figure}
\includegraphics[width=\linewidth]{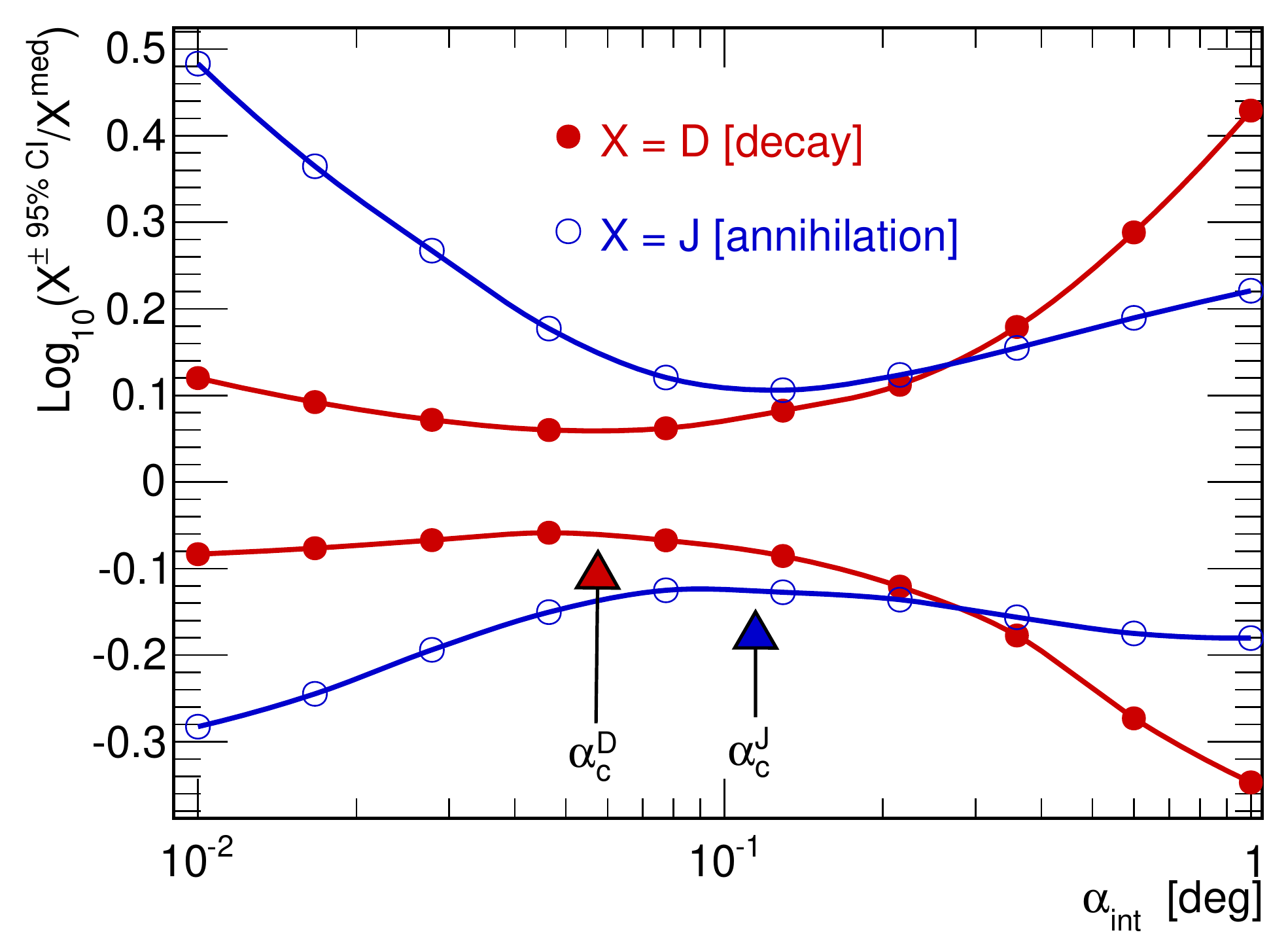}
\caption{Ratio $X^{\pm 95 \% \rm CI}/X^{\rm median}$ for annihilating
  DM ($X=J$, blue empty circles) and decaying DM ($X=D$, red filled
  circles). The arrows show the position of the {\it optimal} angles $\alpha_{c}^J$ (Eq.~(\ref{eq:alphacJ})) and $\alpha_{c}^D$ (Eq.~(\ref{eq:alphacD})), for this peculiar mock classical dSph galaxy. These angles coincide with the pinch of the CIs for which the uncertainty on $X$ is minimised.}
\label{fig:crit_angle}
\end{figure}

For the $D$-factor, we also find a similar optimal integration angle, namely
\begin{equation}
  \alpha_{c}^{D} \approx \frac{r_{s}^{\star}}{d} = \frac{\alpha_{c}^J}{2}\,.
  \label{eq:alphacD}
\end{equation}
The factor two difference with the annihilation case presumably comes from the non-squared DM profile
involved in the $D$-factor w.r.t. to the $J$-factor.
This is illustrated, for a mock classical dSph analysed in the \textit{maximum
  knowledge} setup, in Fig.~\ref{fig:crit_angle} which shows the
ratios of the upper and lower 95\% CIs to the median values of both
the $J$- (blue empty circles) and $D$-factors (red filled circles).

%________________________________________________________________________
\section{Impact of anisotropy parametrisation on the dark matter density profile and mass reconstruction}
\label{app:bimodality}

\begin{figure*}
\includegraphics[width=0.33\linewidth]{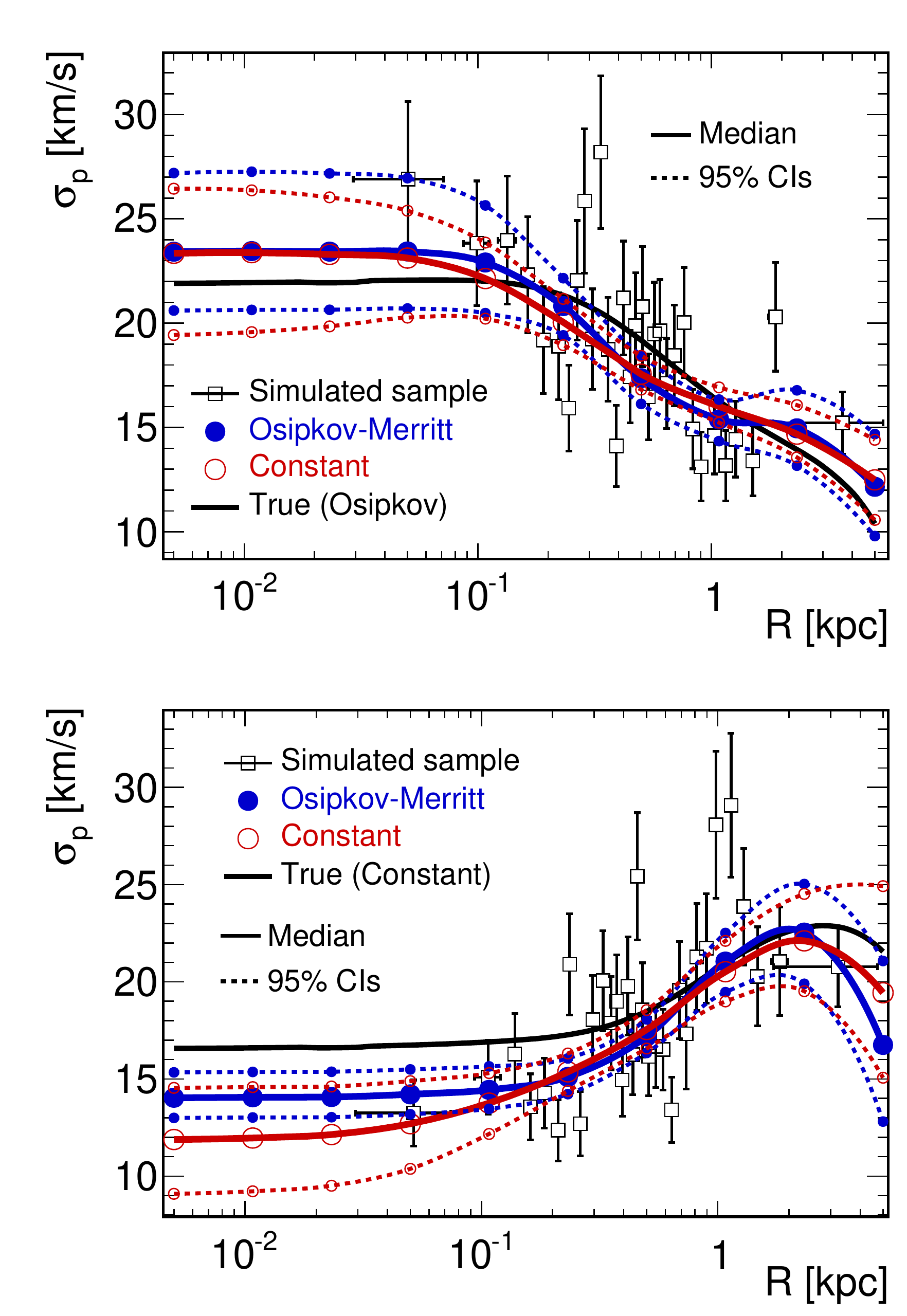}
\includegraphics[width=0.33\linewidth]{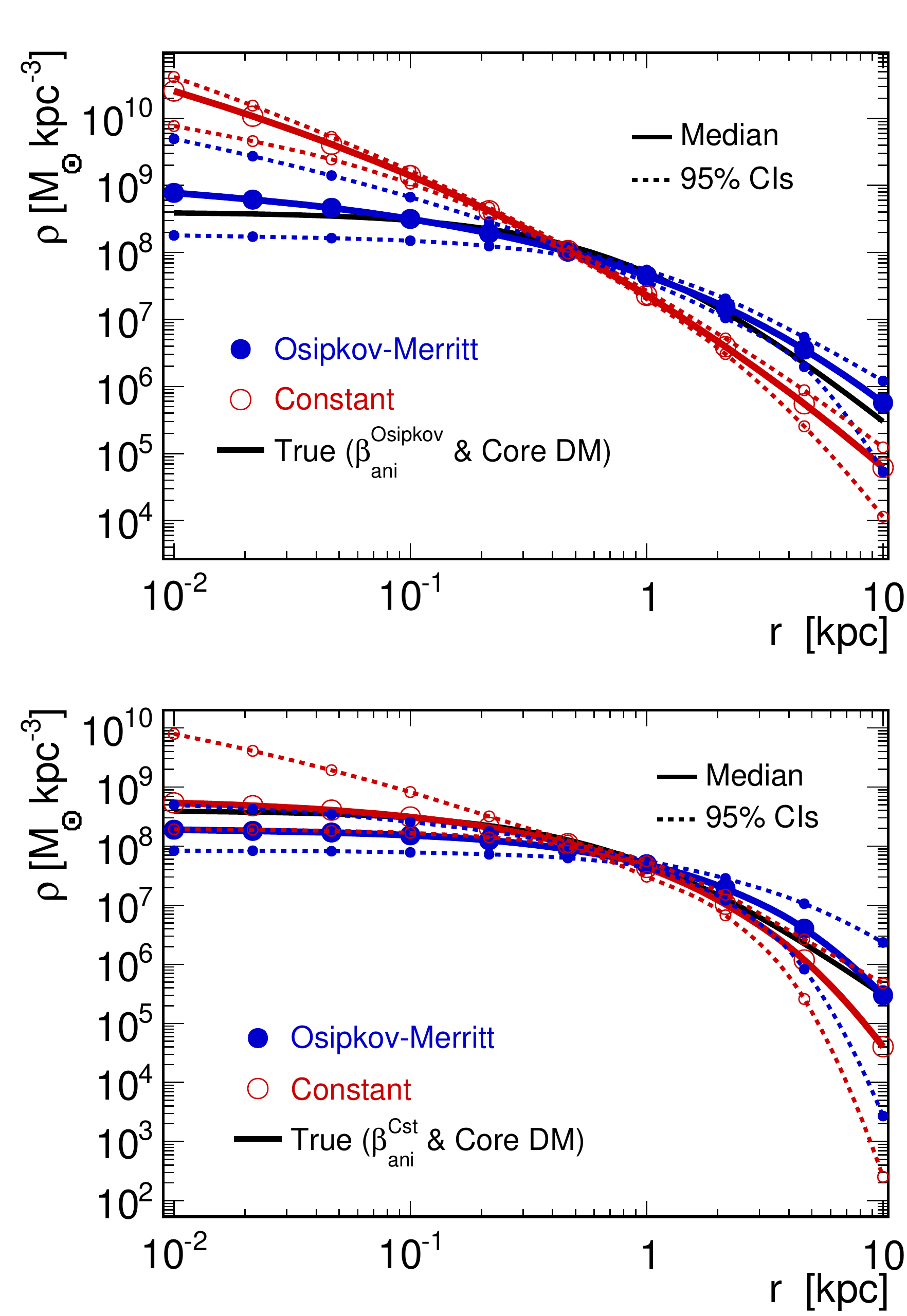}
\includegraphics[width=0.33\linewidth]{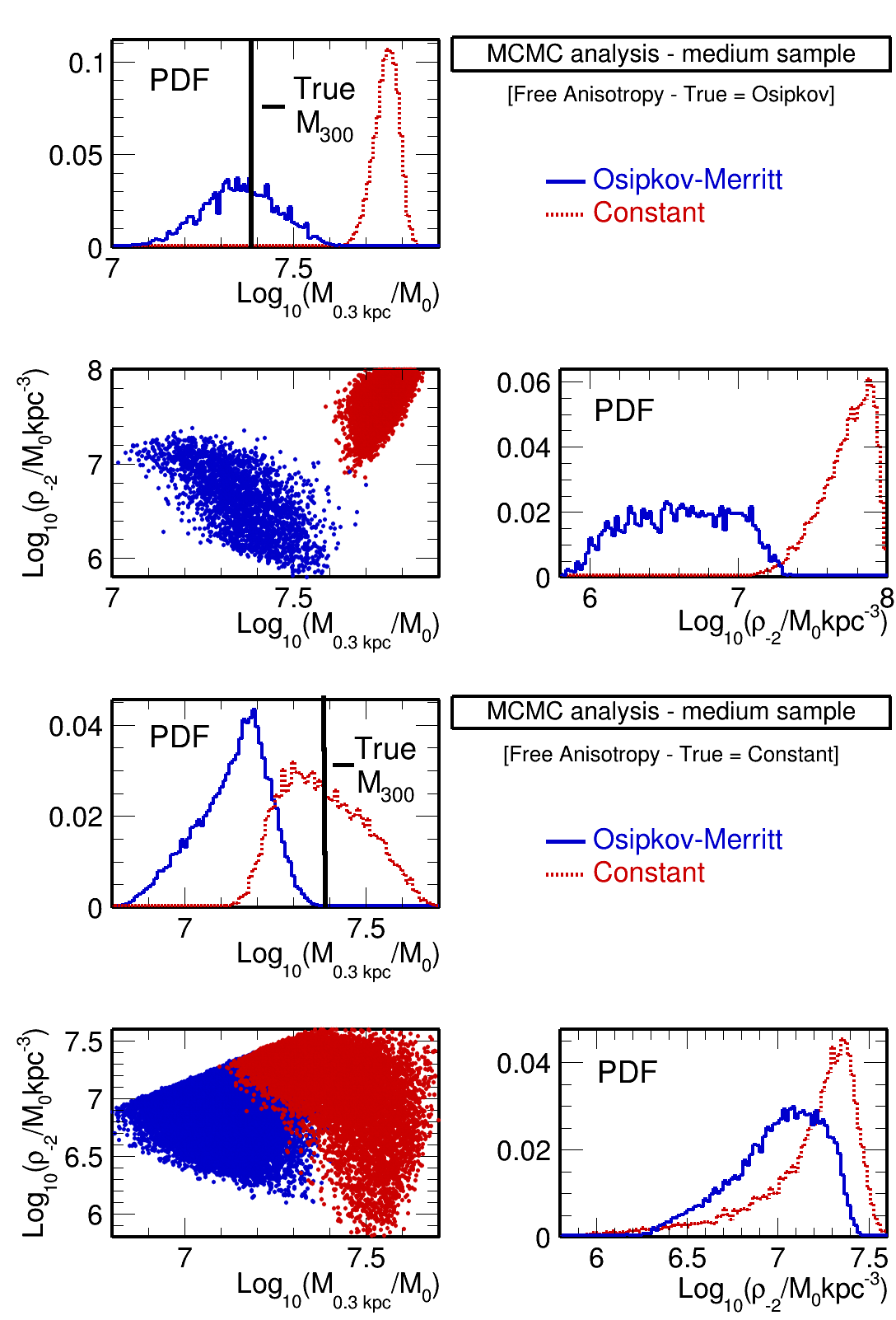}
\caption{Velocity dispersion profile (left), DM density profile
  (middle) and PDFs of $M_{300}$ (mass at $r = 300$ pc) and
  $\rho_{-2}$ (right), reconstructed for a mock cored classical dSph galaxy
  with an Osipkov-Merritt ({\em top}) or a constant ({\em bottom})
  velocity anisotropy. For each model, the Jeans analysis is run using either an Osipkov-Merritt (blue) or a constant (red) anisotropy. Both give similar fits to the velocity dispersion profiles, but using the wrong anisotropy profile leads to strong biases on both the DM density profile and the mass at 300 pc.}
\label{fig:bimodality}
\end{figure*}

This appendix further explores the impact of using the wrong
velocity anisotropy parametrisation on the DM density profile and mass
reconstruction. We refer the reader to section \ref{sub:beta_r} for the details
of the analysis. We show in Fig.~\ref{fig:bimodality} the velocity
dispersion profile (left), the DM density profile (middle) and the PDFs and correlation (right) of the mass at $r=300$ pc
$M_{300}$ and of $\rho_{-2}$, for two typical mock
classical dSph galaxies. In the top row, the model follows an
Osipkov-Merritt parametrisation while in the bottom row the model has a constant anisotropy. In each case, the Jeans analysis is run using either a constant or an Osipkov-Merritt anisotropy.

The fit to the velocity dispersion profile (Fig.~\ref{fig:bimodality}, left) is always satisfactory,
whether we use the right anisotropy parametrisation or not. However, the
effect on the reconstructed DM density profile (middle) can be very strong. For
example, using a constant anisotropy for the model with an
Osipkov-Merritt anisotropy (Fig.~\ref{fig:bimodality}, top
row -- middle column, red empty circles)  leads to a cuspy DM density
profile whereas the true profile is a core (black
solid line). When we look at the PDFs (top row - right column), we see
the distinct populations arising from using either a constant
(red) or an Osipkov-Merritt (blue) profile. The effect is a bit less
pronounced for the model with a constant anisotropy (bottom panels).

Using the wrong anisotropy parametrisation can therefore produce
strong biases on the DM density profile, that in turns, impact both
the $J$ and $D$-factors but also the estimated mass of the object. Using a Baes \& van Hese anisotropy profile allows to mitigate this effect (see section \ref{sub:beta_r}).

%________________________________________________________________________
\section{Results for $D$-factors (decaying DM)}
\label{app:Dplots}

Many DM models correspond to stable particles, thermally produced in
the early Universe. DM could also consist of unstable long-lived
particles, the decay of which could generate $\gamma$ or $X$-rays,
\textit{via} prompt and inverse Compton emissions
\citepads{2013IJMPA..2830040I}. In this scenario, exotic signals have been looked for in
the Milky Way \citepads{2010NuPhB.840..284C}, in M31
\citepads{2008MNRAS.387.1361B}, in clusters of galaxies
\citepads{2010JCAP...12..015D,2012PhRvD..85f3517C,2012PhRvD..86h3506C}
and in dSphs \citepads{2009PhRvD..80b3506E,2010ApJ...720.1174A,2012PhRvD..85f2001A}
using data sets from various instruments (Fermi-LAT, IACTs,
XMM-Newton). Non-detections have led to constraints on the lifetime $\tau$ of the decaying DM particle. 
Note however that an exotic $X$-ray line at 3.5 keV has recently been identified in M31 and in
clusters of galaxies, that could originate from DM decay \citepads{2014ApJ...789...13B,2014arXiv1402.4119B}.
As in the case of annihilation, a careful estimation of the astrophysical $D$-factors and of their uncertainties are required in order to derive those constraints.

The astrophysical $D$-factor corresponds to the integration
along the line of sight of the DM density (where the $J$-factor requires the
DM density \textit{squared}). Compared to the $J$-factor, the
$D$-factor is therefore less sensitive to the uncertainties on the DM density profile.
In this appendix, we briefly review the impact on the $D$-factor of the different
effects discussed in the body of this paper.

\begin{figure}
\includegraphics[width=\linewidth]{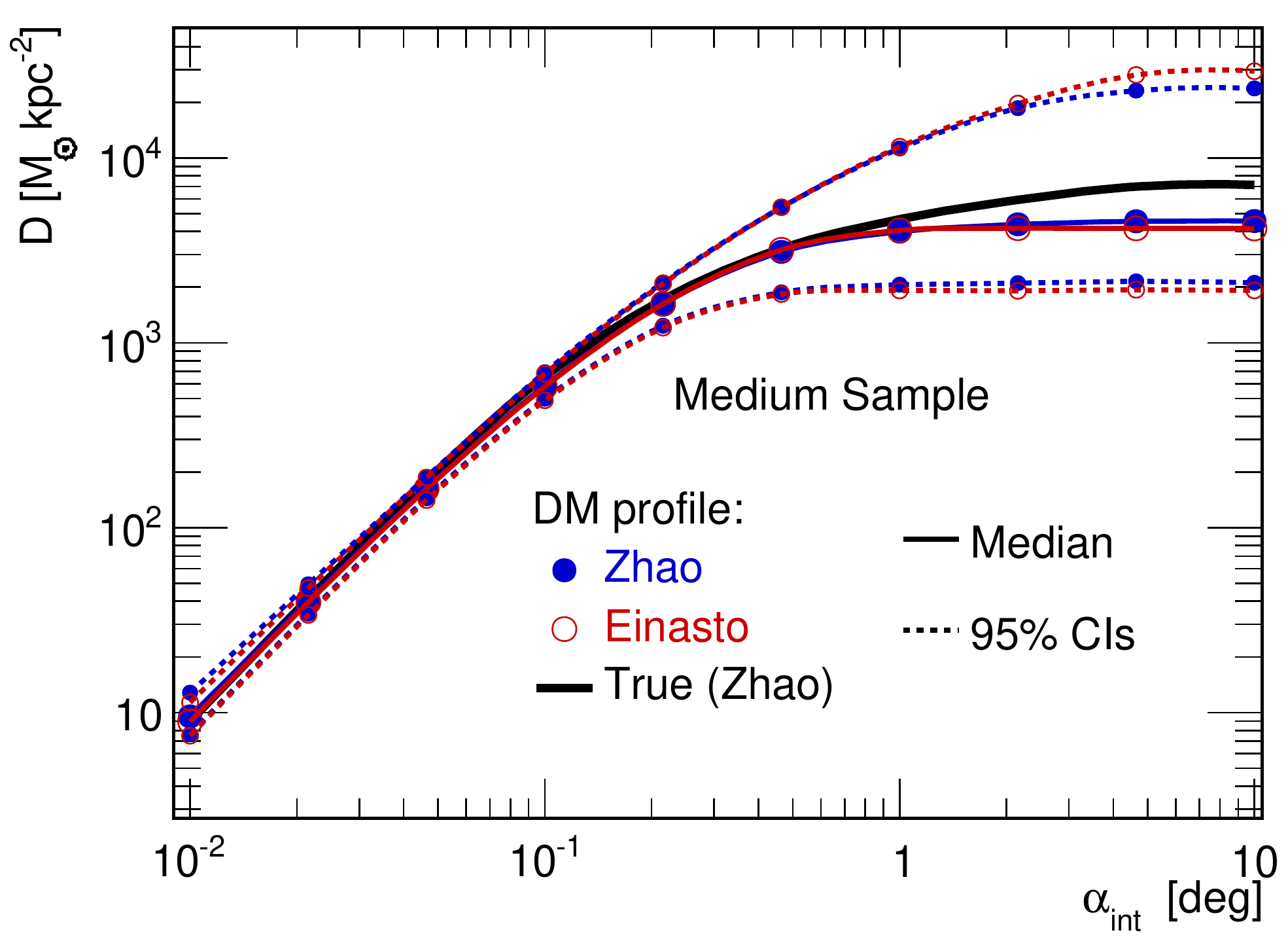}
\caption{$D$-factor median values (solid lines with symbols) and
  95\% CIs (thin dashed lines with symbols) for a mock classical 
  dSph galaxy. MCMC/Jeans analyses with a Zhao (blue filled circles)
  or a Einasto (red empty circles) DM profile give similar results, as for the $J$-factor.}
\label{fig:d_einasto_zhao}
\end{figure}

\paragraph*{DM modelling: maximum knowledge setup}
\begin{itemize}
\item{\textit{Optimal cut} $r_{s} \geq r_{s}^{*}$}. The effect of this
  cut is less pronounced for the $D$-factor than for the $J$-factor:
  for mock ultra-faint dSphs, it slightly reduces the upper CIs
  at small integration angles, and the lower CIs at large angles. This is
  shown, for a typical model, in the bottom right panel of
  Fig. \ref{fig:cut_rs}. As for the $J$-factor, this cut is always
  applied for the $D$-factor determination.
\item{\textit{Zhao vs Einasto}}. Similarly to $J$-factors, using an
  Einasto or a Zhao DM parametrisation gives very comparable
  $D$-factors. Figure \ref{fig:d_einasto_zhao} shows the $D$-factor
  obtained using either a Zhao (blue filled circles) or an Einasto
  (red empty circles) profile, for a mock classical dSph galaxy
  (\textit{maximum knowledge} setup); similar results for median and CI
  values are found in the two cases.
\item{\textit{Sample size}}. The size of the sample also plays a major role
  on the $D$-factor uncertainties. The $D$-factors are less sensitive to the
  DM profile uncertainties than the $J$-factors, and are more
  tightly constrained. Using the $D^{+95\%\; {\rm CI}}/D^{\rm true}$ values obtained for the 64 models from \citetads{2011ApJ...733L..46W}
  (see Table \ref{table:mock_data}), the $D$-factor relative uncertainty at $\alpha_{c}^{D}$ is found (not shown) to be at most $\sim 1.5$ (resp. $\sim 8$) for the mock classical (resp. ultra-faint) dSph galaxies, while the corresponding $J$-factor uncertainty is at most $\sim 3$ (resp. $\sim 25$).
\end{itemize}

\begin{figure}
\includegraphics[width=\linewidth]{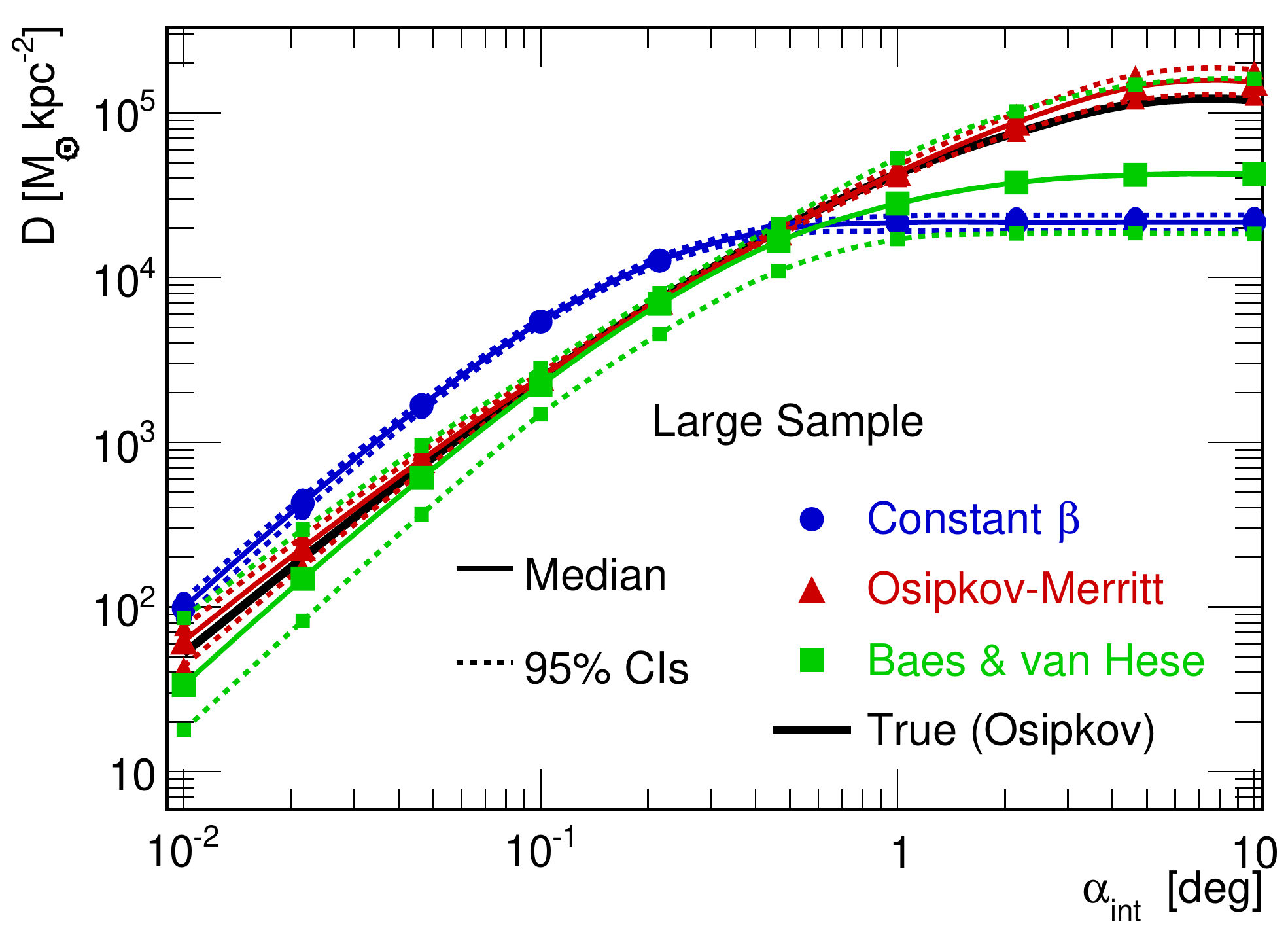}
\caption{Median values (solid lines with symbols) and $95 \%$ CIs
  (dotted lines with symbols) of $D(\alpha_{\rm int})$ for a mock dSph
  (large sample) generated with an Osipkov-Merritt velocity
  anisotropy. The true $D(\alpha_{\rm int})$ is given in solid black. The $D$-factor has been reconstructed using different
  anisotropy prescriptions: i) constant (blue circles), ii)
  Osipkov-Merritt (i.e., the correct parametrisation, red triangles)
  and iii) Baes \& van Hese (green
  squares).}
\label{fig:d_baes}
\end{figure}

\paragraph*{Anisotropy profile}
\begin{itemize}
\item{\textit{Optimal cut} $\alpha \geq 0.12$}. We found in
  Sect.~\ref{sec:anisotropy} that when a constant velocity anisotropy is free to
  vary in the Jeans analysis, the cut $\alpha \geq 0.12$ on the
  Einasto shape parameter significantly reduces the uncertainties on
  the $J$-factor. For the $D$-factor, the cut has no strong effect:
  this quantity is less sensitive to the steepness of the DM profile
  in the inner parts. Nevertheless, for consistency with the $J$-factor
  reconstruction, we also implement this cut for $D$-factor calculations.
\item{$\beta_{\rm ani}^{\rm Cst}$ \textit{vs} $\beta_{\rm ani}^{\rm
    Osipkov}$}. Results similar to the $J$-factor ones are
  found when using the wrong anisotropy profile parametrisation: the
  $D$-factor reconstruction can be strongly biased. We show in
  Fig. \ref{fig:d_baes} the $D$-factor obtained for a mock dSph (large
  sample) generated with an Osipkov-Merritt anisotropy, using either a
  constant (blue circles) or an Osipkov-Merritt (red triangles) model in
  the Jeans analysis. The reconstructed $D$-factor is compatible with the true one (black
  solid line) only when the right parametrisation is used.

\item{$\beta_{\rm ani}^{\rm Baes}$ \textit{analysis}}. Using the more general Baes \& van Hese anisotropy profile (Eq. \ref{eq:beta_baes}) allows to mitigate these biases for medium and large samples. This profile leads to larger CIs (typically, they are $\sim 25\%$ larger at $\alpha_{c}$), but less biased median values, as shown in green squares in Fig. \ref{fig:d_baes}. For mock ultra-faint dSph galaxies, the $D$-factors obtained with the three anisotropy profiles used here are compatible. We also advocate the use of this general parametrisation for the $D$-factor determination.
\end{itemize}

\begin{figure}
\includegraphics[width=0.5\linewidth]{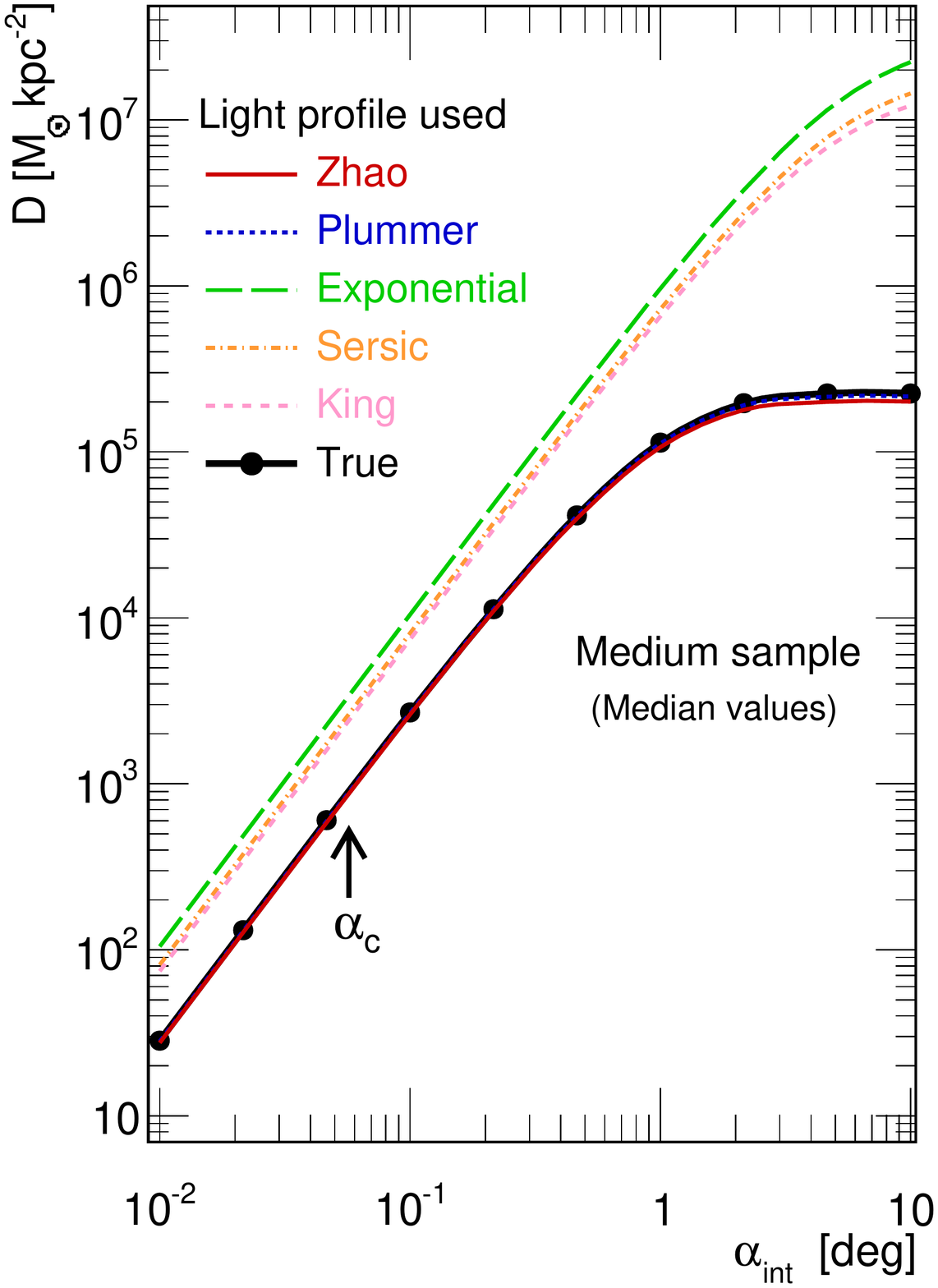}
\includegraphics[width=0.5\linewidth]{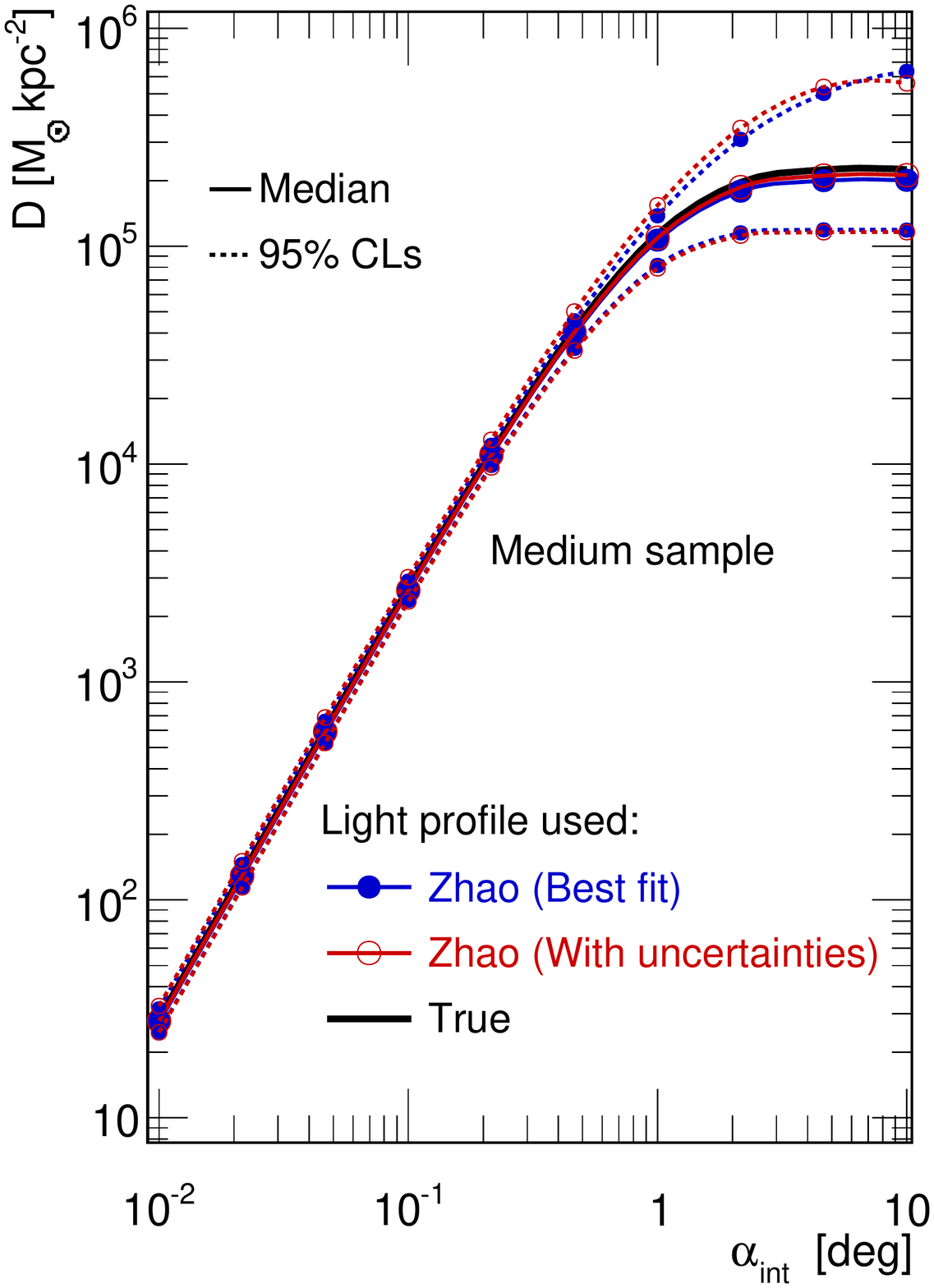}
\caption{{\em Left panel:} $D(\alpha_{\rm int})$ obtained when
  using each of the five best-fit light profiles in the Jeans modelling
  |see \S\ref{sec:diff_light_prof}. {\em Right panel:} Propagation to $D(\alpha_{\rm int})$
  of the uncertainties on a Zhao light profile fit. The same mock dSph galaxy (with medium-size
  velocity dispersion sample) has been used for the two plots.}
\label{fig:I_D}
\end{figure}

\paragraph*{Light profile}
The conclusions reached in
  section \ref{sec:diff_light_prof} for $J$-factors hold for
  $D$-factors: the light profile parametrisation plays a significant
  role in the $D$-factor reconstruction, whereas propagating the light
  profile uncertainties has a weak effect only. This is illustrated in
  Fig.~\ref{fig:I_D}, for the same model as in Fig.~\ref{fig:I_J}: we
  show the $D$-factors obtained for a mock classical dSph galaxy using the
  five different light profile parametrisations of section
  \ref{sec:diff_light_prof} (left panel), as well as the effect of
  propagating the light profile uncertainties (right panel). For a
  Jeans analysis dedicated to $D$-factor determination, we therefore
  advocate the use of a very general light profile parametrisation.

\begin{figure}
\includegraphics[width=\linewidth]{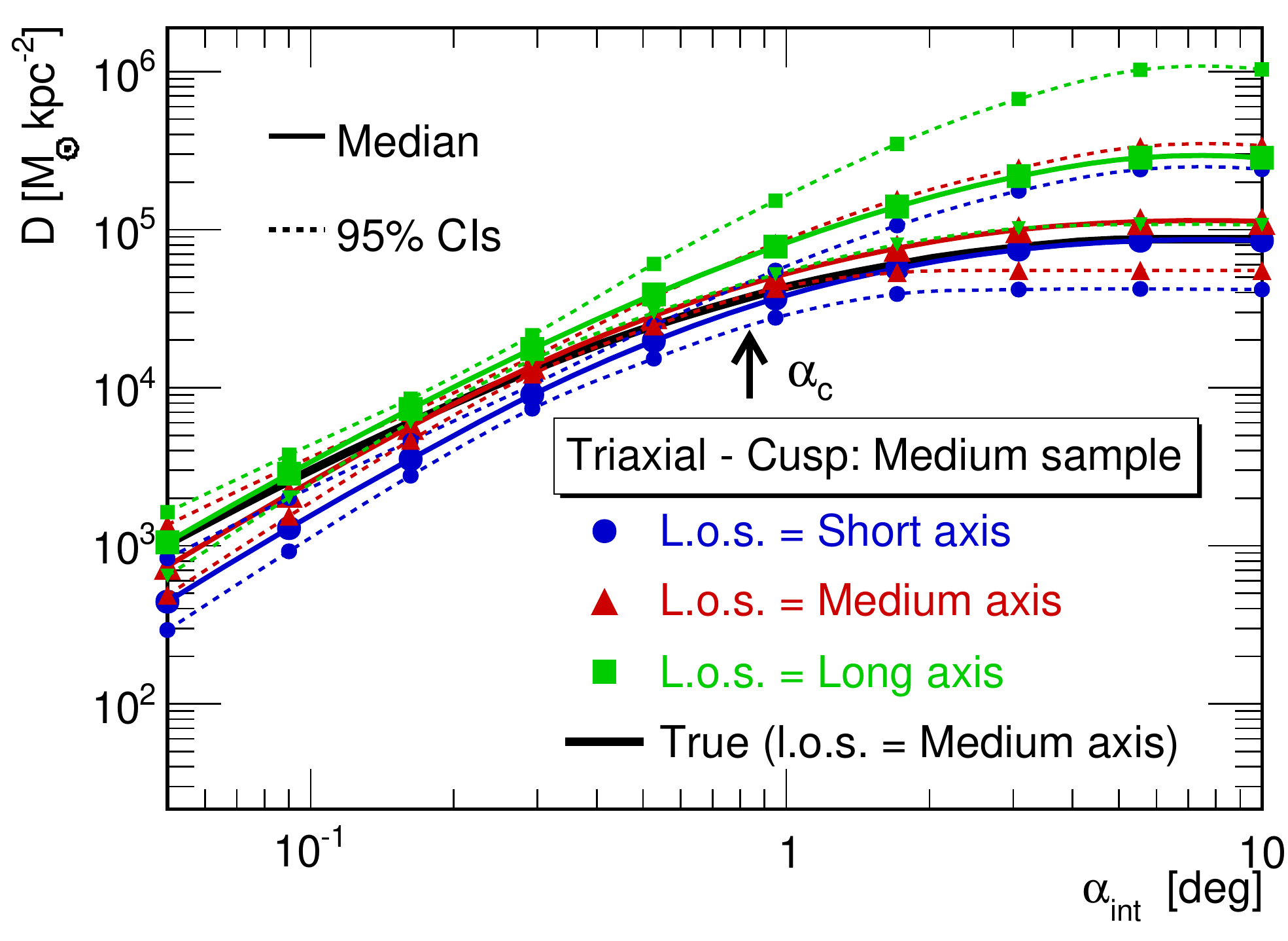}
\caption{Median (solid lines) and $95\%$ CIs (dotted lines) $D$ values reconstructed with the spherical Jeans analysis on a mock classical triaxial dSph galaxy (cuspy profile), for three l.o.s. orientations.}
\label{fig:d_triax}
\end{figure}

\paragraph*{DM halo triaxiality}
\begin{itemize}
\item{\textit{Projection effects}}. For a triaxial halo, the
  l.o.s. orientation with respect to the principal axes of the halo
  plays a role when computing the true $D$-factor, just as for the
  $J$-factor. The same $30\%$ difference (shown in
  Fig.~\ref{fig:true_j_triax} for $J$, not repeated here) appears at very low
  integration angles depending whether the l.o.s. is aligned with the short or the long axis of the halo.
\item{\textit{Triaxiality-induced bias}}. Figure
  \ref{fig:d_triax} shows the $D$-factors reconstructed for a mock
  classical (cuspy) dSph galaxy, with the short (blue circles), medium
  (red triangles) and long axis (green squares) aligned along the line
  of sight. They are compared to the true value in black solid line
  (for the intermediate axis l.o.s. orientation). The same systematic
  shift as observed for the $J$-factors (Fig. \ref{fig:j_triaxial})
  appears between the three $D$-factors, due to projection-induced
  effects of the velocity dispersion profiles
  (Fig. \ref{fig:dispproftriax}). Factors of $\sim 2$ systematic
  uncertainties must then be accounted for in the $D$-factor
  determination.
\end{itemize}

\label{lastpage}
%________________________________________________________________________
\bibliography{dsphs_physics}

\begin{thebibliography}{}

\bibitem[\protect\citeauthoryear{{Abramowski} et~al.,}{{Abramowski}
  et~al.}{2011}]{2011PhRvL.106p1301A}
{Abramowski} A.,  et~al., 2011, Physical Review Letters, 106, 161301

\bibitem[\protect\citeauthoryear{{Acciari}, {Arlen}, {Aune} \& {et
  al.}}{{Acciari} et~al.}{2010}]{2010ApJ...720.1174A}
{Acciari} V.~A.,  {Arlen} T.,  {Aune} T.,    {et al.} 2010, \apj, 720, 1174

\bibitem[\protect\citeauthoryear{{Ackermann} \& {Fermi-LAT
  Collaboration}}{{Ackermann} \& {Fermi-LAT
  Collaboration}}{2011}]{2011PhRvL.107x1302A}
{Ackermann} {Fermi-LAT Collaboration} 2011, Physical Review Letters, 107,
  241302

\bibitem[\protect\citeauthoryear{{Ackermann} \& {Fermi-LAT
  Collaboration}}{{Ackermann} \& {Fermi-LAT
  Collaboration}}{2014}]{2014PhRvD..89d2001A}
{Ackermann} {Fermi-LAT Collaboration} 2014, \prd, 89, 042001

\bibitem[\protect\citeauthoryear{{Ackermann} \& {Fermi LAT
  Collaboration}}{{Ackermann} \& {Fermi LAT
  Collaboration}}{2010}]{2010JCAP...05..025A}
{Ackermann} M.,  {Fermi LAT Collaboration} 2010, \jcap, 5, 25

\bibitem[\protect\citeauthoryear{{Ackermann} \& {Fermi LAT
  Collaboration}}{{Ackermann} \& {Fermi LAT
  Collaboration}}{2012}]{2012ApJ...747..121A}
{Ackermann} M.,  {Fermi LAT Collaboration} 2012, \apj, 747, 121

\bibitem[\protect\citeauthoryear{{Agnello} \& {Evans}}{{Agnello} \&
  {Evans}}{2012}]{2012ApJ...754L..39A}
{Agnello} A.,  {Evans} N.~W.,  2012, \apjl, 754, L39

\bibitem[\protect\citeauthoryear{{Aliu} et~al.,}{{Aliu}
  et~al.}{2012}]{2012PhRvD..85f2001A}
{Aliu} E.,  et~al., 2012, \prd, 85, 062001

\bibitem[\protect\citeauthoryear{{Amorisco} \& {Evans}}{{Amorisco} \&
  {Evans}}{2011}]{2011MNRAS.411.2118A}
{Amorisco} N.~C.,  {Evans} N.~W.,  2011, \mnras, 411, 2118

\bibitem[\protect\citeauthoryear{{Amorisco} \& {Evans}}{{Amorisco} \&
  {Evans}}{2012}]{2012MNRAS.419..184A}
{Amorisco} N.~C.,  {Evans} N.~W.,  2012, \mnras, 419, 184

\bibitem[\protect\citeauthoryear{{An} \& {Evans}}{{An} \&
  {Evans}}{2006}]{2006ApJ...642..752A}
{An} J.~H.,  {Evans} N.~W.,  2006, \apj, 642, 752

\bibitem[\protect\citeauthoryear{{Arlen}, {Aune}, {Beilicke} et~al.,}{{Arlen}
  et~al.}{2012}]{2012ApJ...757..123A}
{Arlen} T.,  {Aune} T.,  {Beilicke}   et~al., 2012, \apj, 757, 123

\bibitem[\protect\citeauthoryear{{Baes} \& {van Hese}}{{Baes} \& {van
  Hese}}{2007}]{2007A&A...471..419B}
{Baes} M.,  {van Hese} E.,  2007, \aap, 471, 419

\bibitem[\protect\citeauthoryear{{Bailin} \& {Steinmetz}}{{Bailin} \&
  {Steinmetz}}{2005}]{2005ApJ...627..647B}
{Bailin} J.,  {Steinmetz} M.,  2005, \apj, 627, 647

\bibitem[\protect\citeauthoryear{{Battaglia}, {Helmi} \&
  {Breddels}}{{Battaglia} et~al.}{2013}]{2013NewAR..57...52B}
{Battaglia} G.,  {Helmi} A.,    {Breddels} M.,  2013, \nar, 57, 52

\bibitem[\protect\citeauthoryear{{Bergstr{\"o}m} \& {Hooper}}{{Bergstr{\"o}m}
  \& {Hooper}}{2006}]{2006PhRvD..73f3510B}
{Bergstr{\"o}m} L.,  {Hooper} D.,  2006, \prd, 73, 063510

\bibitem[\protect\citeauthoryear{{Bergstr{\"o}m}, {Ullio} \&
  {Buckley}}{{Bergstr{\"o}m} et~al.}{1998}]{1998APh.....9..137B}
{Bergstr{\"o}m} L.,  {Ullio} P.,    {Buckley} J.~H.,  1998, Astroparticle
  Physics, 9, 137

\bibitem[\protect\citeauthoryear{{Bett}, {Eke}, {Frenk}, {Jenkins}, {Helly} \&
  {Navarro}}{{Bett} et~al.}{2007}]{2007MNRAS.376..215B}
{Bett} P.,  {Eke} V.,  {Frenk} C.~S.,  {Jenkins} A.,  {Helly} J.,    {Navarro}
  J.,  2007, \mnras, 376, 215

\bibitem[\protect\citeauthoryear{{Binney} \& {Tremaine}}{{Binney} \&
  {Tremaine}}{2008}]{2008gady.book.....B}
{Binney} J.,  {Tremaine} S.,  2008, {Galactic Dynamics: Second Edition}.
Princeton University Press

\bibitem[\protect\citeauthoryear{{Bode}, {Ostriker} \& {Turok}}{{Bode}
  et~al.}{2001}]{2001ApJ...556...93B}
{Bode} P.,  {Ostriker} J.~P.,    {Turok} N.,  2001, \apj, 556, 93

\bibitem[\protect\citeauthoryear{{Boyarsky}, {Iakubovskyi}, {Ruchayskiy} \&
  {Savchenko}}{{Boyarsky} et~al.}{2008}]{2008MNRAS.387.1361B}
{Boyarsky} A.,  {Iakubovskyi} D.,  {Ruchayskiy} O.,    {Savchenko} V.,  2008,
  \mnras, 387, 1361

\bibitem[\protect\citeauthoryear{{Boyarsky}, {Ruchayskiy}, {Iakubovskyi} \&
  {Franse}}{{Boyarsky} et~al.}{2014}]{2014arXiv1402.4119B}
{Boyarsky} A.,  {Ruchayskiy} O.,  {Iakubovskyi} D.,    {Franse} J.,  2014,
  ArXiv:1402.4119

\bibitem[\protect\citeauthoryear{{Breddels}, {Helmi}, {van den Bosch}, {van de
  Ven} \& {Battaglia}}{{Breddels} et~al.}{2013}]{2013MNRAS.433.3173B}
{Breddels} M.~A.,  {Helmi} A.,  {van den Bosch} R.~C.~E.,  {van de Ven} G.,
  {Battaglia} G.,  2013, \mnras, 433, 3173

\bibitem[\protect\citeauthoryear{{Bringmann}, {Doro} \& {Fornasa}}{{Bringmann}
  et~al.}{2009}]{2009JCAP...01..016B}
{Bringmann} T.,  {Doro} M.,    {Fornasa} M.,  2009, Journal of Cosmology and
  Astro-Particle Physics, 1, 16

\bibitem[\protect\citeauthoryear{{Brooks} \& {Zolotov}}{{Brooks} \&
  {Zolotov}}{2014}]{2014ApJ...786...87B}
{Brooks} A.~M.,  {Zolotov} A.,  2014, \apj, 786, 87

\bibitem[\protect\citeauthoryear{{Bulbul}, {Markevitch}, {Foster}, {Smith},
  {Loewenstein} \& {Randall}}{{Bulbul} et~al.}{2014}]{2014ApJ...789...13B}
{Bulbul} E.,  {Markevitch} M.,  {Foster} A.,  {Smith} R.~K.,  {Loewenstein} M.,
     {Randall} S.~W.,  2014, \apj, 789, 13

\bibitem[\protect\citeauthoryear{{Cappellari}}{{Cappellari}}{2008}]{2008MNRAS.390...71C}
{Cappellari} M.,  2008, \mnras, 390, 71

\bibitem[\protect\citeauthoryear{{Charbonnier}, {Combet}, {Daniel}, {Funk},
  {Hinton}, {Maurin}, {Power}, {Read}, {Sarkar}, {Walker} \&
  {Wilkinson}}{{Charbonnier} et~al.}{2011}]{2011MNRAS.418.1526C}
{Charbonnier} A.,  {Combet} C.,  {Daniel} M.,  {Funk} S.,  {Hinton} J.~A.,
  {Maurin} D.,  {Power} C.,  {Read} J.~I.,  {Sarkar} S.,  {Walker} M.~G.,
  {Wilkinson} M.~I.,  2011, \mnras, 418, 1526

\bibitem[\protect\citeauthoryear{{Charbonnier}, {Combet} \&
  {Maurin}}{{Charbonnier} et~al.}{2012}]{2012CoPhC.183..656C}
{Charbonnier} A.,  {Combet} C.,    {Maurin} D.,  2012, Computer Physics
  Communications, 183, 656

\bibitem[\protect\citeauthoryear{{Ciotti} \& {Morganti}}{{Ciotti} \&
  {Morganti}}{2010}]{2010MNRAS.408.1070C}
{Ciotti} L.,  {Morganti} L.,  2010, \mnras, 408, 1070

\bibitem[\protect\citeauthoryear{{Cirelli}, {Moulin}, {Panci}, {Serpico} \&
  {Viana}}{{Cirelli} et~al.}{2012}]{2012PhRvD..86h3506C}
{Cirelli} M.,  {Moulin} E.,  {Panci} P.,  {Serpico} P.~D.,    {Viana} A.,
  2012, \prd, 86, 083506

\bibitem[\protect\citeauthoryear{{Cirelli}, {Panci} \& {Serpico}}{{Cirelli}
  et~al.}{2010}]{2010NuPhB.840..284C}
{Cirelli} M.,  {Panci} P.,    {Serpico} P.~D.,  2010, Nuclear Physics B, 840,
  284

\bibitem[\protect\citeauthoryear{{Combet}, {Maurin}, {Nezri}, {Pointecouteau},
  {Hinton} \& {White}}{{Combet} et~al.}{2012}]{2012PhRvD..85f3517C}
{Combet} C.,  {Maurin} D.,  {Nezri} E.,  {Pointecouteau} E.,  {Hinton} J.~A.,
   {White} R.,  2012, \prd, 85, 063517

\bibitem[\protect\citeauthoryear{{Cuddeford}}{{Cuddeford}}{1991}]{1991MNRAS.253..414C}
{Cuddeford} P.,  1991, \mnras, 253, 414

\bibitem[\protect\citeauthoryear{{Daylan}, {Finkbeiner}, {Hooper}, {Linden},
  {Portillo}, {Rodd} \& {Slatyer}}{{Daylan} et~al.}{2014}]{2014arXiv1402.6703D}
{Daylan} T.,  {Finkbeiner} D.~P.,  {Hooper} D.,  {Linden} T.,  {Portillo}
  S.~K.~N.,  {Rodd} N.~L.,    {Slatyer} T.~R.,  2014, ArXiv e-prints

\bibitem[\protect\citeauthoryear{{Dehnen}}{{Dehnen}}{2009}]{2009MNRAS.395.1079D}
{Dehnen} W.,  2009, \mnras, 395, 1079

\bibitem[\protect\citeauthoryear{{Diemand}, {Kuhlen} \& {Madau}}{{Diemand}
  et~al.}{2007}]{2007ApJ...657..262D}
{Diemand} J.,  {Kuhlen} M.,    {Madau} P.,  2007, \apj, 657, 262

\bibitem[\protect\citeauthoryear{{Diemand}, {Moore} \& {Stadel}}{{Diemand}
  et~al.}{2004}]{2004MNRAS.352..535D}
{Diemand} J.,  {Moore} B.,    {Stadel} J.,  2004, \mnras, 352, 535

\bibitem[\protect\citeauthoryear{{Dugger}, {Jeltema} \& {Profumo}}{{Dugger}
  et~al.}{2010}]{2010JCAP...12..015D}
{Dugger} L.,  {Jeltema} T.~E.,    {Profumo} S.,  2010, \jcap, 12, 15

\bibitem[\protect\citeauthoryear{{Essig}, {Sehgal} \& {Strigari}}{{Essig}
  et~al.}{2009}]{2009PhRvD..80b3506E}
{Essig} R.,  {Sehgal} N.,    {Strigari} L.~E.,  2009, \prd, 80, 023506

\bibitem[\protect\citeauthoryear{{Essig}, {Sehgal}, {Strigari}, {Geha} \&
  {Simon}}{{Essig} et~al.}{2010}]{2010PhRvD..82l3503E}
{Essig} R.,  {Sehgal} N.,  {Strigari} L.~E.,  {Geha} M.,    {Simon} J.~D.,
  2010, \prd, 82, 123503

\bibitem[\protect\citeauthoryear{{Evans}, {An} \& {Walker}}{{Evans}
  et~al.}{2009}]{2009MNRAS.393L..50E}
{Evans} N.~W.,  {An} J.,    {Walker} M.~G.,  2009, \mnras, 393, L50

\bibitem[\protect\citeauthoryear{{Evans}, {Ferrer} \& {Sarkar}}{{Evans}
  et~al.}{2004}]{2004PhRvD..69l3501E}
{Evans} N.~W.,  {Ferrer} F.,    {Sarkar} S.,  2004, \prd, 69, 123501

\bibitem[\protect\citeauthoryear{{Feng}}{{Feng}}{2010}]{2010ARA&A..48..495F}
{Feng} J.~L.,  2010, \araa, 48, 495

\bibitem[\protect\citeauthoryear{{Frenk}, {White}, {Davis} \&
  {Efstathiou}}{{Frenk} et~al.}{1988}]{1988ApJ...327..507F}
{Frenk} C.~S.,  {White} S.~D.~M.,  {Davis} M.,    {Efstathiou} G.,  1988, \apj,
  327, 507

\bibitem[\protect\citeauthoryear{{Geringer-Sameth} \&
  {Koushiappas}}{{Geringer-Sameth} \&
  {Koushiappas}}{2011}]{2011PhRvL.107x1303G}
{Geringer-Sameth} A.,  {Koushiappas} S.~M.,  2011, Physical Review Letters,
  107, 241303

\bibitem[\protect\citeauthoryear{{Governato}, {Brook}, {Mayer}, {Brooks},
  {Rhee}, {Wadsley}, {Jonsson}, {Willman}, {Stinson}, {Quinn} \&
  {Madau}}{{Governato} et~al.}{2010}]{2010Natur.463..203G}
{Governato} F.,  {Brook} C.,  {Mayer} L.,  {Brooks} A.,  {Rhee} G.,  {Wadsley}
  J.,  {Jonsson} P.,  {Willman} B.,  {Stinson} G.,  {Quinn} T.,    {Madau} P.,
  2010, \nat, 463, 203

\bibitem[\protect\citeauthoryear{{Hastings}}{{Hastings}}{1970}]{hastings70}
{Hastings} W.~K.,  1970, \textit{Biometrika}, 57, 97

\bibitem[\protect\citeauthoryear{{Hayashi} \& {Chiba}}{{Hayashi} \&
  {Chiba}}{2012}]{2012ApJ...755..145H}
{Hayashi} K.,  {Chiba} M.,  2012, \apj, 755, 145

\bibitem[\protect\citeauthoryear{{Hernquist}}{{Hernquist}}{1990}]{1990ApJ...356..359H}
{Hernquist} L.,  1990, \apj, 356, 359

\bibitem[\protect\citeauthoryear{{Hooper} \& {Goodenough}}{{Hooper} \&
  {Goodenough}}{2011}]{2011PhLB..697..412H}
{Hooper} D.,  {Goodenough} L.,  2011, Physics Letters B, 697, 412

\bibitem[\protect\citeauthoryear{{Ibarra}, {Tran} \& {Weniger}}{{Ibarra}
  et~al.}{2013}]{2013IJMPA..2830040I}
{Ibarra} A.,  {Tran} D.,    {Weniger} C.,  2013, International Journal of
  Modern Physics A, 28, 30040

\bibitem[\protect\citeauthoryear{{Irwin} \& {Hatzidimitriou}}{{Irwin} \&
  {Hatzidimitriou}}{1995}]{1995MNRAS.277.1354I}
{Irwin} M.,  {Hatzidimitriou} D.,  1995, \mnras, 277, 1354

\bibitem[\protect\citeauthoryear{{Jardel} \& {Gebhardt}}{{Jardel} \&
  {Gebhardt}}{2012}]{2012ApJ...746...89J}
{Jardel} J.~R.,  {Gebhardt} K.,  2012, \apj, 746, 89

\bibitem[\protect\citeauthoryear{{King}}{{King}}{1962}]{1962AJ.....67..471K}
{King} I.,  1962, \aj, 67, 471

\bibitem[\protect\citeauthoryear{{Kuhlen}, {Diemand} \& {Madau}}{{Kuhlen}
  et~al.}{2007}]{2007ApJ...671.1135K}
{Kuhlen} M.,  {Diemand} J.,    {Madau} P.,  2007, \apj, 671, 1135

\bibitem[\protect\citeauthoryear{{Lake}}{{Lake}}{1990}]{1990Natur.346...39L}
{Lake} G.,  1990, \nat, 346, 39

\bibitem[\protect\citeauthoryear{{{\L}okas}}{{{\L}okas}}{2001}]{2001MNRAS.327L..21L}
{{\L}okas} E.~L.,  2001, \mnras, 327, L21

\bibitem[\protect\citeauthoryear{{Long} \& {Mao}}{{Long} \&
  {Mao}}{2010}]{2010MNRAS.405..301L}
{Long} R.~J.,  {Mao} S.,  2010, \mnras, 405, 301

\bibitem[\protect\citeauthoryear{{Mamon}, {Biviano} \& {Bou{\'e}}}{{Mamon}
  et~al.}{2013}]{2013MNRAS.429.3079M}
{Mamon} G.~A.,  {Biviano} A.,    {Bou{\'e}} G.,  2013, \mnras, 429, 3079

\bibitem[\protect\citeauthoryear{{Mamon} \& {{\L}okas}}{{Mamon} \&
  {{\L}okas}}{2005}]{2005MNRAS.363..705M}
{Mamon} G.~A.,  {{\L}okas} E.~L.,  2005, \mnras, 363, 705

\bibitem[\protect\citeauthoryear{{Mamon} \& {{\L}okas}}{{Mamon} \&
  {{\L}okas}}{2006}]{2006MNRAS.370.1582M}
{Mamon} G.~A.,  {{\L}okas} E.~L.,  2006, \mnras, 370, 1582

\bibitem[\protect\citeauthoryear{{Martin}, {de Jong} \& {Rix}}{{Martin}
  et~al.}{2008}]{2008ApJ...684.1075M}
{Martin} N.~F.,  {de Jong} J.~T.~A.,    {Rix} H.,  2008, \apj, 684, 1075

\bibitem[\protect\citeauthoryear{{Martinez}}{{Martinez}}{2013}]{2013arXiv1309.2641M}
{Martinez} G.~D.,  2013, ArXiv:1309.2641

\bibitem[\protect\citeauthoryear{{Martinez}, {Bullock}, {Kaplinghat},
  {Strigari} \& {Trotta}}{{Martinez} et~al.}{2009}]{2009JCAP...06..014M}
{Martinez} G.~D.,  {Bullock} J.~S.,  {Kaplinghat} M.,  {Strigari} L.~E.,
  {Trotta} R.,  2009, Journal of Cosmology and Astro-Particle Physics, 6, 14

\bibitem[\protect\citeauthoryear{{Mateo}}{{Mateo}}{1998}]{1998ARA&A..36..435M}
{Mateo} M.~L.,  1998, \araa, 36, 435

\bibitem[\protect\citeauthoryear{{Maurin}, {Combet}, {Nezri} \&
  {Pointecouteau}}{{Maurin} et~al.}{2012}]{2012A&A...547A..16M}
{Maurin} D.,  {Combet} C.,  {Nezri} E.,    {Pointecouteau} E.,  2012, \aap,
  547, A16

\bibitem[\protect\citeauthoryear{{Merritt}}{{Merritt}}{1985}]{1985AJ.....90.1027M}
{Merritt} D.,  1985, \aj, 90, 1027

\bibitem[\protect\citeauthoryear{{Merritt}, {Graham}, {Moore}, {Diemand} \&
  {Terzi{\'c}}}{{Merritt} et~al.}{2006}]{2006AJ....132.2685M}
{Merritt} D.,  {Graham} A.~W.,  {Moore} B.,  {Diemand} J.,    {Terzi{\'c}} B.,
  2006, \aj, 132, 2685

\bibitem[\protect\citeauthoryear{{Metropolis}, {Rosenbluth}, {Rosenbluth},
  {Teller} \& {Teller}}{{Metropolis} et~al.}{1953}]{1953JChPh..21.1087M}
{Metropolis} N.,  {Rosenbluth} A.~W.,  {Rosenbluth} M.~N.,  {Teller} A.~H.,
  {Teller} E.,  1953, \jcp, 21, 1087

\bibitem[\protect\citeauthoryear{{Mollitor}, {Nezri} \& {Teyssier}}{{Mollitor}
  et~al.}{2014}]{2014arXiv1405.4318M}
{Mollitor} P.,  {Nezri} E.,    {Teyssier} R.,  2014, ArXiv:1405.4318

\bibitem[\protect\citeauthoryear{{Mu{\~n}oz-Cuartas}, {Macci{\`o}},
  {Gottl{\"o}ber} \& {Dutton}}{{Mu{\~n}oz-Cuartas}
  et~al.}{2011}]{2011MNRAS.411..584M}
{Mu{\~n}oz-Cuartas} J.~C.,  {Macci{\`o}} A.~V.,  {Gottl{\"o}ber} S.,
  {Dutton} A.~A.,  2011, \mnras, 411, 584

\bibitem[\protect\citeauthoryear{{Navarro}, {Hayashi}, {Power}, {Jenkins},
  {Frenk}, {White}, {Springel}, {Stadel} \& {Quinn}}{{Navarro}
  et~al.}{2004}]{2004MNRAS.349.1039N}
{Navarro} J.~F.,  {Hayashi} E.,  {Power} C.,  {Jenkins} A.~R.,  {Frenk} C.~S.,
  {White} S.~D.~M.,  {Springel} V.,  {Stadel} J.,    {Quinn} T.~R.,  2004,
  \mnras, 349, 1039

\bibitem[\protect\citeauthoryear{Neal}{Neal}{1993}]{Neal93}
Neal R.~M.,  1993, Probabilistic Inference Using {Markov Chain Monte Carlo}
  Methods, \url{http://cosmologist.info/Neal93}

\bibitem[\protect\citeauthoryear{{Nezri}, {White}, {Combet}, {Hinton}, {Maurin}
  \& {Pointecouteau}}{{Nezri} et~al.}{2012}]{2012MNRAS.425..477N}
{Nezri} E.,  {White} R.,  {Combet} C.,  {Hinton} J.~A.,  {Maurin} D.,
  {Pointecouteau} E.,  2012, \mnras, 425, 477

\bibitem[\protect\citeauthoryear{{Nieto}, {Aleksi{\'c}}, {Barrio}, {Contreras},
  {Doro}, {Lombardi}, {Mirabal}, {Moralejo}, {Pardo}, {Rico}, {Zandanel} \&
  {for the MAGIC Collaboration}}{{Nieto} et~al.}{2011}]{2011arXiv1109.5935N}
{Nieto} D.,  {Aleksi{\'c}} J.,  {Barrio} J.~A.,  {Contreras} J.~L.,  {Doro} M.,
   {Lombardi} S.,  {Mirabal} N.,  {Moralejo} A.,  {Pardo} S.,  {Rico} J.,
  {Zandanel} F.,    {for the MAGIC Collaboration} 2011, arXiv:1109.5935

\bibitem[\protect\citeauthoryear{{Osipkov}}{{Osipkov}}{1979}]{1979PAZh....5...77O}
{Osipkov} L.~P.,  1979, Pisma v Astronomicheskii Zhurnal, 5, 77

\bibitem[\protect\citeauthoryear{{Palomares-Ruiz} \&
  {Siegal-Gaskins}}{{Palomares-Ruiz} \&
  {Siegal-Gaskins}}{2010}]{2010JCAP...07..023P}
{Palomares-Ruiz} S.,  {Siegal-Gaskins} J.~M.,  2010, \jcap, 7, 23

\bibitem[\protect\citeauthoryear{Pieri et~al.,}{Pieri
  et~al.}{2009}]{2009A&A...496..351P}
Pieri L.,  et~al., 2009, \aap, 496, 351

\bibitem[\protect\citeauthoryear{{Pieri}, {Lattanzi} \& {Silk}}{{Pieri}
  et~al.}{2009}]{2009MNRAS.399.2033P}
{Pieri} L.,  {Lattanzi} M.,    {Silk} J.,  2009, \mnras, 399, 2033

\bibitem[\protect\citeauthoryear{{Plummer}}{{Plummer}}{1911}]{1911MNRAS..71..460P}
{Plummer} H.~C.,  1911, \mnras, 71, 460

\bibitem[\protect\citeauthoryear{{Pontzen} \& {Governato}}{{Pontzen} \&
  {Governato}}{2012}]{2012MNRAS.421.3464P}
{Pontzen} A.,  {Governato} F.,  2012, \mnras, 421, 3464

\bibitem[\protect\citeauthoryear{{Prugniel} \& {Simien}}{{Prugniel} \&
  {Simien}}{1997}]{1997A&A...321..111P}
{Prugniel} P.,  {Simien} F.,  1997, \aap, 321, 111

\bibitem[\protect\citeauthoryear{{Putze}}{{Putze}}{2011}]{2011ICRC....6..260P}
{Putze} A.,  2011, International Cosmic Ray Conference, 6, 260

\bibitem[\protect\citeauthoryear{Putze \& Derome}{Putze \&
  Derome}{2014}]{Putze:2014aba}
Putze A.,  Derome L.,  2014, Phys.Dark Univ.

\bibitem[\protect\citeauthoryear{{Putze}, {Derome} \& {Maurin}}{{Putze}
  et~al.}{2010}]{2010A&A...516A..66P}
{Putze} A.,  {Derome} L.,    {Maurin} D.,  2010, \aap, 516, A66

\bibitem[\protect\citeauthoryear{{Putze}, {Derome}, {Maurin}, {Perotto} \&
  {Taillet}}{{Putze} et~al.}{2009}]{2009A&A...497..991P}
{Putze} A.,  {Derome} L.,  {Maurin} D.,  {Perotto} L.,    {Taillet} R.,  2009,
  \aap, 497, 991

\bibitem[\protect\citeauthoryear{{Putze}, {Maurin} \& {Donato}}{{Putze}
  et~al.}{2011}]{2011A&A...526A.101P}
{Putze} A.,  {Maurin} D.,    {Donato} F.,  2011, \aap, 526, A101

\bibitem[\protect\citeauthoryear{{Richardson} \& {Fairbairn}}{{Richardson} \&
  {Fairbairn}}{2013}]{2013MNRAS.432.3361R}
{Richardson} T.,  {Fairbairn} M.,  2013, \mnras, 432, 3361

\bibitem[\protect\citeauthoryear{{Richardson} \& {Fairbairn}}{{Richardson} \&
  {Fairbairn}}{2014}]{2014MNRAS.441.1584R}
{Richardson} T.,  {Fairbairn} M.,  2014, \mnras, 441, 1584

\bibitem[\protect\citeauthoryear{{Richardson}, {Spolyar} \&
  {Lehnert}}{{Richardson} et~al.}{2014}]{2014MNRAS.440.1680R}
{Richardson} T.~D.,  {Spolyar} D.,    {Lehnert} M.~D.,  2014, \mnras, 440, 1680

\bibitem[\protect\citeauthoryear{{S{\'a}nchez-Conde} \&
  {Prada}}{{S{\'a}nchez-Conde} \& {Prada}}{2014}]{2014MNRAS.442.2271S}
{S{\'a}nchez-Conde} M.~A.,  {Prada} F.,  2014, \mnras, 442, 2271

\bibitem[\protect\citeauthoryear{{S{\'a}nchez-Conde}, {Prada}, {{\L}okas},
  {G{\'o}mez}, {Wojtak} \& {Moles}}{{S{\'a}nchez-Conde}
  et~al.}{2007}]{2007PhRvD..76l3509S}
{S{\'a}nchez-Conde} M.~A.,  {Prada} F.,  {{\L}okas} E.~L.,  {G{\'o}mez} M.~E.,
  {Wojtak} R.,    {Moles} M.,  2007, \prd, 76, 123509

\bibitem[\protect\citeauthoryear{{Schneider}, {Frenk} \& {Cole}}{{Schneider}
  et~al.}{2012}]{2012JCAP...05..030S}
{Schneider} M.~D.,  {Frenk} C.~S.,    {Cole} S.,  2012, \jcap, 5, 30

\bibitem[\protect\citeauthoryear{{Schwarzschild}}{{Schwarzschild}}{1979}]{1979ApJ...232..236S}
{Schwarzschild} M.,  1979, \apj, 232, 236

\bibitem[\protect\citeauthoryear{{Sersic}}{{Sersic}}{1968}]{1968adga.book.....S}
{Sersic} J.~L.,  1968, {Atlas de galaxias australes}.
Observatorio astronomico, Universidad de Cordoba

\bibitem[\protect\citeauthoryear{{Spergel} \& {Steinhardt}}{{Spergel} \&
  {Steinhardt}}{2000}]{2000PhRvL..84.3760S}
{Spergel} D.~N.,  {Steinhardt} P.~J.,  2000, Physical Review Letters, 84, 3760

\bibitem[\protect\citeauthoryear{{Springel}, {Wang}, {Vogelsberger}, {Ludlow},
  {Jenkins}, {Helmi}, {Navarro}, {Frenk} \& {White}}{{Springel}
  et~al.}{2008}]{2008MNRAS.391.1685S}
{Springel} V.,  {Wang} J.,  {Vogelsberger} M.,  {Ludlow} A.,  {Jenkins} A.,
  {Helmi} A.,  {Navarro} J.~F.,  {Frenk} C.~S.,    {White} S.~D.~M.,  2008,
  \mnras, 391, 1685

\bibitem[\protect\citeauthoryear{{Strigari}}{{Strigari}}{2013}]{2013PhR...531....1S}
{Strigari} L.~E.,  2013, \physrep, 531, 1

\bibitem[\protect\citeauthoryear{{Strigari}, {Bullock}, {Kaplinghat},
  {Kravtsov}, {Gnedin}, {Abazajian} \& {Klypin}}{{Strigari}
  et~al.}{2006}]{2006ApJ...652..306S}
{Strigari} L.~E.,  {Bullock} J.~S.,  {Kaplinghat} M.,  {Kravtsov} A.~V.,
  {Gnedin} O.~Y.,  {Abazajian} K.,    {Klypin} A.~A.,  2006, \apj, 652, 306

\bibitem[\protect\citeauthoryear{{Strigari}, {Bullock}, {Kaplinghat}, {Simon},
  {Geha}, {Willman} \& {Walker}}{{Strigari} et~al.}{2008}]{2008Natur.454.1096S}
{Strigari} L.~E.,  {Bullock} J.~S.,  {Kaplinghat} M.,  {Simon} J.~D.,  {Geha}
  M.,  {Willman} B.,    {Walker} M.~G.,  2008, \nat, 454, 1096

\bibitem[\protect\citeauthoryear{{Strigari}, {Koushiappas}, {Bullock} \&
  {Kaplinghat}}{{Strigari} et~al.}{2007}]{2007PhRvD..75h3526S}
{Strigari} L.~E.,  {Koushiappas} S.~M.,  {Bullock} J.~S.,    {Kaplinghat} M.,
  2007, \prd, 75, 083526

\bibitem[\protect\citeauthoryear{{Strigari}, {Koushiappas}, {Bullock},
  {Kaplinghat}, {Simon}, {Geha} \& {Willman}}{{Strigari}
  et~al.}{2008}]{2008ApJ...678..614S}
{Strigari} L.~E.,  {Koushiappas} S.~M.,  {Bullock} J.~S.,  {Kaplinghat} M.,
  {Simon} J.~D.,  {Geha} M.,    {Willman} B.,  2008, \apj, 678, 614

\bibitem[\protect\citeauthoryear{{Syer} \& {Tremaine}}{{Syer} \&
  {Tremaine}}{1996}]{1996MNRAS.282..223S}
{Syer} D.,  {Tremaine} S.,  1996, \mnras, 282, 223

\bibitem[\protect\citeauthoryear{{Vera-Ciro}, {Sales}, {Helmi} \&
  {Navarro}}{{Vera-Ciro} et~al.}{2014}]{2014MNRAS.439.2863V}
{Vera-Ciro} C.~A.,  {Sales} L.~V.,  {Helmi} A.,    {Navarro} J.~F.,  2014,
  \mnras, 439, 2863

\bibitem[\protect\citeauthoryear{{Vogelsberger} \& {White}}{{Vogelsberger} \&
  {White}}{2011}]{2011MNRAS.413.1419V}
{Vogelsberger} M.,  {White} S.~D.~M.,  2011, \mnras, 413, 1419

\bibitem[\protect\citeauthoryear{{Walker}}{{Walker}}{2013}]{2013pss5.book.1039W}
{Walker} M.,  2013, {Dark Matter in the Galactic Dwarf Spheroidal Satellites}.
{Oswalt}, T.~D. and {Gilmore}, G., p.~1039

\bibitem[\protect\citeauthoryear{{Walker}, {Combet}, {Hinton}, {Maurin} \&
  {Wilkinson}}{{Walker} et~al.}{2011}]{2011ApJ...733L..46W}
{Walker} M.~G.,  {Combet} C.,  {Hinton} J.~A.,  {Maurin} D.,    {Wilkinson}
  M.~I.,  2011, \apjl, 733, L46

\bibitem[\protect\citeauthoryear{{Walker}, {Mateo}, {Olszewski}, {Bernstein},
  {Wang} \& {Woodroofe}}{{Walker} et~al.}{2006}]{2006AJ....131.2114W}
{Walker} M.~G.,  {Mateo} M.,  {Olszewski} E.~W.,  {Bernstein} R.,  {Wang} X.,
   {Woodroofe} M.,  2006, \aj, 131, 2114

\bibitem[\protect\citeauthoryear{{Walker}, {Mateo}, {Olszewski},
  {Pe{\~n}arrubia}, {Wyn Evans} \& {Gilmore}}{{Walker}
  et~al.}{2009}]{2009ApJ...704.1274W}
{Walker} M.~G.,  {Mateo} M.,  {Olszewski} E.~W.,  {Pe{\~n}arrubia} J.,  {Wyn
  Evans} N.,    {Gilmore} G.,  2009, \apj, 704, 1274

\bibitem[\protect\citeauthoryear{{Walker}, {Mateo}, {Olszewski},
  {Pe{\~n}arrubia}, {Wyn Evans} \& {Gilmore}}{{Walker}
  et~al.}{2010}]{2010ApJ...710..886W}
{Walker} M.~G.,  {Mateo} M.,  {Olszewski} E.~W.,  {Pe{\~n}arrubia} J.,  {Wyn
  Evans} N.,    {Gilmore} G.,  2010, \apj, 710, 886

\bibitem[\protect\citeauthoryear{{Walker} \& {Pe{\~n}arrubia}}{{Walker} \&
  {Pe{\~n}arrubia}}{2011}]{2011ApJ...742...20W}
{Walker} M.~G.,  {Pe{\~n}arrubia} J.,  2011, \apj, 742, 20

\bibitem[\protect\citeauthoryear{{Weniger}}{{Weniger}}{2012}]{2012JCAP...08..007W}
{Weniger} C.,  2012, \jcap, 8, 7

\bibitem[\protect\citeauthoryear{{White} \& {Rees}}{{White} \&
  {Rees}}{1978}]{1978MNRAS.183..341W}
{White} S.~D.~M.,  {Rees} M.~J.,  1978, \mnras, 183, 341

\bibitem[\protect\citeauthoryear{{Wolf}, {Martinez}, {Bullock}, {Kaplinghat},
  {Geha}, {Mu{\~n}oz}, {Simon} \& {Avedo}}{{Wolf}
  et~al.}{2010}]{2010MNRAS.406.1220W}
{Wolf} J.,  {Martinez} G.~D.,  {Bullock} J.~S.,  {Kaplinghat} M.,  {Geha} M.,
  {Mu{\~n}oz} R.~R.,  {Simon} J.~D.,    {Avedo} F.~F.,  2010, \mnras, 406, 1220

\bibitem[\protect\citeauthoryear{{Zhao}}{{Zhao}}{1996}]{1996MNRAS.278..488Z}
{Zhao} H.,  1996, \mnras, 278, 488

\end{thebibliography}
\end{document}